\documentstyle[preprint,aps,eqsecnum]{revtex}
\tightenlines
\begin{document}
\draft

\newcommand{\beq}{\begin{equation}}
\newcommand{\eeq}{\end{equation}}
\newcommand{\bea}{\begin{eqnarray}}
\newcommand{\eea}{\end{eqnarray}}
\newcommand{\cir}{{\buildrel \circ \over =}}

\title{A New Parametrization for Tetrad Gravity.}

\author{Luca Lusanna}

\address{
Sezione INFN di Firenze\\
L.go E.Fermi 2 (Arcetri)\\
50125 Firenze, Italy\\
E-mail LUSANNA@FI.INFN.IT}

\author{and}

\author{Stefano Russo}

\address
{Condominio dei Pioppi 16\\
6916 Grancia (Lugano)\\
Svizzera}

\maketitle
\begin{abstract}

A new version of tetrad gravity in globally hyperbolic, asymptotically
flat at spatial infinity spacetimes with Cauchy surfaces diffeomorphic
to $R^3$ is obtained by using a new parametrization of arbitrary
cotetrads to define a set of configurational variables to be used in
the ADM metric action. Seven of the fourteen first class constraints
have the form of the vanishing of canonical momenta. A comparison is
made with other models of tetrad gravity and with the ADM canonical
formalism for metric gravity.


\vskip 1truecm
\noindent \today
\vskip 1truecm

\end{abstract}

\newpage

\vfill\eject

\section
{Introduction}

In this paper we develop a new parametrization of arbitrary cotetrads,
whose use implies a simplified form of some of the constraints of
tetrad gravity. This will open the possibility to restart the study of
the canonical reduction of tetrad gravity. Our motivation is the
attempt to arrive at a unified description of the four interactions
based on Dirac-Bergmann theory of constraints\cite{dirac}, which is
the main tool for the Hamiltonian formulation of both gauge theories
and general relativity. Therefore, we shall study general relativity
from the canonical point of view generalizing to it all the results
already obtained in the canonical study of gauge theories in a
systematic way, since neither a complete reduction of gravity with an
identification of the physical canonical degrees of freedom of the
gravitational field nor a detailed study of its Hamiltonian group of
gauge transformations, whose infinitesimal generators are the first
class constraints, has ever been pushed till the end in an explicit
way.

The research program aiming to express the special relativistic
strong, weak and electromagnetic interactions in terms of Dirac's
observables \cite{dirac,dira} is in an advanced  stage of
development\cite{re}. This program is based on the Shanmugadhasan
canonical transformations \cite{sha}: if a system has first class
constraints at the Hamiltonian level \footnote{So that its dynamics is
restricted to a presymplectic submanifold of phase space.}, then, at
least locally, one can find a canonical basis with as many new momenta
as first class constraints (Abelianization of first class
constraints), with their conjugate canonical variables as Abelianized
gauge variables and with the remaining pairs of canonical variables as
pairs of canonically conjugate Dirac's observables \footnote{Canonical
basis of physical variables adapted to the chosen Abelianization; they
give a trivialization of the BRST construction of observables.}.
Putting equal to zero the Abelianized gauge variables  defines a local
gauge of the model. If a system with constraints admits one (or more)
{\it global} Shanmugadhasan canonical transformations, one obtains one
(or more) privileged {\it global gauges} in which the physical Dirac
observables are globally defined and globally separated from the gauge
degrees of freedom \footnote{For systems with a compact configuration
space this is in general impossible.}. These privileged gauges (when
they exist) can be called {\it generalized Coulomb or radiation
gauges}. Second class constraints, when present, are also taken into
account by the Shanmugadhasan canonical transformation\cite{sha}.

The problem of how to covariantize this kind  of canonical reduction
is solved by using Dirac reformulation (see the book in
Ref.\cite{dirac}) of classical field theory on spacelike hypersurfaces
foliating \footnote{The foliation is defined by an embedding $R\times
\Sigma \rightarrow M^4$, $(\tau ,\vec
\sigma ) \mapsto z^{\mu}(\tau ,\vec \sigma )$, with $\Sigma$ an abstract
3-surface diffeomorphic to $R^3$: this is the classical basis of
Tomonaga-Schwinger quantum field theory.} Minkowski spacetime $M^4$.
In this way one gets parametrized Minkowski field theory with a
covariant 3+1 splitting of flat spacetime and already in a form suited
to the transition to general relativity in its ADM canonical
formulation
\footnote{See also Ref.\cite{kuchar} , where a theoretical study of
this problem is done in curved spacetimes.}. The price is that one has
to add as new configuration variables  the points $z^{\mu}(\tau ,\vec
\sigma )$ of the spacelike hypersurface $\Sigma_{\tau}$ \footnote{The
only ones carrying Lorentz indices; the scalar parameter $\tau$ labels
the leaves of the foliation and $\vec \sigma$ are curvilinear
coordinates on $\Sigma_{\tau}$.} and then define the fields on
$\Sigma_{\tau}$ so that they know  the hypersurface $\Sigma_{\tau}$ of
$\tau$-simultaneity \footnote{For a Klein-Gordon field $\phi (x)$,
this new field is $\tilde \phi (\tau ,\vec \sigma )=\phi (z(\tau ,\vec
\sigma ))$: it contains the non-local information about the
embedding.}. Then one rewrites the Lagrangian of the given isolated
system in the form required by the coupling to an external
gravitational field, makes the previous 3+1 splitting of Minkowski
spacetime and interpretes all the fields of the system as the new
fields on $\Sigma_{\tau}$ (they are Lorentz scalars, having only
surface indices). Instead of considering the 4-metric as describing a
gravitational field \footnote{Therefore as an independent field as it
is done in metric gravity, where one adds the Hilbert action to the
action for the matter fields.}, here one replaces the 4-metric with
the the induced metric $g_{ AB}[z]
=z^{(\mu )}_{A}\eta_{(\mu )(\nu )}z^{(\nu )}_{B}$ on
$\Sigma_{\tau}$ \footnote{A functional of $z^{(\mu )}$; here we use
the notation $\sigma^{A}=(\tau ,\sigma^{r})$; $(\mu )$ is a flat
Minkowski index; $z^{(\mu )}_{A}=
\partial z^{(\mu )}/\partial \sigma^{A}$ are flat cotetrad fields on Minkowski
spacetime with the $z^{(\mu )}_r$'s tangent to $\Sigma_{\tau}$.} and
considers the embedding coordinates $z^{(\mu )}(\tau ,\vec \sigma )$
as independent fields \footnote{This is not possible in metric
gravity, because in curved spacetimes $z^{\mu}_{A}\not= \partial
z^{\mu}/\partial \sigma^{A}$ are not tetrad fields since the holonomic
coordinates $z^{\mu}(\tau ,\vec \sigma )$ do not exist.}. From this
Lagrangian, besides a Lorentz-scalar form of the constraints of the
given system, we get four extra primary first class constraints ${\cal
H}_{\mu}(\tau ,\vec \sigma ) \approx 0$ implying the independence of
the description from the choice of the foliation with spacelike
hypersufaces. Therefore the embedding variables $z^{(\mu )}(\tau ,\vec
\sigma )$ are the {\it gauge} variables associated with this kind of
general covariance. In special relativity, it is convenient to restrict ourselves to
arbitrary spacelike hyperplanes $z^{(\mu )} (\tau ,\vec
\sigma )=x^{(\mu )}_s(\tau )+b^{(\mu )}_{r}(\tau ) \sigma^{r}$. Since
they are described by only 10 variables \footnote{An origin $x^{(\mu
)}_s(\tau )$ and, on it, three orthogonal spacelike unit vectors
$b^{\mu}_r(\tau )$ generating the fixed constant timelike unit normal
to the hyperplane.}, we remain only with 10 first class constraints
determining the 10 variables conjugate to the hyperplane
\footnote{They are a 4-momentum $p^{(\mu )}_s$ and the six independent
degrees of freedom hidden in a spin tensor $S^{(\mu )(\nu )}_s$.} in
terms of the variables of the system.

If we now consider only the set of configurations of the isolated system with
 timelike \footnote{$\epsilon p^2_s > 0$; $\epsilon =\pm 1$ according to the chosen
 convention for the Lorentz signature of the metric $\eta^{(\mu )(\nu )}
 =\epsilon (+---)$.} 4-momenta, we can
restrict the description to the so-called {\it Wigner hyperplanes}
orthogonal to $p^{(\mu )}_s$ itself. To get this result, we must boost
at rest all the variables with Lorentz indices by using the standard
Wigner boost $L^{(\mu )}{}
_{(\nu )}(p_s,{\buildrel \circ \over p}_s)$ for timelike Poincar\'e orbits, and
then add the gauge-fixings $b^{(\mu )}_{\check r}(\tau )-L^{(\mu
)}{}_{\check r}(p_s, {\buildrel \circ \over p}_s)\approx 0$. Since
these gauge-fixings depend on $p^{(\mu )}_s$, the final canonical
variables, apart $p^{(\mu )}_s$ itself, are of 3 types: i) there is a
non-covariant {\it external} center-of-mass variable ${\tilde x}^{(\mu
)} (\tau )$ \footnote{It is only covariant under the little group of
timelike Poincar\'e orbits like the Newton-Wigner position operator.};
ii) all the 3-vector variables become Wigner spin 1 3-vectors
\footnote{Boosts in $M^4$ induce Wigner rotations on them.}; iii) all the
other variables are Lorentz scalars. Only  four 1st class constraints
are left: one of them identifies the invariant mass of the isolated
system, to be used as Hamiltonian, while the other three are the
rest-frame conditions implying the vanishing of the {\it internal}
(i.e. inside the Wigner hyperplane) total 3-momentum.

We obtain in this way a new kind of instant form of the dynamics (see
Ref.\cite{dira2}), the  {\it Wigner-covariant 1-time rest-frame
instant form}\cite{lus1,re} with a universal breaking of Lorentz
covariance independent from the isolated system under investigation.
It is the special relativistic generalization of the non-relativistic
separation of the center of mass from the relative motions [$H={{
{\vec P}^2}\over {2M}}+H_{rel}$].

 As shown in Refs.\cite{lus1,lusa}, the rest-frame instant form
 of dynamics automatically gives a
physical ultraviolet cutoff in the spirit of Dirac and Yukawa: it is
the {\it M$\o$ller radius}\cite{mol} $\rho
=\sqrt{-\epsilon W^2}/\epsilon P^2=|\vec S|/\sqrt{\epsilon P^2}$
\footnote{$W^2=-\epsilon P^2{\vec S}^2$ is the Pauli-Lubanski Casimir
when $\epsilon P^2 > 0$.}, namely the classical intrinsic radius of
the worldtube, around the covariant non-canonical Fokker-Pryce center
of inertia $Y^{(\mu )}$, inside which the non-covariance of the
canonical center of mass ${\tilde x}^{(\mu )}$ is concentrated. At the
quantum level $\rho$ becomes the Compton wavelength of the isolated
system multiplied its spin eigenvalue $\sqrt{s(s+1)}$ , $\rho \mapsto
\hat \rho = \sqrt{s(s+1)} \hbar /M=\sqrt{s(s+1)} \lambda_M$ with
$M=\sqrt{\epsilon P^2}$ the invariant mass and $\lambda_M=\hbar /M$
its Compton wavelength. Therefore, the {\it criticism} to classical
relativistic physics, based on quantum {\it pair production}, concerns
the testing of distances where, due to the Lorentz signature of
spacetime, one has intrinsic classical covariance problems: it is
impossible to localize the canonical center of mass ${\tilde x}^{(\mu
)}$ of the system in a frame independent way. Let us remember
\cite{lus1} that $\rho$ is also a remnant in flat Minkowski spacetime
of the {\it energy conditions} of general relativity: since the
M$\o$ller non-canonical, non-covariant center of energy has its
non-covariance localized inside the same worldtube with radius $\rho$
(it was discovered in this way) \cite{mol}, it turns out that for an
extended relativistic system with the material radius smaller than its
intrinsic radius $\rho$ one has: i) its peripheral rotation velocity
can exceed the velocity of light; ii) its classical energy density
cannot be positive definite everywhere in every frame.

Now, the real relevant point is that this ultraviolet cutoff
determined by $\rho$ exists also in Einstein's general relativity
(which is not power counting renormalizable) in the case of
asymptotically flat spacetimes, taking into account the Poincar\'e
Casimirs of its asymptotic ADM Poincar\'e charges \footnote{When
supertranslations are eliminated with suitable boundary conditions;
let us remark that Einstein and Wheeler use closed universes because
they don't want to introduce boundary conditions, but in this way they
loose Poincar\'e charges and the possibility to make contact  with
particle physics and to define spin.} at spatial infinity. See
Ref.\cite{restmg} for the definition of the rest-frame instant form of
ADM metric gravity.

Moreover, the extended Heisenberg relations  of string
theory\cite{ven}, i.e. $\triangle x ={{\hbar}\over {\triangle
p}}+{{\triangle p}\over {T_{cs}}}= {{\hbar}\over {\triangle
p}}+{{\hbar \triangle p}\over {L^2_{cs}}}$ implying the lower bound
$\triangle x > L_{cs}=\sqrt{\hbar /T_{cs}}$ due to the $y+1/y$
structure, have a counterpart in the quantization of the M$\o$ller
radius\cite{lus1}: if we ask that, also at the quantum level, one
cannot test the inside of the worldtube, we must ask $\triangle x >
\hat \rho$ and this is the lower bound implied by the modified
uncertainty relation $\triangle x ={{\hbar}\over {\triangle
p}}+{{\hbar \triangle p}\over {{\hat \rho}^2}}$. This could imply that
the center-of-mass canonical non-covariant  3-coordinate $\vec
z=\sqrt{\epsilon P^2}({\vec {\tilde x}}-{{\vec P}\over
{P^{(o)}}}{\tilde x}^{(o)})$ \cite{lus1} cannot become a self-adjoint
operator. See Hegerfeldt's theorems (quoted in Refs.\cite{lusa,lus1})
and his interpretation pointing at the impossibility of a good
localization of relativistic particles \footnote{Experimentally one
determines only a worldtube in spacetime emerging from the interaction
region.}. Since the eigenfunctions of the canonical center-of-mass
operator are playing the role of the wave function of the universe,
one could also say that the center-of-mass variable has not to be
quantized, because it lies on the classical macroscopic side of
Copenhagen's interpretation and, moreover, because, in the spirit of
Mach's principle that only relative motions can be observed, no one
can observe it (it is only used to define a decoupled {\it point
particle clock}). On the other hand, if one rejects the canonical
non-covariant center of mass in favor of the covariant non-canonical
Fokker-Pryce center of inertia $Y^{(\mu )}$, $\{ Y^{(\mu )},Y^{(\nu )}
\}
\not= 0$, one could invoke the philosophy of quantum groups to quantize
$Y^{(\mu )}$ to get some kind of quantum plane for the center-of-mass
description. Let us remark that the quantization of the square root
Hamiltonian done in Ref.\cite{lam} is consistent with this
problematic.

In conclusion, the best set of canonical coordinates adapted to the
constraints and to the geometry of Poincar\'e orbits and naturally
predisposed to the coupling to canonical tetrad gravity is emerging
for the electromagnetic, weak and strong interactions with matter
described either by fermion fields or by relativistic particles with a
definite sign of the energy. Therefore, we can begin to think how to
quantize the standard model in the Wigner-covariant Coulomb gauge in
the rest-frame instant form with the M\"oller radius as a ultraviolet
cutoff.

Since our aim is to arrive at a unified description of the four
interactions, in this paper we put the basis for the canonical
reduction to Dirac's observables of tetrad gravity (more natural than
metric gravity for the coupling to fermion fields) and for exploring
the connection of Dirac's observables with Bergmann's definition of
observables and the problem of time in general relativity
\cite{be,ish,kuchar1}.

Our approach to tetrad gravity (see Refs.
\cite{weyl,dirr,schw,kib,tetr,char,clay,maluf,hen1,hen2,hen3,hen4} for
the existing versions of the theory) utilizes the ADM action of metric
gravity with the 4-metric expressed in terms of arbitrary cotetrads,
which are parametrized in a particular way in terms of the parameters
of special Wigner boosts\footnote{This is syggested by the rest-frame
Wigner-covariant instant form approach.} and cotetrads adapted to
$\Sigma_{\tau}$\footnote{Which, in turn, depend on cotriads on
$\Sigma_{\tau}$ and on lapse and shift functions.}. The introduction
of this new parametrization of arbitrary cotetrads in the ADM
Lagrangian allows to get a new Lagrangian for tetrad gravity. Then we
study the associated Hamiltonian formulation, we identify its fourteen
first class constraints and we evaluate their Poisson brackets.

We shall restrict ourselves to the simplest class of spacetimes to
have some chance to have a well posed formulation of tetrad gravity,
which hopefully will allow to arrive at the end of the canonical
reduction. Refs. \cite{naka,oneil,blee} are used for the background in
differential geometry. A spacetime is a time-oriented
pseudo-Riemannian (or Lorentzian) 4-manifold $(M^4,{}^4g)$ with
signature $\epsilon \, (+---)$ ($\epsilon =\pm 1$) and with a choice
of time orientation
\footnote{I.e. there exists a continuous, nowhere vanishing timelike
vector field which is used to separate the non-spacelike vectors at
each point of $M^4$ in either future- or past-directed vectors.}. In
Appendix A we give a review of notions on 4-dimensional
pseudo-Riemannian manifolds, tetrads on them and triads on
3-manifolds, which unifies many results, scattered in the literature,
needed not only for a well posed formulation of tetrad gravity but
also for the further study of its canonical reduction. Also a review
of the action principles used in metric and tetrad gravity is given in
Appendix A for completeness.

Our spacetimes are assumed to be:

i) {\it Globally hyperbolic} 4-manifolds, i.e. topologically they are
$M^4\approx R\times \Sigma$, so to have a well posed Cauchy problem
(with $\Sigma$ the abstract model of Cauchy surface) at least till
when no singularity develops in $M^4$ (see the singularity theorems).
Therefore, these spacetimes admit regular foliations with orientable,
complete, non-intersecting spacelike 3-manifolds: the leaves of the
foliation are the embeddings $i_{\tau}:\Sigma \rightarrow
\Sigma_{\tau} \subset M^4$, $\vec \sigma \mapsto z^{\mu}(\tau ,\vec \sigma )$,
where $\vec \sigma =\{ \sigma^r \}$, r=1,2,3, are local coordinates in a chart
of the $C^{\infty}$-atlas of the abstract 3-manifold $\Sigma$ and $\tau :M^4
\rightarrow R$, $z^{\mu} \mapsto \tau (z^{\mu})$, is a global timelike
future-oriented function labelling the leaves (surfaces of simultaneity). In
this way, one obtains 3+1 splittings of $M^4$ and the possibility of a
Hamiltonian formulation.

ii) {\it Asymptotically flat at spatial infinity}, so to have the
possibility to define asymptotic Poincar\'e charges
\cite{adm,reg,reg1,reg2,reg3,ash,restmg}: they allow the definition of
a M$\o$ller radius in general relativity and are a bridge towards a
future soldering with the theory of elementary particles in Minkowski
spacetime defined as irreducible representation of its kinematical,
globally implemented Poincar\'e group according to Wigner. We will not
compactify space infinity at a point like in the spi approach of
Ref.\cite{ash}.

iii) Since we want to be able to introduce Dirac fermion fields, our
spacetimes $M^4$ must admit a {\it spinor (or spin)
structure}\cite{wald}. Since we consider non-compact space- and
time-orientable spacetimes, spinors can be defined if and only if they
are {\it parallelizable} \cite{ger}. This means that we have trivial
principal frame bundle $L(M^4)=M^4\times GL(4,R)$ with GL(4,R) as
structure group and trivial orthonormal frame bundle $F(M^4)=M^4\times
SO(3,1)$; the fibers of $F(M^4)$ are the disjoint union of four
components and $F_o(M^4)=M^4\times L^{\uparrow}_{+}$ (with projection
$\pi: F_o(M^4)\rightarrow M^4$) corresponds to the proper subgroup
$L^{\uparrow}_{+}\subset SO(3,1)$ of the Lorentz group. Therefore,
global frames (tetrads) and coframes (cotetrads) exist. A spin
structure for $F_o(M^4)$ is, in this case, the trivial spin principal
SL(2,C)-bundle $S(M^4)=M^4\times SL(2,C)$ (with projection
$\pi_s:S(M^4)\rightarrow M^4$) and a map $\lambda : S(M^4)\rightarrow
F_o(M^4)$ such that $\pi (\lambda (p))=\pi_s(p)\in M^4$ for all $p\in
S(M^4)$ and $\lambda (pA)=\lambda (p) \Lambda (A)$ for all $p\in
S(M^4)$, $A\in SL(2,C)$, with $\Lambda :SL(2,C)\rightarrow
L^{\uparrow}_{+}$ the universal covering homomorphism. Then, Dirac
fields are defined as cross sections of a bundle associated with
$S(M^4)$ \cite{blee}. Since $M^4 \approx R\times
\Sigma$ is time- and space-oriented, the hypersurfaces $\Sigma_{\tau}$ of
simultaneity are necessarily space-oriented and are parallelizable (as every
3-manifold\cite{ger}): therefore, global triads and cotriads exist. $F(\Sigma
_{\tau})=\Sigma_{\tau}\times SO(3)$ is the trivial orthonormal frame
SO(3)-bundle and, since one has $\pi_1(SO(3))=\pi_1(L^{\uparrow}_{+})=Z_2$ for
the first homotopy group, one can
define SU(2) spinors on $\Sigma_{\tau}$ \cite{spinor,spinor1}.

iv) The non-compact parallelizable simultaneity 3-manifolds (the
Cauchy surfaces) $\Sigma_{\tau}$ are assumed to be {\it topologically
trivial, geodesically complete} \footnote{So that the Hopf-Rinow
theorem\cite{oneil} assures metric completeness of the Riemannian
3-manifold $(\Sigma_{\tau},{}^3g)$.} and, finally, {\it diffeomorphic
to $R^3$}. These 3-manifolds have the same manifold structure as
Euclidean spaces \cite{oneil}: a) the geodesic exponential map
$Exp_p:T_p\Sigma_{\tau}\rightarrow \Sigma
_{\tau}$ is a diffeomorphism (Hadamard theorem); b) the sectional curvature is
less or equal  zero everywhere; c) they have no {\it conjugate locus}
\footnote{I.e. there are no pairs of conjugate Jacobi points (intersection
points of distinct geodesics through them) on any geodesic.} and no
{\it cut locus} \footnote{I.e. no closed geodesics through any
point.}. In these manifolds two points determine a line, so that the
{\it static} tidal forces in $\Sigma_{\tau}$ due to the 3-curvature
tensor are repulsive; instead in $M^4$ the tidal forces due to the
4-curvature tensor are attractive, since they describe gravitation,
which is always attractive, and this implies that the sectional
4-curvature of timelike tangent planes must be negative (this is the
source of the singularity theorems) \cite{oneil}. In 3-manifolds not
of this class one has to give a physical (topological) interpretation
of  {\it static} quantities like the two quoted loci. In particular,
these 3-manifolds have {\it global charts} inherited by $R^3$ through
the diffeomorphism. Given a Cauchy surface $\Sigma_{\tau_o}$ of this
type and a set of Cauchy data for the gravitational field (and for
matter, if present), the Hamiltonian evolution we are going to
describe will be valid from $\tau_o$ till $\tau_o+
\triangle \tau$, where the interval $\triangle \tau$ is determined by the
appearance of either conjugate points on $\Sigma_{\tau_o+\triangle \tau}$
or 4-dimensional singularities in $M^4$ on its slice $\Sigma_{\tau_o+\triangle
\tau}$.

v) Like in Yang-Mills case \cite{lusa}, the 3-spin-connection on the
orthogonal frame SO(3)-bundle (and therefore triads and cotriads) will
have to be restricted to {\it suited weighted Sobolev spaces} to avoid
Gribov ambiguities. In turn, this implies the {\it absence of
isometries} of the non-compact Riemannian 3-manifold
$(\Sigma_{\tau},{}^3g)$ (see for instance the review paper in Ref.
\cite{cho}). All the problems of the boundary conditions on lapse and
shift functions and on cotriads will be studied in connection with the
Poincar\'e charges in a future paper see however Ref.\cite{restmg} for
the case of metric gravity).

Diffeomorphisms on $\Sigma_{\tau}$ ($Diff\, \Sigma_{\tau}$) will be
interpreted in the passive way, following Ref.\cite{be}, in accord
with the Hamiltonian point of view that infinitesimal diffeomorphisms
are generated by taking the Poisson bracket with the first class
supermomentum constraints \footnote{Passive diffeomorphisms are also
named {\it pseudo-diffeomorphisms}.}. The Lagrangian approach based on
the Hilbert action, connects general covariance with the invariance of
the action under spacetime diffeomorphisms ($Diff\, M^4$) extended to
4-tensors. Therefore, the moduli space (or superspace or space of
4-geometries) is the space $Riem\, M^4/Diff\, M^4$ \cite{whe}, where
$Riem\, M^4$ is the space of Lorentzian 4-metrics; as shown in
Refs.\cite{fis,ing}, superspace, in general, is not a manifold
\footnote{It is a stratified manifold with singularities\cite{arms}.}
due to the existence (in Sobolev spaces) of 4-metrics and 4-geometries
with isometries. See Ref.\cite{giul} for the study of great
diffeomorphisms, which are connected with the existence of disjoint
components of the diffeomorphism group
\footnote{In Ref.\cite{lusa} there is the analogous discussion of the
connection of winding number with the great gauge transformations.}.
Instead, in the ADM Hamiltonian formulation of metric
gravity\cite{adm} space diffeomorphisms are replaced by $Diff\,
\Sigma_{\tau}$ \footnote{Or better by their induced action on 3-tensors
generated by the supermomentum constraints.}, while time
diffeomorphisms are distorted to the transformations generated by the
superhamiltonian 1st class constraint\cite{wa,ish,beig} and by the
momenta conjugate to the lapse and shift functions. In the
Lichnerowicz-York conformal approach to canonical reduction
\cite{conf,york} (see Refs.\cite{cho,yoyo,ciuf} for reviews), one
defines, in the case of closed 3-manifolds, the conformal superspace
as the space of conformal 3-geometries \footnote{Namely the space of
conformal 3-metrics modulo $Diff\,
\Sigma_{\tau}$ or, equivalently, as $Riem\, \Sigma_{\tau}$ (the space of
Riemannian 3-metrics) modulo $Diff\, \Sigma_{\tau}$ and conformal
transformations ${}^3g \mapsto \phi^4\, {}^3g$ ($\phi > 0$).}, because
in this approach gravitational dynamics is regarded as the time
evolution of conformal 3-geometry \footnote{The momentum conjugate to
the conformal factor $\phi$ is replaced by York time
\cite{york,qadir}, i.e.  the trace of the extrinsic curvature of
$\Sigma_{\tau}$.}. See Ref.\cite{restmg} for  the interpretation of
the gauge transformations generated by the superhamiltonian
constraint: they perform the transition from an allowed 3+1 splitting
of spacetime to another one so that the theory is independent from its
choice like it happens in parametrized Minkowski theories. Moreover,
the Hamiltonian group of gauge transformations of the ADM theory has 8
(and not 4) generators, because, besides the superhamiltonian and
supermomentum constraints, there are the four primary first class
constraints giving the vanishing of the canonical momenta conjugate to
the lapse and shift functions \footnote{Whose gauge nature is
connected with the gauge nature (conventionality) of simultaneity
\cite{simul} and of the standards of time and length.}. A preliminary
discussion of these problems and of general covariance versus Dirac's
observables has been given in Ref.\cite{russo2}
\footnote{As also recently noted in Ref.\cite{but} the problem of
observables is still open in canonical gravity.}.

The same happens in tetrad gravity, where there are 14 first class
constraints. As we shall see, in our formulation the Hamiltonian gauge
group contains: i) a $R^3\times SO(3)$ subgroup replacing the usual
Lorentz subgroup due to our parametrization which Abelianizes Lorentz
boosts; ii) $Diff\,
\Sigma_{\tau}$ in the sense of the pseudo-diffeomorphisms generated by the
supermomentum constraints; iii) the gauge transformations generated by
a superhamiltonian 1st class constraint; iv) the gauge transformations
generated by the momenta conjugate to the lapse and shift functions.
In the paper\cite{russo2}
 we begun to extract Dirac's observables starting from the symplectic
action of infinitesimal diffeomorphisms in $Diff\, \Sigma_{\tau}$,
ignoring the problems on the structure in large of the component of
$Diff\, \Sigma_{\tau}$ connected to the identity when a differential
structure is posed on it. Although such global properties can be
studied in Yang-Mills theory \footnote{Since the group of gauge
transformations is a Hilbert-Lie group.}, as shown in Ref.\cite{lusa},
and can be applied to the SO(3) gauge transformations of cotriads
\footnote{In our approach the Lorentz boosts are automatically Abelianized.},
one has that SO(3) gauge transformations and $Diff\, \Sigma_{\tau}$ do
not commute. Therefore, in tetrad gravity the group of SO(3) gauge
transformations is an invariant subgroup of a larger group, the group
of automorphisms of the SO(3) frame bundle, containing also $Diff\,
\Sigma_{\tau}$ and again the global situation in the large is of
difficult control \footnote{$Diff\, \Sigma
_{\tau}$ is an inductive limit of Hilbert-Lie groups \cite{sch}, but the
global properties of its group manifold are not well understood.}.
However, these are topics for future papers.

In Section II  the new parametrization of cotetrads is defined.

In Section III such parametrized cotetrads are inserted in the ADM
metric action to generate  a new Lagrangian for tetrad  gravity. The
Hamiltonian formulation is developed with the identification of
fourteen first class constraints and with the evaluation of their
Poisson brackets. The comparison with other formulations of tetrad
gravity is done.

In Section IV there is a comparison with ADM canonical metric gravity.

In the Conclusions the next step, namely the identification of the
Dirac observables with respect to the gauge transfomations generated
by thirteen constraints (only the superhamitonian constraint is not
treated) is delineated.

Appendix A is devoted to a review of 4-dimensional pseudo-Riemannian
and 3-dimensional Riemannian manifolds asymptotically flat at spatial
infinity, of the tetrad and triad formalisms and of the Lagrangians
used for general relativity.

Ref.\cite{russo1} contains an enlarged version of this paper: i) more
review material is included in its Section II; ii) there is an
Appendix A with the explicit expression of 4-tensors and of the
geodesic equation and also with a review on the congruences of
timelike worldlines; iii) there is an Appendix B with the Hamiltonian
expression of 4-tensors.

\vfill\eject

\section{New Parametrization of $\Sigma_{\tau}$-Adapted  Tetrads.}

As said in the Introduction and in Appendix A, to which we refer for
the notations and the definitions, let our globally hyperbolic
spacetime $M^4$ be foliated with spacelike Cauchy hypersurfaces
$\Sigma_{\tau}$, obtained with the embeddings $i_{\tau}:\Sigma
\rightarrow \Sigma_{\tau} \subset M^4$, $\vec \sigma \mapsto
x^{\mu}=z^{\mu}(\tau ,\vec
\sigma )$, of a 3-manifold $\Sigma$ in $M^4$ \footnote{$\tau :M^4\rightarrow R$ is a
global, timelike, future-oriented function labelling the leaves of the
foliation; $x^{\mu}$ are local coordinates in a chart of $M^4$; $\vec
\sigma =\{ \sigma^r \}$, r=1,2,3, are local coordinates in a chart of
$\Sigma$, which is diffeomorphic to $R^3$; we shall use the notation
$\sigma^A=(\sigma^{\tau}=\tau ;\vec \sigma )$, $A=\tau ,r$, and
$z^{\mu}(\sigma )=z^{\mu}(\tau ,\vec \sigma )$.}.

In the family of $\Sigma_{\tau}$-adapted frames and coframes on $M^4$,
we can select special tetrads and cotetrads ${}^4_{(\Sigma )}{\check
E}_{(\alpha )}$ and ${}^4_{(\Sigma )}{\check \theta}^{(\alpha )}$ also
adapted to a given set of triads ${}^3e^r_{(a)}$ and cotriads
${}^3e^{(a)}_r={}^3e_{(a)r}$ on $\Sigma_{\tau}$

\begin{eqnarray}
{}^4_{(\Sigma )}{\check E}^{\mu}_{(\alpha )}&=&\lbrace {}^4_{(\Sigma )}{\check
E}^{\mu}_{(o)}=l^{\mu}= {\hat b}^{\mu}_l={1\over N}(b^{\mu}
_{\tau}-N^rb^{\mu}_r);
\,\, {}^4_{(\Sigma )}{\check E}^{\mu}_{(a)}={}^3e^s_{(a)}
b^{\mu}_s \rbrace ,\nonumber \\
{}^4_{(\Sigma )}{\check E}_{\mu}^{(\alpha )}&=&\lbrace {}^4_{(\Sigma )}{\check
E}_{\mu}^{(o)}=\epsilon l_{\mu}= {\hat b}^l_{\mu}= N b^{\tau}
_{\mu};\,\, {}^4_{(\Sigma )}{\check E}_{\mu}^{(a)}={}^3e_s^{(a)}
{\hat b}^s_{\mu}\rbrace ,\nonumber \\
&&{}\nonumber \\
{}^4_{(\Sigma )}{\check E}^{\mu}_{(\alpha )}&& {}^4g_{\mu\nu}\,\,\,
{}^4_{(\Sigma )}{\check E}^{\nu}_{(\beta )} = {}^4\eta_{(\alpha )(\beta )},
\label{II1}
\end{eqnarray}

\noindent where $b^{\mu}_r$ and $b_{\mu}^r$ are defined in Eqs.(\ref{a3}) and
$l^{\mu}(\tau ,\vec \sigma )$ is the unit normal to $\Sigma_{\tau}$ at
$\vec \sigma$. $N$ and $N^r$ are the standard lapse and shift
functions.

The components of these tetrads and cotetrads in the holonomic basis
defined in Subsection 1 of Appendix A are respectively (see
Refs.\cite{dirr,hen1}; ${}^4_{(\Sigma )}{\check {\tilde E}}^{(o)}_r=0$
is the {\it Schwinger time gauge condition}\cite{schw})

\begin{eqnarray}
{}^4_{(\Sigma )}{\check {\tilde E}}^A_{(\alpha )}&=&{}^4_{(\Sigma )}{\check E}
^{\mu}_{(\alpha )}\, b^A_{\mu},\quad \Rightarrow {}^4_{(\Sigma )}{\check
{\tilde E}}^A_{(o)}=\epsilon l^A,\nonumber \\
&&{}^4_{(\Sigma )}{\check {\tilde E}}^{\tau}_{(o)}={1\over N},\quad\quad
{}^4_{(\Sigma )}{\check {\tilde E}}^{\tau}_{(a)}=0,\nonumber \\
&&{}^4_{(\Sigma )}{\check {\tilde E}}^r_{(o)}=-{{N^r}\over N},\quad\quad
{}^4_{(\Sigma )}{\check {\tilde E}}^r_{(a)}={}^3e^r_{(a)};\nonumber \\
{}^4_{(\Sigma )}{\check {\tilde E}}_A^{(\alpha )}&=&{}^4_{(\Sigma )}{\check E}
_{\mu}^{(\alpha )}\, b^{\mu}_A,\quad \Rightarrow {}^4_{(\Sigma )}{\check
{\tilde E}}_A^{(o)} = l_A,\nonumber \\
&&{}^4_{(\Sigma )}{\check {\tilde E}}^{(o)}_{\tau}=N,\quad\quad
{}^4_{(\Sigma )}{\check {\tilde E}}^{(a)}_{\tau}=N^r\, {}^3e^{(a)}_r=N^{(a)},
\nonumber \\
&&{}^4_{(\Sigma )}{\check {\tilde E}}^{(o)}_r=0,\quad\quad
{}^4_{(\Sigma )}{\check {\tilde E}}^{(a)}_r={}^3e^{(a)}_r,\nonumber \\
&&{}\nonumber \\
&&{}^4_{(\Sigma )}{\check E}^A_{(\alpha )}\, {}^4g_{AB}\, {}^4_{(\Sigma
)}{\check E}^B_{(\beta )}={}^4\eta_{(\alpha )(\beta )}.
\label{II2}
\end{eqnarray}

With the cotetrads ${}^4_{(\Sigma )}{\check E}^{(\alpha )}_{\mu}(z(\sigma ))$
we can build the vector ${\buildrel \circ
\over V}^{(\alpha )}=l^{\mu}(z(\sigma ))\, {}^4_{(\Sigma )}{\check E}^{(\alpha
)}_{\mu}(z(\sigma ))= (1 ;\vec 0)$: it is the same unit timelike
future-pointing Minkowski 4-vector in the tangent plane of each point
$z^{\mu}(\sigma )=z^{\mu}(\tau ,\vec \sigma )\in \Sigma_{\tau}\subset M^4$
for every $\tau$ and $\vec \sigma$; we have ${\buildrel \circ \over V}
^{(\alpha )}\, {}^4\eta_{(\alpha )(\beta )}\, {\buildrel \circ \over V}
^{(\beta )}=\epsilon$.

Let ${}^4E^{\mu}_{(\alpha )}(z)$ and ${}^4E^{(\alpha )}_{\mu}(z)$ be arbitrary
tetrads and cotetrads on $M^4$. Let us define the point-dependent Minkowski
4-vector $V^{(\alpha )}(z(\sigma ))=l^{\mu}(z(\sigma ))\, {}^4E^{(\alpha )}
_{\mu}(z(\sigma ))$ (assumed to be future-pointing), which satisfies
$V^{(\alpha )}(z(\sigma ))\, {}^4\eta_{(\alpha )(\beta )}\, V^{(\beta
)} (z(\sigma ))=\epsilon$, so that $V^{(\alpha )}(z(\sigma ))=(
V^{(o)}(z(\sigma ))=+\sqrt{1+\sum_rV^{(r) 2}(z(\sigma ))};
V^{(r)}(z(\sigma )) {\buildrel {def}\over =}\, \varphi^{(r)}(\sigma )
)$ : therefore, the point-dependent Minkowski 4-vector $V^{(\alpha )}
(z(\sigma ))$ depends only on the three functions
$\varphi^{(r)}(\sigma )$ \footnote{One has $\varphi^{(r)}(\sigma
)=-\epsilon
\varphi_{(r)}(\sigma )$ since ${}^4\eta_{rs}=-\epsilon \,
\delta_{rs}$; having the Euclidean signature (+++) for both $\epsilon
=\pm 1$, we shall define the Kronecker delta as $\delta
^{(i)(j)}=\delta^{(i)}_{(j)}=\delta_{(i)(j)}$.}.
If in each tangent plane we introduce the point-dependent Lorentz
transformation

\begin{eqnarray}
L^{(\alpha )}{}_{(\beta )}(V(z(\sigma ));{\buildrel \circ \over V})&=&\delta
^{(\alpha )}_{(\beta )}+2\epsilon V^{(\alpha )}(z(\sigma )){\buildrel \circ
\over V}_{(\beta )}-\epsilon { {(V^{(\alpha )}(z(\sigma ))+{\buildrel \circ
\over V}^{(\alpha )})(V_{(\beta )}(z(\sigma ))+{\buildrel \circ \over V}
_{(\beta )})}\over {1+V^{(o)}(z(\sigma ))}}=\nonumber \\
&=& \left( \begin{array}{cc} V^{(o)} & -\epsilon V_{(j)} \\
V^{(i)} & \delta^{(i)}_{(j)}-\epsilon {{V^{(i)}V_{(j)}}\over {1+
V^{(o)} }} \end{array} \right) (z(\sigma )),
\label{II3}
\end{eqnarray}

\noindent which is the standard Wigner boost for timelike Poincar\'e orbits
[see Ref.\cite{longhi}], one has by construction

\begin{equation}
V^{(\alpha )}(z(\sigma )) = L^{(\alpha )}{}_{(\beta )}(V(z(\sigma
));{\buildrel \circ \over V})\, {\buildrel \circ \over V}^{(\beta )}
{\buildrel {def} \over =}\, l^{\mu}(z(\sigma ))\, {}^4E^{(\alpha
)}_{\mu} (z(\sigma )).
\label{II4}
\end{equation}

We shall define our class of  {\it arbitrary cotretads}
${}^4E^{(\alpha )}_{\mu}(z(\sigma ))$ on $M^4$ starting from the
special $\Sigma_{\tau}$- and cotriad-adapted cotetrads ${}^4_{(\Sigma
)}{\check E}^{(\alpha )}_{\mu} (z(\sigma ))$ by means of the formula

\begin{equation}
{}^4E^{(\alpha )}_{\mu}(z(\sigma ))=L^{(\alpha )}{}_{(\beta )}(V(z(\sigma ));
{\buildrel \circ \over V})\, {}^4_{(\Sigma )}{\check E}^{(\beta )}_{\mu}
(z(\sigma )).
\label{II5}
\end{equation}

\noindent Let us remark that with this definition we are putting equal to zero,
by convention, the angles of an arbitrary 3-rotation of
$b^s_{\mu}(z(\sigma ))$ \footnote{I.e. of the choice of the three axes
tangent to $\Sigma_{\tau}$.} inside ${}^4_{(\Sigma )}{\check
E}^{(\alpha )}_{\mu}(z(\sigma ))$.

Since $\varphi^{(a)}(\sigma )=V^{(a)}(z(\sigma ))=l^{\mu}(z(\sigma
))\, {}^4E^{(a)}_{\mu}(z(\sigma ))$ are the three parameters of the
Wigner boost \footnote{$\varphi^{(a)}=\bar \gamma\beta^{(a)}$, $\bar
\gamma
=\sqrt{1+\sum_{(c)}\varphi
^{(c) 2}}$, $\beta^{(a)}=\varphi^{(a)}/ \sqrt{1+\sum_{(c)}\varphi^{(c) 2}}$.},
the previous equation can be rewritten in the following form [remembering
that $\varphi^{(a)}=-\epsilon \varphi_{(a)}$]

\begin{equation}
\left( \begin{array}{l} {}^4E^{(o)}_{\mu}\\ {}^4E^{(a)}_{\mu} \end{array}
\right) (z(\sigma ))=\left( \begin{array}{cc}  \sqrt{1+\sum_{(c)}
\varphi^{(c) 2}} &-\epsilon \varphi_{(b)}\\  \varphi^{(a)} &
\delta^{(a)}_{(b)}-\epsilon {{\varphi^{(a)}\varphi_{(b)} }\over {1+
\sqrt{1+\sum_{(c)}\varphi^{(c) 2}} }} \end{array} \right) (z(\sigma ))
\left( \begin{array}{l} l_{\mu} \\ {}^3e^{(b)}_s\, b^s_{\mu} \end{array}
\right) (\sigma ).
\label{II6}
\end{equation}

If we go to holonomic bases, ${}^4E^{(\alpha )}_A(z(\sigma ))={}^4E^{(\alpha )}
_{\mu}(z(\sigma ))\, b^{\mu}_A(\sigma )$ and ${}^4_{(\Sigma )}{\check {\tilde
E}}^{(\alpha )}_A(z(\sigma ))={}^4_{(\Sigma )}{\check E}^{(\alpha )}_{\mu}
(z(\sigma ))\, b^{\mu}_A(\sigma )$, one has

\begin{eqnarray}
\left( \begin{array}{l} {}^4E^{(o)}_A \\ {}^4E^{(a)}_A \end{array} \right) &=&
\left( \begin{array}{cc}  \sqrt{1+\sum_{(c)}
\varphi^{(c) 2}} &-\epsilon \varphi_{(b)}\\  \varphi^{(a)} &
\delta^{(a)}_{(b)}-\epsilon {{\varphi^{(a)}\varphi_{(b)} }\over {1+
\sqrt{1+\sum_{(c)}\varphi^{(c) 2}} }} \end{array} \right) \times\nonumber \\
&&\left( \begin{array}{l} {}^4_{(\Sigma )}{\check {\tilde E}}^{(o)}_A=(N;\vec
0)\\ {}^4_{(\Sigma )}{\check {\tilde E}}^{(b)}_A=(N^{(b)}={}^3e^{(b)}_rN^r;
{}^3e^{(b)}_r) \end{array} \right) ,
\label{II7}
\end{eqnarray}

\noindent so that we get that the cotetrad in the holonomic basis can be
expressed in terms of N, $N^{(a)}={}^3e^{(a)}_sN^s=N_{(a)}$, $\varphi^{(a)}$ and
${}^3e^{(a)}_r$ [${}^3g_{rs}=\sum_{(a)} {}^3e_{(a)r}\, {}^3e_{(a)s}$]

\begin{eqnarray}
&&{}^4E^{(o)}_{\tau}(z(\sigma ))= \sqrt{1+\sum_{(c)}\varphi^{(c) 2}
(\sigma )}\, N(\sigma )+\sum_{(a)}\varphi^{(a)}(\sigma ) N^{(a)}(\sigma ),
\nonumber \\
&&{}^4E^{(o)}_r(z(\sigma ))= \sum_{(a)}\varphi^{(a)}(\sigma )\,
{}^3e^{(a)}_r(\sigma ),\nonumber \\
&&{}^4E^{(a)}_{\tau}(z(\sigma ))= \varphi^{(a)}(\sigma )N(\sigma )+
\sum_{(b)}[\delta^{(a)}_{(b)}-\epsilon {{\varphi^{(a)}(\sigma )\varphi_{(b)}
(\sigma )}\over {1+\sqrt{1+\sum_{(c)}\varphi^{(c) 2}(\sigma )} }}]N^{(b)}
(\sigma ),\nonumber \\
&&{}^4E^{(a)}_r(z(\sigma ))=\sum_{(b)}[\delta^{(a)}_{(b)}-\epsilon
{{\varphi^{(a)}(\sigma )\varphi_{(b)}
(\sigma )}\over {1+\sqrt{1+\sum_{(c)}\varphi^{(c) 2}(\sigma )} }}] {}^3e
^{(b)}_r(\sigma ),\nonumber \\
 &&{}\nonumber \\
 &&{}\nonumber \\
 \Rightarrow {}^4g_{AB}&=&{}^4E^{(\alpha )}_A\,
{}^4\eta_{(\alpha )(\beta )}\, {}^4E^{(\beta )}_B= {}^4_{(\Sigma
)}{\check E}^{(\alpha )}_A\, {}^4\eta_{(\alpha )(\beta )}\,
{}^4_{(\Sigma )}{\check E}^{(\beta )}_B=\nonumber \\
 &=&\epsilon \left( \begin{array}{cc} (N^2- {}^3g_{rs}N^rN^s) &
-{}^3g_{st}N^t\\ -{}^3g_{rt}N^t & -{}^3g
_{rs} \end{array} \right) ,
\label{II8}
\end{eqnarray}

\noindent with the last line in accord with Eqs.(\ref{a3}); we have used
$L^T {}^4\eta L={}^4\eta$, valid for every Lorentz transformation. We find
$L^{-1}(V,{\buildrel \circ \over V})={}^4\eta L^T(V,{\buildrel \circ \over V})
{}^4\eta=L(V,{\buildrel \circ \over V}){|}_{\varphi^{(a)}\mapsto -\varphi^{(a)}
}$ and [${}^4E^A_{(\alpha )}={}^4E^{\mu}_{(\alpha )}b^A_{\mu}$,
${}^4_{(\Sigma )}{\check {\tilde E}}^A_{(\alpha )}={}^4_{(\Sigma )}{\check E}
^{\mu}_{(\alpha )}b^A_{\mu}$]

\begin{eqnarray}
\left( \begin{array}{l} {}^4E^{\mu}_{(o)}\\
{}^4E^{\mu}_{(a)}\end{array}
\right) &=&\left( \begin{array}{cc}  \sqrt{1+\sum_{(c)}
\varphi^{(c) 2}} &- \varphi^{(b)}\\ \epsilon \varphi_{(a)} &
\delta_{(a)}^{(b)}-\epsilon {{\varphi_{(a)}\varphi^{(b)} }\over {1+
\sqrt{1+\sum_{(c)}\varphi^{(c) 2}} }} \end{array} \right) \,
\left( \begin{array}{l} l^{\mu} \\ b^{\mu}_s\, {}^3e^s_{(b)} \end{array}
\right) , \nonumber \\
 &&{}\nonumber \\
 \left( \begin{array}{l} {}^4E^A_{(o)} \\ {}^4E^A_{(a)} \end{array} \right)
&=&\left( \begin{array}{cc}  \sqrt{1+\sum_{(c)}
\varphi^{(c) 2}} &- \varphi^{(b)}\\ \epsilon \varphi^{(a)} &
\delta_{(a)}^{(b)}-\epsilon {{\varphi_{(a)}\varphi^{(b)} }\over {1+
\sqrt{1+\sum_{(c)}\varphi^{(c) 2}} }} \end{array} \right) \,
\left( \begin{array}{l} {}^4_{(\Sigma )}{\check {\tilde E}}^A_{(o)}=(1/N;
-N^r/N) \\ {}^4_{(\Sigma )}{\check {\tilde E}}^A_{(b)}=(0; {}^3e^r_{(b)})
\end{array} \right) ,\nonumber \\
 &&{}\nonumber \\
 &&{}\nonumber \\
 &&{}^4E^{\tau}_{(o)}(z(\sigma ))= \sqrt{1+\sum_{(c)}\varphi^{(c) 2}
(\sigma )} {1\over {N(\sigma )}},\nonumber \\
&&{}^4E^r_{(o)}(z(\sigma )=- \sqrt{1+\sum_{(c)}\varphi^{(c) 2}
(\sigma )} {{N^r(\sigma )}\over {N(\sigma )}}-\varphi^{(b)}(\sigma )
\, {}^3e^r_{(b)}(\sigma ),\nonumber \\
&&{}^4E^{\tau}_{(a)}(z(\sigma ))=\epsilon {{\varphi_{(a)}(\sigma )}\over
{N(\sigma )}},\nonumber \\
&&{}^4E^r_{(a)}(z(\sigma ))=-\epsilon \varphi_{(a)}(\sigma ) {{N^r(\sigma )}
\over {N(\sigma )}}+\sum_{(b)}[\delta_{(a)}^{(b)}-\epsilon
{{\varphi_{(a)}(\sigma )\varphi^{(b)}(\sigma )}\over {1+\sqrt{1+\sum_{(c)}
\varphi^{(c) 2}(\sigma )}}}]\, {}^3e^r_{(b)}(\sigma ),\nonumber \\
 &&{}\nonumber \\
 \Rightarrow {}^4g^{AB}&=&{}^4E^A_{(\alpha )}\, {}^4\eta^{(\alpha
)(\beta )}\, {}^4E^B_{(\beta )}= {}^4_{(\Sigma )}{\check E}_{(\alpha
)}^A\, {}^4\eta_{(\alpha )(\beta )}\, {}^4_{(\Sigma )}{\check
E}_{(\beta )}^B=\nonumber \\
 &=&\epsilon \left( \begin{array}{cc} {1\over {N^2}} & -
{{N^s}\over {N^2}} \\ - {{N^r}\over {N^2}} &
- ({}^3g^{rs}-{{N^rN^s}\over {N^2}})\end{array} \right) ,
\label{II9}
\end{eqnarray}

\noindent with the last line in accord with Eqs.(\ref{a3}).

From ${}^4_{(\Sigma )}{\check {\tilde E}}^{(\alpha )}_A(z(\sigma ))=(L^{-1})
^{(\alpha )}{}_{(\beta )}(V(z(\sigma ));{\buildrel \circ \over V})\, {}^4E
^{(\beta )}_A(z(\sigma ))$ and ${}^4_{(\Sigma )}{\check {\tilde E}}^A_{(\alpha
)}(z(\sigma ))={}^4E^A_{(\beta )}\, (L^{-1})^{(\beta )}{}_{(\alpha )}
(V(z(\sigma ));{\buildrel \circ \over V})$ it turns out\cite{longhi} that the
flat indices $(a)$ of the adapted tetrads ${}^4_{(\Sigma )}{\check E}^{\mu}
_{(a)}$ and of the triads ${}^3e^r_{(a)}$ and cotriads ${}^3e_r^{(a)}$ on
$\Sigma_{\tau}$ transform as Wigner spin 1 indices under
point-dependent SO(3) Wigner rotations $R^{(a)}{}_{(b)}(V(z(\sigma
));\Lambda (z(\sigma )))$ associated with Lorentz transformations
$\Lambda^{(\alpha )}{}_{(\beta )}(z)$ in the tangent plane to $M^4$ in
the same point \footnote{$R^{(\alpha )}{}_{(\beta )} (\Lambda
(z(\sigma ));V(z(\sigma )))=[L({\buildrel \circ \over V};V(z(\sigma ))
)\, \Lambda^{-1}(z(\sigma ))\, L(\Lambda (z(\sigma ))V(z(\sigma
));{\buildrel \circ \over V})]^{(\alpha )}{}_{(\beta )}=\left(
\begin{array}{cc} 1&0\\ 0& R^{(a)}{}_{(b)}(V(z(\sigma ));\Lambda
(z(\sigma )))\end{array}\right)$.}. Instead the index ${(o)}$ of the
adapted tetrads ${}^4_{(\Sigma )}{\check E}
^{\mu}_{(o)}$ is a local Lorentz scalar in each point. Therefore, the adapted
tetrads in the holonomic basis should be denoted as ${}^4_{(\Sigma
)}{\check {\tilde E}}^A_{(\bar \alpha )}$, with $(\bar o)$ and
$A=(\tau ,r)$ Lorentz scalar indices and with $(\bar a)$ Wigner spin 1
indices; we shall go on with the indices $(o),(a)$ without the overbar
for the sake of simplicity. In this way the tangent planes to
$\Sigma_{\tau}$ in $M^4$ are described in a Wigner covariant way,
reminiscent of the flat rest-frame covariant instant form of dynamics
introduced in Minkowski spacetime in Ref.\cite{lus1}. Similar
conclusions are reached independently in Ref.\cite{ltt} in the
framework of non-linear Poincar\'e gauge theory \footnote{The vector
fields $e_{\alpha}$ and the 1-forms $\theta^{\alpha}$ of that paper
correspond to $X_{\tilde A}$ and $\theta^{\tilde A}$ in Eq.(\ref{a5})
respectively.}.

Therefore, an arbitrary tetrad field, namely a (in general
non-geodesic) congruence of observers' timelike worldlines with
4-velocity field $u^A(\tau ,\vec \sigma )={}^4E^A_{(o)}(\tau ,\vec
\sigma )$, can be obtained with a pointwise Wigner boost from the
special surface-forming timelike congruence whose 4-velocity field is
the normal to $\Sigma_{\tau}$ $l^A(\tau ,\vec \sigma )
=\epsilon \, {}^4_{(\Sigma )}{\check {\tilde E}}^A_{(o)}(\tau ,\vec \sigma )$
\footnote{It is associated with the 3+1 splitting of $M^4$ with leaves
$\Sigma_{\tau}$.}.

We can invert Eqs.(\ref{II9}) to get N, $N^r={}^3e^r_{(a)}N^{(a)}$,
$\varphi
^{(a)}$ and ${}^3e^r_{(a)}$ in terms of the tetrads ${}^4E^A_{(\alpha )}$

\begin{eqnarray}
N&=&  \frac{1}{ \sqrt{[{}^4E^{\tau}_{(o)}]^2
          -\sum_{(c)} [{}^4E^{\tau}_{(c)}]^2}}.
 \nonumber \\
N^r&=& - \frac{{}^4E^{\tau}_{(o)}\, {}^4E^{r}_{(0)} -\sum_{(c)}{}^4E^{\tau}
_{(c)}\, {}^4E^r_{(c)}}
{[{}^4E^{\tau}_{(0)}]^2-\sum_{(c)}[{}^4E^{\tau}_{(c)}]^2 }\nonumber \\
\varphi_{(a)}&=&\frac{\epsilon~ {}^4E^{\tau}_{(a)}  }{
          \sqrt{[{}^4E^{\tau}_{(o)}]^2
          -\sum_{(c)} [{}^4E^{\tau}_{(c)}]^2}}\nonumber\\
{}^3e^r_{(a)}&=&\sum_{(b)} B_{(a)(b)}
    \big( {}^4E^r_{(b)} + N^r~ {}^4E^\tau_{(b)} \big)
\nonumber \\
&&{}\nonumber \\
&& B_{(a)(b)} = \delta_{(a)(b)}
    -\frac{ {}^4E^{\tau}_{(a)} {}^4E^{\tau}_{(b)}  }{
           {}^4E^{\tau}_{(0)} \big[  {}^4E^{\tau}_{(0)}
       + \sqrt{[{}^4E^{\tau}_{(0)}]^2
          -\sum_{(c)} [{}^4E^{\tau}_{(c)}]^2}] }.
\label{II10}
\end{eqnarray}

If ${}^3e^{-1}=det\, ({}^3e^r_{(a)})$, then from the orthonormality
condition we get ${}^3e_{(a)r}= {}^3e ({}^3e^s_{(b)}\,
{}^3e^t_{(c)}-{}^3e^t_{(b)}\, {}^3e^s_{(c)})$ \footnote{With
$(a),(b),(c)$ and $r,s,t$ cyclic.} and it allows to express the
cotriads in terms of the tetrads ${}^4E^A_{(\alpha )}$. Therefore,
given the tetrads ${}^4E^A_{(\alpha )}$ (or equivalently the cotetrads
${}^4E_A^{(\alpha )}$) on $M^4$, an equivalent set of variables with
the local Lorentz covariance replaced with local Wigner covariance are
the lapse N, the shifts $N^{(a)}=N_{(a)}={}^3e_{(a)r}N^r$, the
Wigner-boost parameters $\varphi^{(a)}=-\epsilon \varphi_{(a)}$ and
either the triads ${}^3e^r_{(a)}$ or the cotriads ${}^3e_{(a)r}$.

\vfill\eject

\section
{The Lagrangian and the Hamiltonian in the New Variables.}

\subsection{The Lagrangian Formulation.}

As said in Subsection 4 of Appendix A, we can get an action principle
for tetrad gravity starting  from  the ADM action $S_{ADM}$
(\ref{a29}):

\bea
S_{ADM} &=& -\epsilon {{c^3}\over {16\pi G}} \int_{U} d^4x\,
\sqrt{{}^4g} [{}^3R+{}^3K_{\mu\nu}\, {}^3K^{\mu\nu} -({}^3K)^2]=\nonumber \\
 &=&-\epsilon k\int_{\triangle \tau}d\tau  \, \int d^3\sigma \, \lbrace
\sqrt{\gamma} N\, [{}^3R+{}^3K_{rs}\, {}^3K^{rs}-({}^3K)^2]\rbrace (\tau ,\vec
\sigma ).
\label{III1}
\eea

Its independent variables in metric gravity have now the following
expression in terms of N, $N^{(a)}=N_{(a)}={}^3e^r_{(a)}N_r$,
$\varphi^{(a)}=-\epsilon \varphi_{(a)}$, ${}^3e^{(a)}_r={}^3e_{(a)r}$
\footnote{$\gamma =det\, ({}^3g_{rs})= ({}^3e)^2=(det\, (e_{(a)r}))^2$.}

\begin{eqnarray}
N,&&{}\quad\quad
N_r={}^3e_r^{(a)}N_{(a)}={}^3e_{(a)r}N_{(a)},\nonumber \\
{}^3g_{rs}&=&{}^3e^{(a)}_r\, \delta_{(a)(b)}\, {}^3e^{(b)}_s={}^3e_{(a)r}\,
{}^3e_{(a)s},
\label{III2}
\end{eqnarray}

\noindent so that the line element of $M^4$ becomes

\bea
ds^2 &=&\epsilon (N^2- N_{(a)}N_{(a)})(d\tau )^2-2\epsilon N_{(a)}\,
{}^3e_{(a)r} d\tau d\sigma^r-
\epsilon \, {}^3e_{(a)r}\, {}^3e_{(a)s} d\sigma^rd\sigma^s=\nonumber \\
 &=&\epsilon \Big[ N^2
(d\tau )^2-({}^3e_{(a)r}d\sigma^r +N_{(a)}d\tau
)({}^3e_{(a)s}d\sigma^s +N_{(a)}d\tau )\Big].
\label{III3}
\eea

The extrinsic curvature takes the form
\footnote{$N_{(a)|r}={}^3e^s_{(a)}N_{s|r}=
\partial_r N_{(a)}-\epsilon_{(a)(b)(c)}\, {}^3\omega_{r(b)}N_{(c)}$ from
Eq.(\ref{a22}).}

\begin{eqnarray}
{}^3K_{rs}&=& {\hat b}^{\mu}_r{\hat b}^{\nu}_s {}^3K_{\mu\nu}={1\over {2N}}
(N_{r | s}+N_{s | r}-\partial_{\tau} {}^3g_{rs})=\nonumber \\
&=&{1\over {2N}} ({}^3e_{(a)r} \delta^w_s+{}^3e_{(a)s}\delta^w_r)(N_{(a) | w}-
\partial_{\tau}\, {}^3e_{(a)w}),\nonumber \\
{}^3K_{r(a)}&=&{}^3K_{rs}\, {}^3e^s_{(a)}={1\over {2N}}(\delta_{(a)(b)}\delta
^w_r+{}^3e^w_{(a)}\, {}^3e_{(b)r})(N_{(b)|w}-\partial_{\tau}\, {}^3e_{(c)w}),
\nonumber \\
{}^3K&=&{1\over N}\, {}^3e^r_{(a)} (N_{(a)|r}-\partial_{\tau}\, {}^3e_{(a)r}),
\label{III4}
\end{eqnarray}

\noindent so that the ADM action in the new variables is (from now on we shall
use the notation $k = {{c^3}\over {16\pi G}}$)

\begin{eqnarray}
{\hat S}_{ADMT}&=&\int d\tau {\hat L}_{ADMT}=\nonumber \\
&=&-\epsilon k \int d\tau d^3\sigma \lbrace N\, {}^3e\, \epsilon_{(a)(b)(c)}\,
{}^3e^r_{(a)}\, {}^3e^s_{(b)}\, {}^3\Omega_{rs(c)}+\nonumber \\
&+&{{{}^3e}\over {2N}} ({}^3G_o^{-1})_{(a)(b)(c)(d)} {}^3e^r_{(b)}(N_{(a) | r}-
\partial_{\tau}\, {}^3e_{(a)r})\, {}^3e^s_{(d)}(N_{(c) | s}-\partial_{\tau}
\, {}^3e_{(c) \ s})\rbrace,
\label{III5}
\end{eqnarray}

\noindent where we introduced the flat inverse
Wheeler-DeWitt supermetric

\begin{equation}
({}^3G_o^{-1})_{(a)(b)(c)(d)}=\delta_{(a)(c)}\delta_{(b)(d)}+\delta_{(a)(d)}
\delta_{(b)(c)}-2\delta_{(a)(b)}\delta_{(c)(d)}.
\label{III6}
\end{equation}

\noindent The flat supermetric is

\begin{eqnarray}
{}^3G_{o(a)(b)(c)(d)}&=&{}^3G_{o(b)(a)(c)(d)}={}^3G_{o(a)(b)(d)(c)}=
{}^3G_{o(c)(d)(a)(b)}=\nonumber \\
&=&\delta_{(a)(c)}\delta_{(b)(d)}+\delta_{(a)(d)}\delta_{(b)(c)}-\delta
_{(a)(b)}\delta_{(c)(d)},\nonumber \\
&&{}\nonumber \\
&&{1\over 2}\, {}^3G_{o(a)(b)(e)(f)}\, {1\over 2}\, {}^3G^{-1}_{o(e)(f)(c)(d)}
={1\over 2}[\delta_{(a)(c)}\delta_{(b)(d)}+\delta_{(a)(d)}\delta_{(b)(c)}].
\label{III7}
\end{eqnarray}

The new action does not depend on the 3 boost variables
$\varphi^{(a)}$ \footnote{Like the Higgs model Lagrangian in the
unitary gauge does not depend on some of the Higgs
fields\cite{lv1,lv2}.}, contains lapse N and modified shifts $N_{(a)}$
as Lagrange multipliers, and is a functional independent from the
second time derivatives of the fields.

Instead of deriving its Euler-Lagrange equations we shall study its
Hamiltonian formulation.

\subsection{The Hamiltonian Formulation.}

The canonical momenta and the Poisson brackets are

\begin{eqnarray}
&&{\tilde \pi}^{\vec \varphi}_{(a)}(\tau ,\vec \sigma )={{\delta {\hat S}
_{ADMT}}\over {\delta \partial_{\tau} \varphi_{(a)}(\tau ,\vec \sigma )}}=0,
\nonumber \\
&&{\tilde \pi}^N(\tau ,\vec \sigma )={{\delta {\hat S}_{ADMT}}\over {\delta
\partial_{\tau} N(\tau ,\vec \sigma )}}=0,\nonumber \\
&&{\tilde \pi}^{\vec N}_{(a)}(\tau ,\vec \sigma )={{\delta {\hat S}_{ADMT}}
\over {\delta \partial_{\tau} N_{(a)}(\tau ,\vec \sigma )}}=0,\nonumber \\
&&{}^3{\tilde \pi}^r_{(a)}(\tau ,\vec \sigma )={{\delta {\hat S}_{ADMT}}
\over {\delta \partial_{\tau} {}^3e_{(a)r}(\tau ,\vec \sigma )}}=
[{{\epsilon k{}^3e}\over
N} ({}^3G^{-1}_o)_{(a)(b)(c)(d)}\, {}^3e_{(b)}^r\,
{}^3e^s_{(d)}\, (N_{(c) | s}-\partial_{\tau}\, {}^3e_{(c)s})](\tau ,\vec
\sigma )=\nonumber \\
&&=2\epsilon k [{}^3e ({}^3K^{rs}-{}^3e^r_{(c)}\, {}^3e^s_{(c)}\, {}^3K)
{}^3e_{(a)s}](\tau ,\vec \sigma ),\nonumber \\
&&{}\nonumber \\
&&\lbrace N(\tau ,\vec \sigma ),{\tilde \pi}^N(\tau ,{\vec \sigma}^{'} )
\rbrace = \delta^3(\vec \sigma ,{\vec \sigma}^{'}),\nonumber \\
&&\lbrace N_{(a)}(\tau ,\vec \sigma ),{\tilde \pi}^{\vec N}_{(b)}(\tau ,{\vec
\sigma}^{'} )\rbrace =\delta_{(a)(b)} \delta^3(\vec \sigma ,{\vec \sigma}^{'}),
\nonumber \\
&&\lbrace \varphi_{(a)}(\tau ,\vec \sigma ),{\tilde \pi}^{\vec \varphi}_{(b)}
(\tau ,{\vec \sigma}^{'} )\rbrace = \delta_{(a)(b)} \delta^3(\vec \sigma ,
{\vec \sigma}^{'}),\nonumber \\
&&\lbrace {}^3e_{(a)r}(\tau ,\vec \sigma ),{}^3{\tilde \pi}^s_{(b)}(\tau ,
{\vec \sigma}^{'} )\rbrace =\delta_{(a)(b)} \delta^s_r \delta^3(\vec \sigma ,
{\vec \sigma}^{'}),\nonumber \\
&&{}\nonumber \\
&&\lbrace {}^3e^r_{(a)}(\tau ,\vec \sigma),{}^3{\tilde \pi}^s_{(b)}(\tau ,
{\vec \sigma}^{'})\rbrace =-{}^3e^r_{(b)}(\tau ,\vec \sigma )\, {}^3e^s
_{(a)}(\tau ,\vec \sigma ) \delta^3(\vec \sigma ,{\vec \sigma}^{'}),
\nonumber \\
&&\lbrace {}^3e(\tau ,\vec \sigma ), {}^3{\tilde \pi}^r_{(a)}(\tau ,{\vec
\sigma}^{'})\rbrace ={}^3e(\tau ,\vec \sigma )\, {}^3e^r_{(a)}(\tau ,\vec
\sigma )\, \delta^3(\vec \sigma ,{\vec \sigma}^{'}),
\label{III8}
\end{eqnarray}

\noindent where the Dirac delta distribution is a density of weight -1
\footnote{It behaves as $\sqrt{\gamma (\tau ,\vec \sigma )}$, because we have the
${\vec \sigma}^{'}$-reparametrization invariant result $\int
d^3\sigma^{'}
\delta^3(\vec \sigma ,{\vec \sigma}^{'}) f({\vec \sigma}^{'})=f(\vec \sigma )$.}.
The momentum ${}^3{\tilde \pi}^r_{(a)}$ is a density of weight -1.

Besides the seven primary constraints

\bea
&&{\tilde \pi}^{\vec \varphi}_{(a)} (\tau ,\vec \sigma )\approx
0,\nonumber \\
 &&{\tilde \pi}^N(\tau ,\vec \sigma )\approx 0,\nonumber \\
 && {\tilde \pi}^{\vec N}_{(a)}(\tau ,\vec \sigma )\approx 0,
\label{III9}
\eea

\noindent there are the following three primary constraints (the generators of
the inner rotations)

\begin{eqnarray}
{}^3{\tilde M}_{(a)}(\tau ,\vec \sigma )&=&\epsilon_{(a)(b)(c)}\, {}^3e_{(b)r}
(\tau ,\vec \sigma )\, {}^3{\tilde \pi}^r_{(c)}(\tau ,\vec \sigma )={1\over 2}
\epsilon_{(a)(b)(c)}\, {}^3{\tilde M}_{(b)(c)}(\tau ,\vec \sigma )\approx 0,
\nonumber \\
&\Rightarrow& {}^3{\tilde M}_{(a)(b)}(\tau ,\vec \sigma )=\epsilon_{(a)(b)(c)}
\, {}^3{\tilde M}_{(c)}(\tau ,\vec \sigma )=\nonumber \\
&=&{}^3e_{(a)r}(\tau ,\vec \sigma )\,
{}^3{\tilde \pi}^r_{(b)}(\tau ,\vec \sigma )-{}^3e_{(b)r}(\tau ,\vec \sigma )\,
{}^3{\tilde \pi}^r_{(a)}(\tau ,\vec \sigma )\approx 0.
\label{III10}
\end{eqnarray}

By using Eqs.(\ref{III7}) and (\ref{III8}) we get the following
inversion

\begin{eqnarray}
{}^3e^r_{(a)}&(& N_{(b) | r}-\partial_{\tau}\, {}^3e_{(b)r})+{}^3e^r_{(b)}
(N_{(a) | r}-\partial_{\tau}\, {}^3e_{(a)r})=\nonumber \\
&=&{{\epsilon N}\over {2k\, {}^3e}}\,
{}^3G_{o(a)(b)(c)(d)}\, {}^3e_{(c)r}\, {}^3{\tilde \pi}^r_{(d)},
\label{III11}
\end{eqnarray}

\noindent so that, even if, due to the degeneracy associated with the first class
constraints, this equation cannot be solved for $\partial_{\tau}\,
{}^3e_{(a)r}$ , we can get the phase space expression of the extrinsic
curvature without using the Hamilton equations

\begin{eqnarray}
{}^3K_{rs}&=&{{\epsilon}\over {4k\, {}^3e}}\, {}^3G_{o(a)(b)(c)(d)}\, {}^3e
_{(a)r}\, {}^3e_{(b)s}\, {}^3e_{(c)u}\, {}^3{\tilde \pi}^u_{(d)},\nonumber \\
{}^3K&=& -{{\epsilon}\over {2k\sqrt{\gamma}}} {}^3{\tilde \Pi}=-{{\epsilon}
\over {4k\, {}^3e}} {}^3e_{(a)r}\, {}^3{\tilde \pi}^r_{(a)}.
\label{III12}
\end{eqnarray}

Since at the Lagrangian level the primary constraints are identically zero,
we have

\begin{eqnarray}
{}^3{\tilde \pi}^r_{(a)}&=&{}^3e^r_{(b)}\, {}^3e_{(b)s}\, {}^3{\tilde \pi}^s
_{(a)}={1\over 2}{}^3e^r_{(b)}[{}^3e_{(a)s}\, {}^3{\tilde \pi}^s_{(b)}+{}^3e
_{(b)s}\, {}^3{\tilde \pi}^s_{(a)}]-{1\over 2}{}^3{\tilde M}_{(a)(b)}\,
{}^3e^r_{(b)}\equiv \nonumber \\
&\equiv& {1\over 2}{}^3e^r_{(b)}[{}^3e_{(a)s}\, {}^3{\tilde \pi}^s_{(b)}+{}^3e
_{(b)s}\, {}^3{\tilde \pi}^s_{(a)}],\nonumber \\
&&{}\nonumber \\
{}^3{\tilde \pi}^r_{(a)} &\partial_{\tau}& {}^3e_{(a)r}\equiv
{1\over 2}[{}^3e_{(a)s}\, {}^3{\tilde \pi}^s_{(b)}+{}^3e
_{(b)s}\, {}^3{\tilde \pi}^s_{(a)}]{}^3e^r_{(b)}\, \partial_{\tau}\, {}^3e
_{(a)r}\equiv \nonumber \\
&\equiv& {}^3{\tilde \pi}^r_{(a)}\, N_{(a) | r}-{N\over {4k\, {}^3e}}\,
{}^3G_{o(a)(b)(c)(d)}\,
{}^3e_{(a)s}\, {}^3{\tilde \pi}^s_{(b)}\, {}^3e_{(c)r}\, {}^3{\tilde \pi}^r
_{(d)},
\label{III13}
\end{eqnarray}

\noindent and the canonical Hamiltonian is

\begin{eqnarray}
{\hat H}_{(c)}&=&\int d^3\sigma [{\tilde \pi}^N\partial_{\tau}N+{\tilde
\pi}^{\vec N}_{(a)}\partial_{\tau}N_{(a)}+{\tilde \pi}^{\vec \varphi}_{(a)}
\partial_{\tau}\varphi_{(a)}+{}^3{\tilde \pi}^r_{(a)}\partial_{\tau}
{}^3e_{(a)r}](\tau ,\vec \sigma ) - {\hat L}_{ADMT}=\nonumber \\
&=&\int_{\Sigma_{\tau}}
d^3\sigma [\epsilon N\, (k\, {}^3e\, \epsilon_{(a)(b)(c)}\, {}^3e^r
_{(a)}\, {}^3e^s_{(b)}\, {}^3\Omega_{rs(c)}-\nonumber \\
&-&{1\over {8k\, {}^3e}}\, {}^3G_{o(a)(b)(c)(d)}\,
{}^3e_{(a)r}\, {}^3{\tilde \pi}^r_{(b)}\, {}^3e_{(c)s}\, {}^3{\tilde \pi}^s
_{(d)})-\nonumber \\
&-& N_{(a)}\, {}^3{\tilde \pi}^r_{(a) | r}](\tau ,\vec \sigma
)+\int_{\partial \Sigma_{\tau}} d^2\Sigma_r[N_{(a)}\, {}^3{\tilde
\pi}^r_{(a)}](\tau ,\vec \sigma ).
\label{III14}
\end{eqnarray}

\noindent In this paper we shall ignore the surface term.

The Dirac Hamiltonian is (the $\lambda (\tau ,\vec \sigma )$'s are
arbitrary Dirac multipliers)

\begin{equation}
{\hat H}_{(D)}={\hat H}_{(c)}+\int d^3\sigma [\lambda_N\, {\tilde
\pi}^N+\lambda^{\vec N}_{(a)}\, {\tilde \pi}^{\vec N}_{(a)}+\lambda^{\vec
\varphi}_{(a)}\, {\tilde \pi}^{\vec \varphi}_{(a)}+\mu_{(a)}\, {}^3{\tilde
M}_{(a)}](\tau ,\vec \sigma ).
\label{III15}
\end{equation}

The $\tau$-constancy of the ten primary constraints
($\partial_{\tau}\, {\tilde \pi}^N(\tau ,\vec
\sigma )\approx 0$ and  $\partial_{\tau}\, {\tilde
\pi}^{\vec N}_{(a)}(\tau ,\vec \sigma )
\approx 0$) generates four secondary constraints

\begin{eqnarray}
{\hat {\cal H}}(\tau ,\vec \sigma )&=& \epsilon
\Big[ k\, {}^3e\, \epsilon_{(a)(b)(c)}\, {}^3e^r
_{(a)}\, {}^3e^s_{(b)}\, {}^3\Omega_{rs(c)}-\nonumber \\
&-&{1\over {8k\, {}^3e}}
{}^3G_{o(a)(b)(c)(d)}\, {}^3e_{(a)r}\, {}^3{\tilde \pi}^r_{(b)}\, {}^3e_{(c)s}\,
{}^3{\tilde \pi}^s_{(d)}\Big] (\tau ,\vec \sigma )=\nonumber \\
&=&\epsilon \Big[ k\, {}^3e\, {}^3R-{1\over {8k\, {}^3e}}
{}^3G_{o(a)(b)(c)(d)}\, {}^3e_{(a)r}\, {}^3{\tilde \pi}^r_{(b)}\, {}^3e_{(c)s}\,
{}^3{\tilde \pi}^s_{(d)}\Big] (\tau ,\vec \sigma )
\approx 0,\nonumber \\
{\hat {\cal H}}_{(a)}(\tau ,\vec \sigma )&=&[\partial_r\, {}^3{\tilde \pi}^r
_{(a)}-\epsilon_{(a)(b)(c)}\, {}^3\omega_{r(b)}\, {}^3{\tilde \pi}^r_{(c)}]
(\tau ,\vec \sigma )={}^3{\tilde \pi}^r_{(a)|r}(\tau ,\vec \sigma )
\approx 0,\nonumber \\
&&{}\nonumber \\
&\Rightarrow& {\hat H}_{(c)}= \int d^3\sigma [ N\, {\hat {\cal H}}-
N_{(a)}\, {\hat {\cal H}}_{(a)}](\tau ,\vec \sigma )\approx 0.
\label{III16}
\end{eqnarray}

It can be checked that the superhamiltonian constraint ${\hat {\cal H}}(\tau ,
\vec \sigma )\approx 0$ coincides with the ADM metric superhamiltonian one
${\tilde {\cal H}}(\tau ,\vec \sigma )\approx 0$ given in
Eqs.(\ref{IV10}) of Section IV, where also the ADM metric
supermomentum constraints is expressed in terms of the tetrad gravity
constraints.

It is convenient to replace the constraints ${\hat {\cal H}}_{(a)}(\tau ,\vec
\sigma )\approx 0$ \footnote{They are of the type of SO(3) Yang-Mills Gauss laws,
because they are the covariant divergence of a vector density.}  with
the 3 constraints generating space pseudo-diffeomorphisms on the
cotriads and their conjugate momenta

\begin{eqnarray}
{}^3{\tilde \Theta}_r(\tau ,\vec \sigma )&=&-[{}^3e_{(a)r}\, {\hat {\cal H}}
_{(a)}+{}^3\omega_{r(a)}\, {}^3{\tilde M}_{(a)}](\tau ,\vec \sigma )=
\nonumber \\
&=&[{}^3{\tilde \pi}^s_{(a)}\, \partial_r\, {}^3e_{(a)s}-\partial_s
({}^3e_{(a)r}\, {}^3{\tilde \pi}^s_{(a)})](\tau ,\vec \sigma )\approx 0,
\nonumber \\
&&{}\nonumber \\ {\hat {\cal H}}_{(a)}(\tau ,\vec \sigma
)&=&-{}^3e^r_{(a)}(\tau ,\vec \sigma ) [{}^3{\tilde \Theta}_r +
{}^3\omega_{r(b)}\, {}^3{\tilde M}_{(b)}] (\tau ,\vec \sigma )\approx
0,\nonumber \\ &&{}\nonumber \\ &\Rightarrow& {\hat H}_{(D)}={\hat
H}^{'}_{(c)}+\int d^3\sigma [\lambda_N{\tilde \pi}^N+\lambda^{\vec
N}_{(a)}{\tilde \pi}^{\vec N}_{(a)}+
\lambda^{\vec \varphi}_{(a)}{\tilde \pi}^{\vec \varphi}_{(a)}+{\hat \mu}_{(a)}
\, {}^3{\tilde M}_{(a)}](\tau ,\vec \sigma ),\nonumber \\
&& {\hat H}^{'}_{(c)}= \int d^3\sigma [ N {\hat {\cal H}}
+N_{(a)}\, {}^3e^r_{(a)}\, {}^3{\tilde \Theta}_r](\tau ,\vec \sigma ),
\nonumber \\
&&{}
\label{III17}
\end{eqnarray}

\noindent where we replaced $[\mu_{(a)}- N_{(b)}\, {}^3e^r_{(b)}\,
{}^3\omega_{r(a)}](\tau ,\vec \sigma )$ with the new Dirac multipliers ${\hat
\mu}_{(a)}(\tau ,\vec \sigma )$.

All the constraints are first class because the only non-identically
vanishing Poisson brackets are

\begin{eqnarray}
\lbrace {}^3{\tilde M}_{(a)}(\tau ,\vec \sigma ),{}^3{\tilde M}_{(b)}
(\tau ,{\vec \sigma}^{'})\rbrace &=&\epsilon_{(a)(b)(c)}\, {}^3{\tilde M}_{(c)}
(\tau ,\vec \sigma ) \delta^3(\vec \sigma ,{\vec \sigma}^{'}),\nonumber \\
\lbrace {}^3{\tilde M}_{(a)}(\tau ,\vec \sigma ),{}^3{\tilde \Theta}_r
(\tau ,{\vec \sigma}^{'})\rbrace &=&{}^3{\tilde M}_{(a)}(\tau ,{\vec \sigma}
^{'})\, {{\partial \delta^3(\vec \sigma ,{\vec \sigma}^{'})}\over {\partial
\sigma^r}},\nonumber \\
\lbrace {}^3{\tilde \Theta}_r(\tau ,\vec \sigma ),{}^3{\tilde \Theta}_s
(\tau ,{\vec \sigma}^{'})\rbrace &=&[{}^3{\tilde \Theta}_r(\tau ,{\vec \sigma}
^{'}) {{\partial}\over {\partial \sigma^s}} +{}^3{\tilde \Theta}_s(\tau ,\vec
\sigma ) {{\partial}\over {\partial \sigma^r}}] \delta^3(\vec \sigma ,{\vec
\sigma}^{'}),\nonumber \\
\lbrace {\hat {\cal H}}(\tau ,\vec \sigma ),{}^3{\tilde \Theta}_r(\tau ,{\vec
\sigma}^{'})\rbrace &=& {\hat {\cal H}}(\tau ,{\vec \sigma}^{'}) {{\partial
\delta^3(\vec \sigma ,{\vec \sigma}^{'})}\over {\partial \sigma^r}},
\nonumber \\
\lbrace {\hat {\cal H}}(\tau ,\vec \sigma ),{\hat {\cal H}}(\tau ,{\vec
\sigma}^{'})\rbrace &=& [ {}^3e^r_{(a)}(\tau ,\vec \sigma )\, {\hat {\cal H}}
_{(a)}(\tau ,\vec \sigma ) +\nonumber \\
&+& {}^3e^r_{(a)}(\tau ,{\vec \sigma}^{'})\,
{\hat {\cal H}}_{(a)}(\tau ,{\vec \sigma}^{'}) ] {{\partial \delta^3(\vec \sigma
,{\vec \sigma}^{'})}\over {\partial \sigma^r}}=\nonumber \\
&=&\{ [{}^3e^r_{(a)}\, {}^3e^s_{(a)}\, [{}^3{\tilde \Theta}_s+{}^3\omega_{s(b)}
\, {}^3{\tilde M}_{(b)}]](\tau ,\vec \sigma ) + \nonumber \\
&+&[{}^3e^r_{(a)}\, {}^3e^s_{(a)}\, [{}^3{\tilde \Theta}_s+{}^3\omega_{s(b)}\,
{}^3{\tilde M}_{(b)}]](\tau ,{\vec \sigma}^{'}) \} \, {{\partial \delta^3(\vec
\sigma ,{\vec \sigma}^{'})}\over {\partial \sigma^r}}.
\label{III18}
\end{eqnarray}

As said at the end of the Introduction, the Hamiltonian gauge group
has the 14 first class constraints as generators of infinitesimal
gauge transformations connected with the identity. In particular
${\tilde \pi}^{\vec \varphi}_{(a)}(\tau ,\vec \sigma )\approx 0$ and
${}^3{\tilde M}_{(a)}(\tau ,\vec \sigma )\approx 0$ are the generators
of the $R^3\times SO(3)$ subgroup replacing the Lorentz subgroup with
our parametrization, while ${}^3{\tilde \Theta}_r(\tau ,\vec \sigma
)\approx 0$ are the generators of the infinitesimal
pseudodiffeomorphisms in $Diff\, \Sigma_{\tau}$.

The Poisson brackets of the cotriads and of their conjugate momenta with the
constraints are [${}^3R=\epsilon_{(a)(b)(c)}\, {}^3e^r_{(a)}\, {}^3e^s_{(b)}\,
{}^3\Omega_{rs(c)}$]

\begin{eqnarray}
&&\lbrace {}^3e_{(a)r}(\tau ,\vec \sigma ),{}^3{\tilde M}_{(b)}(\tau ,{\vec
\sigma}^{'})\rbrace =\epsilon_{(a)(b)(c)}\, {}^3e_{(c)r}(\tau ,\vec \sigma )
\delta^3(\vec \sigma ,{\vec \sigma}^{'}),\nonumber \\
&&\lbrace {}^3e_{(a)r}(\tau ,\vec \sigma ),{}^3{\tilde \Theta}_s(\tau ,{\vec
\sigma}^{'})\rbrace ={{\partial \, {}^3e_{(a)r}(\tau ,\vec \sigma )}\over
{\partial \sigma^s}}\delta^3(\vec \sigma ,{\vec \sigma}^{'})+{}^3e_{(a)s}
(\tau ,\vec \sigma ){{\partial \delta^3(\vec \sigma ,{\vec \sigma}^{'})}\over
{\partial \sigma^r}},\nonumber \\
&&\lbrace {}^3e_{(a)r}(\tau ,\vec \sigma ),{\hat {\cal H}}(\tau ,{\vec \sigma}
^{'})\rbrace =-{{\epsilon}\over {4k}}\Big[ {1\over {{}^3e}}\,
{}^3G_{o(a)(b)(c)(d)}\, {}^3e_{(b)r}
\, {}^3e_{(c)s}\, {}^3{\tilde \pi}^s_{(d)}\Big] (\tau ,\vec \sigma )\delta^3
(\vec \sigma ,{\vec \sigma}^{'}),\nonumber \\
&&{}\nonumber \\
&&\lbrace {}^3{\tilde \pi}^r_{(a)}(\tau ,\vec \sigma ),{}^3{\tilde M}_{(b)}
(\tau ,{\vec \sigma}^{'})\rbrace =\epsilon_{(a)(b)(c)}\, {}^3{\tilde \pi}^r
_{(c)}(\tau ,\vec \sigma )\delta^3(\vec \sigma ,{\vec \sigma}^{'}),\nonumber \\
&&\lbrace {}^3{\tilde \pi}^r_{(a)}(\tau ,\vec \sigma ),{}^3{\tilde \Theta}_s
(\tau ,{\vec \sigma}^{'})\rbrace =- {}^3{\tilde \pi}^r_{(a)}(\tau ,{\vec \sigma}
^{'}){{\partial \delta^3(\vec \sigma ,{\vec \sigma}^{'})}\over {\partial
\sigma^{{'}s}}}+\delta^r_s {{\partial}\over {\partial \sigma^{{'}u}}}
[{}^3{\tilde \pi}^u_{(a)}(\tau ,{\vec \sigma}^{'})\delta^3(\vec \sigma ,
{\vec \sigma}^{'})],\nonumber \\
&&\lbrace {}^3{\tilde \pi}^r_{(a)}(\tau ,\vec \sigma ),{\hat {\cal H}}
(\tau ,{\vec \sigma}^{'})\rbrace =\epsilon  \Big[ 2k\, {}^3e\,  ({}^3R^{rs}-
{1\over 2}\, {}^3g^{rs}\, {}^3R)\, {}^3e_{(a)s}+\nonumber \\
&+&{1\over {4k\, {}^3e}}
{}^3G_{o(a)(b)(c)(d)}\, {}^3{\tilde \pi}^r_{(b)}\, {}^3e_{(c)s}\,
{}^3{\tilde \pi}^s_{(d)}-\nonumber \\
&-&{1\over {8k\, {}^3e}} {}^3e^r_{(a)}\, {}^3G_{o(b)(c)(d)(e)}\,
{}^3e_{(b)u}\, {}^3{\tilde \pi}^u_{(c)}\, {}^3e_{(d)v}\, {}^3{\tilde \pi}^v
_{(e)} \Big] (\tau ,\vec \sigma ) \delta^3(\vec \sigma ,{\vec \sigma}^{'})+
\nonumber \\
&+&2k\, {}^3e(\tau ,\vec \sigma )
\Big[ {}^3\Gamma^w_{uv}({}^3e^u_{(a)}\, {}^3g^{rv}-
{}^3e^r_{(a)}\, {}^3g^{uv})\Big] (\tau ,{\vec \sigma}^{'}){{\partial \delta^3
(\vec \sigma ,{\vec \sigma}^{'})}\over {\partial \sigma^w}}+\nonumber \\
&+&2k\, {}^3e(\tau ,\vec \sigma )\Big[ {}^3e^u_{(a)}\, {}^3g^{rv}-
{}^3e^r_{(a)}\, {}^3g^{uv})\Big] (\tau ,{\vec \sigma}^{'}){{\partial^2\delta
^3(\vec \sigma ,{\vec \sigma}^{'})}\over {\partial \sigma^u\partial \sigma^v}},
\label{III19}
\end{eqnarray}

\noindent where we used\hfill\break
\hfill\break
 $\lbrace {}^3e(\tau ,\vec \sigma ){}^3R(\tau ,\vec
\sigma ),{}^3{\tilde \pi}^r_{(a)}(\tau ,{\vec \sigma}^{'})\rbrace =
-2k\Big[ {}^3e({}^3R^{rs}-{1\over 2}{}^3g^{rs}\, {}^3R){}^3e_{(a)s}\Big] (\tau
,\vec \sigma )\delta^3(\vec \sigma ,{\vec \sigma}^{'})+$\hfill\break
$+2k\, {}^3e(\tau ,\vec \sigma ) \Big[ {}^3\Gamma^w_{uv}({}^3e^u_{(a)}\,
{}^3g^{rv}-{}^3e^r_{(a)}\, {}^3g^{uv})\Big]
(\tau ,{\vec \sigma}^{'}){{\partial \delta^3(\vec \sigma ,{\vec \sigma}^{'})}
\over {\partial \sigma^w}}+$\hfill\break
$+2k\, {}^3e(\tau ,\vec \sigma )\Big[ {}^3e^u_{(a)}\,
{}^3g^{rv}-{}^3e^r_{(a)}\, {}^3g^{uv}\Big] (\tau ,{\vec \sigma}^{'}){{\partial^2
\delta^3(\vec \sigma ,{\vec \sigma}^{'})}\over {\partial \sigma^u\partial
\sigma^v}}$.\hfill\break
\hfill\break

The Hamilton equations associated with the Dirac Hamiltonian
(\ref{III17}) are (see Eqs.(\ref{a25}) for ${}^3R^{uv}$)

\begin{eqnarray}
\partial_{\tau}N(\tau ,\vec \sigma ) &{\buildrel \circ \over =}& \lbrace
N(\tau ,\vec \sigma ),{\hat H}^{'}_{(D)}\rbrace =\lambda_N(\tau ,\vec
\sigma ),\nonumber \\
\partial_{\tau}N_{(a)}(\tau ,\vec \sigma ) &{\buildrel \circ \over =}& \lbrace
N_{(a)}(\tau ,\vec \sigma ),{\hat H}^{'}_{(D)}\rbrace = \lambda^{\vec N}
_{(a)}(\tau ,\vec \sigma ),\nonumber \\
\partial_{\tau}\varphi_{(a)}(\tau ,\vec \sigma ) &{\buildrel \circ \over =}&
\lbrace \varphi_{(a)}(\tau ,\vec \sigma ),{\hat H}^{'}_{(D)}\rbrace =
\lambda^{\vec \varphi}_{(a)}(\tau ,\vec \sigma ),\nonumber \\
\partial_{\tau}\, {}^3e_{(a)r}(\tau ,\vec \sigma ) &{\buildrel \circ \over =}&
\lbrace {}^3e_{(a)r}(\tau ,\vec \sigma ),{\hat H}^{'}_{(D)}\rbrace =
\nonumber \\
&=&- {{\epsilon}\over {4k}}\Big[ {N\over{{}^3e}}\, {}^3G_{o(a)(b)(c)(d)}
\, {}^3e_{(b)r}\, {}^3e_{(c)s}\, {}^3{\tilde \pi}^s_{(d)}\Big] (\tau ,\vec
\sigma )+\nonumber \\
&+&\Big[ N_{(b)}\, {}^3e^s_{(b)}{{\partial \, {}^3e_{(a)r}}\over {\partial
\sigma^s}}+{}^3e_{(a)s} {{\partial}\over {\partial \sigma^r}} (N_{(b)}\,
{}^3e^s_{(b)})\Big] (\tau ,\vec \sigma )+\nonumber \\
&+&\epsilon_{(a)(b)(c)}\, {\hat \mu}_{(b)}(\tau ,\vec \sigma )\, {}^3e_{(c)r}
(\tau ,\vec \sigma ),\nonumber \\
\partial_{\tau}\, {}^3{\tilde \pi}^r_{(a)}(\tau ,\vec \sigma ) &{\buildrel
\circ \over =}& \lbrace {}^3{\tilde \pi}^r_{(a)}(\tau ,\vec \sigma ),{\hat
H}^{'}_{(D)}\rbrace =\nonumber \\
&=&2k\epsilon \Big[ {}^3e N ({}^3R^{rs}-{1\over 2} {}^3g^{rs}\, {}^3R){}^3e
_{(a)s}+{}^3e (N^{|r|s}-{}^3g^{rs}\, N^{|u}{}_{|u}){}^3e_{(a)s}\Big] (\tau
,\vec \sigma )-\nonumber \\
&-&\epsilon {{N(\tau ,\vec \sigma )}\over {8k}}
\Big[ {1\over {{}^3e}}\, {}^3G_{o(a)(b)(c)(d)}\, {}^3{\tilde \pi}
^r_{(b)}\, {}^3e_{(c)s}\, {}^3{\tilde \pi}^s_{(d)}-\nonumber \\
&-&{2\over {{}^3e}}\, {}^3e^r_{(a)}\, {}^3G_{o(b)(c)(d)(e)}\, {}^3e_{(b)u}\,
{}^3{\tilde \pi}^u_{(c)}\, {}^3e_{(d)v}\, {}^3{\tilde \pi}^v_{(e)}\Big]
(\tau ,\vec \sigma )+\nonumber \\
&+&{{\partial}\over {\partial \sigma^s}} \Big[ N_{(b)}\, {}^3e^s_{(b)}\,
{}^3{\tilde \pi}^r_{(a)}\Big] (\tau ,\vec \sigma )-{}^3{\tilde \pi}^u_{(a)}
(\tau ,\vec \sigma ){{\partial}\over {\partial \sigma^u}}\Big[ N_{(b)}\, {}^3e^r
_{(b)}\Big] (\tau ,\vec \sigma )+\nonumber \\
&+&\epsilon_{(a)(b)(c)}\, {\hat \mu}_{(b)}(\tau ,\vec \sigma )\, {}^3{\tilde
\pi}^r_{(c)}(\tau ,\vec \sigma ),\nonumber \\
&&\Downarrow \nonumber \\
\partial_{\tau}\, {}^3e^r_{(a)}(\tau ,\vec \sigma )
&=&-\Big[ {}^3e^r_{(b)}\, {}^3e^s_{(a)} \partial_{\tau}\, {}^3e_{(b)s}\Big]
(\tau ,\vec \sigma )\, {\buildrel \circ \over =}\, \nonumber \\
&{\buildrel \circ \over =}&\,
{{\epsilon}\over {4k}}\Big[ {N\over {{}^3e}}\,
{}^3G_{o(a)(b)(c)(d)}\, {}^3e_{(b)}^r\,
 {}^3e_{(c)s}\, {}^3{\tilde \pi}^s_{(d)}\Big] (\tau ,\vec \sigma )-
\nonumber \\
&-&{}^3e^s_{(a)}\Big[ N_{(c)}\, {}^3e^u_{(c)}\, {}^3e^r_{(b)}
{{\partial \, {}^3e_{(b)s}}\over {\partial
\sigma^u}}+ {{\partial}\over {\partial \sigma^s}} (N_{(c)}\,
{}^3e^r_{(c)})\Big] (\tau ,\vec \sigma )+\nonumber \\
&+&\epsilon_{(a)(b)(c)}\, {\hat \mu}_{(b)}(\tau ,\vec \sigma )\, {}^3e_{(c)}^r
(\tau ,\vec \sigma ),\nonumber \\
\partial_{\tau}\, {}^3e(\tau ,\vec \sigma )
&=&\Big[ {}^3e\, {}^3e^r_{(a)} \partial_{\tau}\, {}^3e_{(a)r}\Big] (\tau ,\vec
\sigma )\, {\buildrel \circ \over =}\nonumber \\
&{\buildrel \circ \over =}&\, {{\epsilon}\over {4k}}\Big[ N\, {}^3e_{(a)s}\,
{}^3{\tilde \pi}^s_{(a)}\Big] (\tau ,\vec \sigma )+\nonumber \\
&+&\Big( {}^3e\, \Big[ N_{(b)}\, {}^3e^s_{(b)}\, {}^3e^r_{(a)}\partial_s\,
{}^3e_{(a)r}+{}^3e^r_{(a)}\, {}^3e_{(a)s} \partial_r(N_{(b)}\, {}^3e^s_{(b)})
\Big] \Big)
(\tau ,\vec \sigma ).
\label{III20}
\end{eqnarray}

From the Hamilton equations and Eqs.(\ref{a25}), (\ref{III12}), we
get

\begin{eqnarray}
\partial_{\tau}\, {}^3g_{rs}(\tau ,\vec \sigma )\, &{\buildrel \circ \over =}\,&
\Big[ N_{r|s}+N_{s|r}-2N\, {}^3K_{rs}\Big] (\tau ,\vec \sigma ),\nonumber \\
\partial_{\tau}\, {}^3K_{rs}(\tau ,\vec \sigma )
&{\buildrel \circ \over =}\,&{1\over {4k}}\, {}^3G_{o(a)(b)(c)(d)} \Big(
{{\epsilon}\over {{}^3e}} \Big[ \partial_v(N_{(m)}\, {}^3e^v_{(m)}\, {}^3e
_{(a)r}\, {}^3e_{(b)s}\, {}^3e_{(c)u}\, {}^3{\tilde \pi}^u_{(d)})+\nonumber \\
&+&{}^3e_{(c)u}\, {}^3{\tilde \pi}^u_{(d)}\Big[ 2k N ({}^3R^{uv}-{1\over 2}\,
{}^3g^{uv}\, {}^3R)+\epsilon (N^{|u|v}-{}^3g^{uv} N^{|l}{}_{|l})\Big]
{}^3e_{(d)v}-\nonumber \\
&-&{N\over {4k\, {}^3e^2}}\Big[ {1\over 2}\, {}^3e_{(a)r}\, {}^3e_{(b)s}\,
{}^3e_{(c)u}\, {}^3G_{o(d)(e)(f)(g)}\, {}^3{\tilde \pi}^u_{(e)}\, {}^3e_{(f)v}\,
{}^3{\tilde \pi}^v_{(g)}-\nonumber \\
&-&{}^3e_{(a)r}\, {}^3e_{(b)s} \delta_{(c)(d)}\, {}^3G_{o(e)(f)(g)(h)}\,
{}^3e_{(e)u}\, {}^3{\tilde \pi}^u_{(f)}\, {}^3e_{(g)v}\, {}^3{\tilde \pi}^v
_{(h)}+\nonumber \\
&+&{}^3e_{(a)r}\, {}^3e_{(b)s} ({}^3e_{(m)v}\, {}^3{\tilde \pi}^v_{(m)}\,
{}^3e_{(c)u}\, {}^3{\tilde \pi}^u_{(d)}+{}^3G_{o(c)(e)(f)(g)}\, {}^3e_{(e)u}\,
{}^3{\tilde \pi}^u_{(d)}\, {}^3e_{(f)v}\, {}^3{\tilde \pi}^v_{(f)})+\nonumber \\
&+&({}^3e_{(a)r}\, {}^3G_{o(b)(e)(f)(g)}\, {}^3e_{(e)s}+{}^3e_{(b)s}\, {}^3G
_{o(a)(e)(f)(g)}\, {}^3e_{(e)r})\nonumber \\
&& {}^3e_{(f)u}\, {}^3{\tilde \pi}^u_{(g)}\,
{}^3e_{(c)v}\, {}^3{\tilde \pi}^v_{(d)}\Big] \Big) (\tau ,\vec \sigma )
,\nonumber \\
\partial_{\tau}\, {}^3K(\tau ,\vec \sigma )\, &{\buildrel \circ \over =}\,&
\Big( {1\over 4} N\, {}^3R+ 4N^{|r}{}_{|r}+\nonumber \\
&+&{N\over {(4k\, {}^3e)^2}} \Big[ ({}^3e_{(a)r}\, {}^3{\tilde \pi}^r_{(a)})^2-
{3\over 2}\, {}^3G_{o(a)(b)(c)(d)}\, {}^3e_{(a)r}\, {}^3{\tilde \pi}^r_{(b)}\,
{}^3e_{(c)s}\, {}^3{\tilde \pi}^s_{(d)}\Big] -\nonumber \\
&-&{{\epsilon}\over {4k\, {}^3e}}\, {}^3{\tilde \pi}^r_{(a)}\Big[ N_{(m)}\,
{}^3e^u_{(m)}\partial_u\, {}^3e_{(a)r}+{}^3e_{(a)u}\partial_r(N_{(m)}\, {}^3e
^u_{(m)})\Big] \Big) (\tau ,\vec \sigma ),\nonumber \\
\partial_{\tau}\, {}^3\omega_{r(a)(b)}(\tau ,\vec \sigma )
&{\buildrel \circ \over =}\,&{{\epsilon N}\over {4k\, {}^3e}} \Big(
(\partial_r\, {}^3e_{(b)s}-\partial_s\, {}^3e_{(b)r}){}^3G_{o(a)(l)(m)(n)}\,
{}^3e^s_{(l)}\, +\nonumber \\
&+&(\partial_s\, {}^3e_{(a)r}-\partial_r\, {}^3e_{(a)s}){}^3G_{o(b)(l)(m)(n)}\,
{}^3e^s_{(l)}+\nonumber \\
&+&(\partial_v\, {}^3e_{(c)u}-\partial_u\, {}^3e_{(c)v})\Big[ {}^3e^v_{(b)}\,
{}^3e_{(c)r}\, {}^3G_{o(a)(l)(m)(n)}\, {}^3e^u_{(l)}+\nonumber \\
&+&{}^3e^u_{(a)}\, {}^3e_{(c)r}\, {}^3G_{o(b)(l)(m)(n)}\, {}^3e^v_{(l)}-
\nonumber \\
&-&{}^3e^u_{(a)}\, {}^3e^v_{(b)}\, {}^3G_{o(c)(l)(m)(n)}\, {}^3e_{(l)r}\Big]
{}^3e_{(m)t}\, {}^3{\tilde \pi}^t_{(n)} \Big) -\nonumber \\
&-&{{\epsilon}\over {4k}}\Big( \Big[ {}^3e^s_{(a)}\, {}^3G_{o(b)(l)(m)(n)}-
{}^3e^s_{(b)}\, {}^3G_{o(a)(l)(m)(n)}\Big] \nonumber \\
&&\Big[ \partial_r({N\over {{}^3e}}\,
{}^3e_{(l)s}\, {}^3e_{(m)t}\, {}^3{\tilde \pi}^t_{(n)})-\partial_s({N\over
{{}^3e}}\, {}^3e_{(l)r}\, {}^3e_{(m)t}\, {}^3{\tilde \pi}^t_{(n)})\Big]+
\nonumber \\
&+&{}^3e^u_{(a)}\, {}^3e^v_{(b)}\, {}^3e_{(c)r}\, {}^3G_{o(c)(l)(m)(n)}
\nonumber \\
&&\Big[
\partial_v({N\over {{}^3e}}\, {}^3e_{(l)u}\, {}^3e_{(m)t}\, {}^3{\tilde \pi}^t
_{(n)})-\partial_u({N\over {{}^3e}}\, {}^3e_{(l)v}\, {}^3e_{(m)t}\,
{}^3{\tilde \pi}^t_{(n)})\Big] \Big)-\nonumber \\
&-&\Big[ (\partial_r\, {}^3e_{(b)s}-\partial_s\, {}^3e_{(b)r}){}^3e^v_{(a)}-
(\partial_r\, {}^3e_{(a)s}-\partial_s\, {}^3e_{(a)r}){}^3e^v_{(b)}\Big]
\nonumber \\
&&\Big[N_{(w)}\, {}^3e^u_{(w)}\, {}^3e^s_{(l)}\partial_u(N_{(w)}\, {}^3e^s
_{(w)})\Big] -\nonumber \\
&-&(\partial_v\, {}^3e_{(c)u}-\partial_u\, {}^3e_{(c)v})\Big[ ({}^3e^v_{(b)}\,
{}^3e^t_{(a)}+{}^3e^u_{(a)}\, {}^3e^t_{(b)}){}^3e_{(c)r}\nonumber \\
&&\Big[ N_{(m)}\, {}^3e^w_{(m)}\, {}^3e^u_{(l)}\partial_w\, {}^3e_{(l)t}+
\partial_t(N_{(w)}\, {}^3e^u_{(w)})\Big] +\nonumber \\
&+&{}^3e^u_{(a)}\, {}^3e^v_{(b)} \Big( N_{(m)}\, {}^3e^w_{(m)}\partial_w\,
{}^3e_{(c)r}+{}^3e_{(c)w}\partial_r(N_{(m)}\, {}^3e^w_{(m)}\Big) \Big] +
\nonumber \\
&+&{}^3e^s_{(a)}\Big( \partial_r(N_{(w)}\, {}^3e^u_{(w)}\partial_u\, {}^3e
_{(b)s}+{}^3e_{(b)u}\partial_s(N_{(w)}\, {}^3e^u_{(w)}) )-\nonumber \\
&-&\partial_s(
N_{(w)}\, {}^3e^u_{(w)}\partial_u\, {}^3e
_{(b)r}+{}^3e_{(b)u}\partial_r(N_{(w)}\, {}^3e^u_{(w)}) )\Big) -\nonumber \\
&-&{}^3e^s_{(b)}\Big( \partial_r(N_{(w)}\, {}^3e^u_{(w)}\partial_u\, {}^3e
_{(a)s}+{}^3e_{(a)u}\partial_s(N_{(w)}\, {}^3e^u_{(w)}) )-\nonumber \\
&-&\partial_s(N_{(w)}\, {}^3e^u_{(w)}\partial_u\, {}^3e
_{(a)r}+{}^3e_{(a)u}\partial_r(N_{(w)}\, {}^3e^u_{(w)}) )\Big) +\nonumber \\
&+&{}^3e^u_{(a)}\, {}^3e^v_{(b)}\, {}^3e_{(c)r}
\Big( \partial_v(N_{(w)}\, {}^3e^t_{(w)}\partial_t\, {}^3e
_{(c)u}+{}^3e_{(c)t}\partial_u(N_{(w)}\, {}^3e^t_{(w)}) )-\nonumber \\
&-&\partial_u(N_{(w)}\, {}^3e^t_{(w)}\partial_t\, {}^3e
_{(c)v}+{}^3e_{(c)t}\partial_v(N_{(w)}\, {}^3e^t_{(w)}) )\Big) +\nonumber \\
&+&\Big( \Big[ (\partial_r\, {}^3e_{(b)s}-\partial_s\, {}^3e_{(b)r})\epsilon
_{(a)(m)(n)}-(\partial_r\, {}^3e_{(a)s}-\partial_s\, {}^3e_{(a)r})\epsilon
_{(b)(m)(n)}\Big] {}^3e^s_{(n)}+\nonumber \\
&+&(\partial_v\, {}^3e_{(c)u}-\partial_u\, {}^3e_{(c)v})\Big[ {}^3e^v_{(b)}\,
{}^3e_{(c)r}\epsilon_{(a)(m)(n)}\, {}^3e^u_{(n)}+\nonumber \\
&+&{}^3e^u_{(a)}\, {}^3e_{(c)r}\epsilon_{(b)(m)(n)}\, {}^3e^v_{(n)}+{}^3e^u
_{(a)}\, {}^3e^v_{(b)}\epsilon_{(c)(m)(n)}\, {}^3e_{(n)r}\Big] \Big) {\hat \mu}
_{(m)}+\nonumber \\
&+&\Big[ {}^3e^s_{(a)}\epsilon_{(b)(m)(n)}-{}^3e^s_{(b)}\epsilon_{(a)(m)(n)}
\Big] \Big[ \partial_r({\hat \mu}_{(m)}\, {}^3e_{(n)s})-\partial_s({\hat \mu}
_{(m)}\, {}^3e_{(n)r})\Big]+\nonumber \\
&+&{}^3e^u_{(a)}\, {}^3e^v_{(b)}\, {}^3e_{(c)r}\epsilon_{(c)(m)(n)}\Big[
\partial_v({\hat \mu}_{(m)}\, {}^3e_{(n)u})-\partial_u({\hat \mu}_{(m)}\,
{}^3e_{(n)v})\Big] .
\label{III21}
\end{eqnarray}

\subsection{Comparison with Other Approaches to Tetrad Gravity.}

Let us consider the canonical transformation

\bea
{}{}&&{\tilde \pi}^N(\tau ,\vec \sigma )
\, dN(\tau ,\vec \sigma )+{\tilde \pi}^{\vec N}_{(a)}(\tau ,\vec \sigma )\,
dN_{(a)}(\tau ,\vec \sigma )+{\tilde \pi}^{\vec \varphi}_{(a)}(\tau ,\vec
\sigma )\, d\varphi_{(a)}(\tau ,\vec \sigma )+\nonumber \\
 &+&{}^3{\tilde \pi}^r_{(a)}(\tau ,
\vec \sigma )\, d{}^3e_{(a)r}(\tau ,\vec \sigma )=
 ={}^4{\tilde \pi}^A_{(\alpha )}
(\tau ,\vec \sigma )\, d{}^4E^{(\alpha )}_A(\tau ,\vec \sigma ),
\label{III22}
\eea

\noindent where ${}^4{\tilde \pi}^A_{(\alpha )}$ \footnote{$\lbrace {}^4E^{(\alpha
)}_A(\tau ,\vec
\sigma ),{}^4{\tilde \pi}^B_{(\beta )}(\tau ,{\vec \sigma}^{'})
\rbrace =\delta^B_A\delta^{(\alpha )}
_{(\beta )}\delta^3(\vec \sigma ,{\vec \sigma}^{'})$.}
would be the canonical momenta if the ADM
action would be considered as a functional of the cotetrads ${}^4E^{(\alpha )}
_A={}^4E^{(\alpha )}_{\mu}\, b^{\mu}_A$ in the holonomic
$\Sigma_{\tau}$-adapted basis, as essentially is done in Refs.\cite{hen3,hen2}.
If $\bar \gamma =\sqrt{1+\sum_{(c)}\varphi^{(c) 2}}$, we have

\begin{eqnarray}
{\tilde \pi}^N&=& (\bar \gamma \, {}^4{\tilde \pi}^{\tau}_{(o)}+\varphi^{(a)}\,
{}^4{\tilde \pi}^{\tau}_{(a)}),\nonumber \\
{\tilde \pi}^{\vec N}_{(a)}&=&-\epsilon
\varphi_{(a)}\, {}^4{\tilde \pi}^{\tau}_{(o)}+
[\delta_{(a)}^{(b)}-\epsilon {{\varphi_{(a)}\varphi^{(b)}}\over
{1+\bar \gamma}}]\,
{}^4{\tilde \pi}^{\tau}_{(b)},\nonumber \\
{\tilde \pi}^{\vec \varphi}_{(a)}&=&({{\epsilon N}\over {\bar \gamma}}\,
\varphi_{(a)}-N_{(a)})\, {}^4{\tilde \pi}^{\tau}_{(o)}-\delta_{(a)}^{(b)}\,
N\, {}^4{\tilde \pi}^{\tau}
_{(b)}-{}^3e_{(a)r}\, {}^4{\tilde \pi}^r_{(o)}-\nonumber \\
&-&{1\over {1+\bar \gamma}}(\delta_{(a)}^{(c)}\varphi^{(b)}+\delta_{(a)}^{(b)}
\varphi^{(c)}+\epsilon
{{\varphi_{(a)}\varphi^{(b)}\varphi^{(c)}}\over {\bar \gamma
(1+\bar \gamma )}})
(N_{(c)}\, {}^4{\tilde \pi}^{\tau}_{(b)}+{}^3e_{(c)r}\, {}^4{\tilde \pi}^r
_{(b)}),\nonumber \\
{}^3{\tilde \pi}^r_{(a)}&=&-\epsilon \varphi_{(a)}\, {}^4{\tilde \pi}^r_{(o)}+
(\delta_{(a)}^{(b)}-\epsilon {{\varphi_{(a)}\varphi^{(b)}}\over
{1+\bar \gamma}})\,
{}^4{\tilde \pi}^r_{(b)},\nonumber \\
&&{}\nonumber \\
{}^4{\tilde \pi}^{\tau}_{(o)}&=&\bar \gamma {\tilde \pi}^N-\varphi^{(a)}\,
{}^3{\tilde \pi}^{\vec N}_{(a)},\nonumber \\
{}^4{\tilde \pi}^{\tau}_{(a)}&=&\epsilon \varphi_{(a)} {\tilde \pi}^N+[\delta
_{(a)}^{(b)}-\epsilon
{{\varphi_{(a)}\varphi^{(b)}}\over {1+\bar \gamma}}]
{\tilde \pi}^{\vec N}_{(b)},\nonumber \\
{}^4{\tilde \pi}^r_{(o)}&=&-\bar \gamma \,
{}^3e^r_{(a)}[\delta_{(a)}^{(b)}-\epsilon
{{\varphi_{(a)}\varphi^{(b)}}\over {1+\bar \gamma}}] {\tilde \pi}^{\vec \varphi}
_{(b)}+\bar \gamma N_{(a)}\, {}^3e^r_{(a)}\, {\tilde \pi}^N+\nonumber \\
&+&{}^3e^r_{(a)} [-N\, \delta_{(a)}^{(b)}-(\delta_{(a)}^{(b)}\varphi^{(c)}-
\delta_{(a)}^{(c)}\varphi^{(b)}){{N_{(c)}}\over
{1+\bar \gamma}}] {\tilde \pi}^{\vec N}
_{(b)}-\nonumber \\
&-&{1\over {1+\bar \gamma}}\,
{}^3e^r_{(a)} [\delta^{(a)(b)}\varphi^{(c)}+{1\over
{\bar \gamma}} \delta^{(b)(c)}\varphi^{(a)}]{}^3e_{(c)s}\,
{}^3{\tilde \pi}^s_{(b)},\nonumber \\
{}^4{\tilde \pi}^r_{(a)}&=&[\delta_{(a)}^{(b)}+\epsilon
{{\varphi_{(a)}\varphi^{(b)}}\over
{\bar \gamma (1+\bar \gamma )}}] {}^3{\tilde \pi}^r_{(b)}-\epsilon
\varphi_{(a)}\, {}^3e^r_{(b)}[\delta_{(b)}^{(c)}-\epsilon
{{\varphi_{(b)}\varphi^{(c)}}\over {1+\bar \gamma}}]{\tilde \pi}
^{\vec \varphi}_{(c)}+\nonumber \\
&+&\epsilon \varphi_{(a)}\, {}^3e^r_{(b)} N_{(b)}{\tilde \pi}^N-\epsilon
\varphi_{(a)}\, {}^3e^r_{(b)}[N\delta^{(b)}_{(c)}+(\delta^{(b)}_{(c)}
\varphi^{(d)}-\delta_{(c)}^{(d)}
\varphi^{(b)}){{N_{(d)}}\over {1+\bar \gamma}}] {\tilde \pi}^{\vec N}_{(c)}-
\nonumber \\
&-&\epsilon
{{\varphi_{(a)}}\over {1+\bar \gamma}} [\varphi^{(b)}\delta^{(c)}_{(d)}+{1\over
{\bar \gamma}} \varphi^{(c)} \delta^{(b)}_{(d)}]{}^3e^r_{(c)}\, {}^3e_{(b)s}\,
{}^3{\tilde \pi}^s_{(d)}.
\label{III23}
\end{eqnarray}

Our canonical transformation (\ref{III23}) allows to consider the
metric ADM Lagrangian as function of the cotetrads ${}^4E^{(\alpha
)}_A={}^4E^{(\alpha )}
_{\mu}\, b_A^{\mu}$ and to find the conjugate momenta ${}^4{\tilde \pi}^A
_{(\alpha )}$. Eqs.(\ref{III23}) show that the four primary constraints, which
contain the informations ${\tilde \pi}^N\approx 0$ and ${\tilde
\pi}^{\vec N}_{(a)}\approx 0$, are ${}^4{\tilde \pi}^{\tau}_{(\alpha
)}\approx 0$. The six primary constraints ${}^4{\tilde M}_{(\alpha
)(\beta )}={}^4E^{(\gamma )}_A[{}^4\eta_{(\alpha )(\gamma )}\,
{}^4{\tilde \pi}^A_{(\beta )}-{}^4\eta_{(\beta )(\gamma )}\,
{}^4{\tilde \pi}^A_{(\alpha )}] \approx 0$, generators of the local
Lorentz transformations in this formulation, have the following
relation with ${\tilde \pi}
^{\vec \varphi}_{(a)}\approx 0$ and ${}^3{\tilde M}_{(a)}\approx 0$

\begin{eqnarray}
{}^4{\tilde M}_{(a)(b)}&=&-\epsilon {}^3{\tilde M}_{(a)(b)}+ (\varphi
_{(a)} {\tilde \pi}^{\vec \varphi}_{(b)}-\varphi_{(b)} {\tilde \pi}^{\vec
\varphi}_{(a)})+\epsilon (\varphi_{(a)}N_{(b)}-\varphi_{(b)}N_{(a)}) {\tilde
\pi}^N-\nonumber \\
&-&(\delta^{(c)}_{(a)}\delta^{(d)}_{(b)}-\delta^{(c)}_{(b)}\delta^{(d)}_{(a)})
[\epsilon N \varphi_{(c)} \delta_{(d)(e)}+\nonumber \\
&+&(\delta_{(c)(f)}+{{\varphi_{(c)}\varphi_{(f)}}\over {1+\bar \gamma}})
(\delta_{(d)(e)}+{{\varphi_{(d)}\varphi_{(e)}}\over {1+\bar \gamma}}) N_{(f)}]
{\tilde \pi}^{\vec N}_{(e)}\approx 0,\nonumber \\
{}^4{\tilde M}_{(a)(o)}&=&-\epsilon \bar \gamma
{\tilde \pi}^{\vec \varphi}_{(a)}-
{{1}\over {1+\bar \gamma}}\, {}^3{\tilde M}_{(a)(b)}\varphi_{(b)}-\epsilon
\bar \gamma (\delta_{(a)(b)}-{{\varphi_{(a)}\varphi_{(b)}}\over {\bar \gamma (1+
\bar \gamma )}})N_{(b)}{\tilde \pi}^N+\nonumber \\
&+& [-\epsilon \bar \gamma N(\delta_{(a)(b)}-{{\varphi_{(a)}\varphi_{(b)}}\over
{\bar \gamma (1+\bar \gamma )}})+
\varphi_{(c)}N_{(c)}\delta_{(a)(b)}-N_{(a)}\varphi
_{(b)}]{\tilde \pi}^{\vec N}_{(b)}\approx 0,\nonumber \\
&&{}\nonumber \\
{}^3{\tilde M}_{(a)(b)}&=&-\epsilon {}^4{\tilde M}_{(a)(b)}+{{\epsilon}\over
{1+\bar \gamma}}
[\varphi_{(a)}\, {}^4{\tilde M}_{(b)(c)}-\varphi_{(b)}\, {}^4{\tilde
M}_{(a)(c)}]\varphi_{(c)}+\nonumber \\
&+& [\varphi_{(a)}\, {}^4{\tilde M}_{(b)(o)}-\varphi_{(b)}\, {}^4
{\tilde M}_{(a)(o)}]- [\varphi_{(a)}\, {}^4E^{\tau}_{(b)}-
\varphi_{(b)}\, {}^4E^{\tau}_{(a)}]{}^4{\tilde \pi}^{\tau}_{(o)}-\nonumber \\
&-&\epsilon [(\delta_{(a)(c)}\delta_{(d)(e)}-\delta_{(a)(e)}\delta_{(b)(c)})
(\delta_{(c)(d)}+{{\varphi_{(c)}\varphi_{(d)}}\over {1+\bar \gamma}})
{}^4E^{\tau}_{(c)}+\nonumber \\
&+&\epsilon ({}^4E^{\tau}_{(o)}+\epsilon
{{\varphi_{(c)}\, {}^4E^{\tau}_{(c)}}\over {1+\bar \gamma }})
(\delta_{(a)(d)}\varphi_{(b)}-\delta_{(b)(d)}\varphi_{(a)})]{}^4{\tilde \pi}
^{\tau}_{(d)}\approx 0,\nonumber \\
{\tilde \pi}^{\vec \varphi}_{(a)}&=&\epsilon (\delta_{(a)(b)}-{{\varphi_{(a)}
\varphi_{(b)}}\over {\bar \gamma (1+\bar \gamma )}})
{}^4{\tilde M}_{(b)(o)}+{{1}
\over {1+\bar \gamma}}\, {}^4{\tilde M}_{(a)(b)} \varphi_{(b)}-\nonumber \\
&-&\epsilon (\delta_{(a)(b)}-{{\varphi_{(a)}\varphi_{(b)}}\over {\bar \gamma (1+
\bar \gamma )}})\,
{}^4E^{\tau}_{(b)}\, {}^4{\tilde \pi}^{\tau}_{(o)}+\nonumber \\
&+&[\epsilon (\delta_{(a)(b)}-{{\varphi_{(a)}\varphi_{(b)}}\over {\bar \gamma
(1+\bar \gamma )}})\,
{}^4E^{\tau}_{(o)}-{{\varphi_{(c)}}\over {1+\bar \gamma}}(\delta
_{(c)(b)}\, {}^4E^{\tau}_{(a)}-\delta_{(c)(a)}\, {}^4E^{\tau}_{(b)})]
{}^4{\tilde \pi}^{\tau}_{(b)}\approx 0.
\label{III24}
\end{eqnarray}

Let us add a comment on the literature on tetrad gravity. The use of
tetrads started with Ref.\cite{weyl}, where vierbeins and spin
connections are used as independent variables in a Palatini form of
the Lagrangian. They were used by Dirac\cite{dirr} for the coupling of
gravity to fermion fields (see also Ref.\cite{hen1}) and here
$\Sigma_{\tau}$-adapted tetrads were introduced. In Ref.\cite{schw}
the reduction of this theory at the Lagrangian level was done by
introducing the so-called {\it time-gauge} ${}^4E
^{(o)}_r=0$ [or ${}^4E^o_{(a)}=0$], which distinguishes the time
coordinate $x^o=const.$ planes; in this paper there is also the
coupling to scalar fields, while in Ref.\cite{kib} the coupling to
Dirac-Maiorana fields is studied. In Ref.\cite{char} there is a
non-metric Lagrangian formulation, see Eq.(\ref{a33}), employing as
basic variables the cotetrads  ${}^4E^{(\alpha )}_{\mu}$, which is
different from our metric Lagrangian and has different primary
constraints; its Hamiltonian formulation is completely developed. See
also Ref.\cite{clay} for a study of the tetrad frame constraint
algebra. In the fourth of Refs.\cite{tetr} cotetrads ${}^4E^{(\alpha
)}_{\mu}$ together with the spin connection ${}^4\omega^{(\alpha
)}_{\mu (\beta )}$ are used as independent variables in a first order
Palatini action
\footnote{See also the Nelson-Regge papers in Refs.\cite{tetr} for a
different approach, the so-called {\it covariant canonical
formalism}.}, while in Ref.\cite{maluf} a first order Lagrangian
reformulation is done for Eq.(\ref{a33}) \footnote{In both these
papers there is a 3+1 decomposition of the tetrads different from our
and, like in Ref.\cite{maluf}, use is done of the Schwinger time gauge
to get free of three boost-like parameters.}.

Instead in most of Refs.\cite{tetr,hen2,hen3} one uses the space components
${}^4E^{(\alpha )}_r$ of cotetrads ${}^4E^{(\alpha )}_{\mu}$, together with
the conjugate momenta ${}^4{\tilde \pi}^r_{(\alpha )}$ inside the ADM
Hamiltonian, in which one puts ${}^3g_{rs}={}^4E_r^{(\alpha )}\, {}^4\eta
_{(\alpha )(\beta )}\, {}^4E^{(\beta )}_s$ and ${}^3{\tilde \Pi}^{rs}=
{1\over 4}{}^4\eta^{(\alpha )(\beta )}[{}^4E^r_{(\alpha )}\, {}^4{\tilde \pi}
^s_{(\beta )}+{}^4E^s_{(\alpha )}\, {}^4{\tilde \pi}^r_{(\beta )}]$. Lapse
and shift functions are treated as Hamiltonian multipliers and there
is no worked out Lagrangian formulation. In Ref.\cite{hen4} it is
shown how to go from the space components ${}^4E^{(\alpha )}_r$ to
cotriads ${}^3e_{(a)r}$ by using the {\it time gauge} on a surface
$x^0=const.$; here it is introduced for the first time the concept of
parameters of Lorentz boosts \footnote{If they are put equal to zero,
one recovers {\it Schwinger's time gauge}.}, which was our starting
point to arrive at the identification of the Wigner boost parameters
$\varphi_{(a)}$. Finally in Ref.\cite{gold} there is a 3+1
decomposition of tetrads and cotetrads in which some boost-like
parameters have been fixed (it is a Schwinger time gauge) so that one
can arrive at a Lagrangian (different from ours) depending only on
lapse, shift and cotriads.

In Ref.\cite{hen4} there is another canonical transformation from cotriads
and their conjugate momenta to a new canonical basis containing densitized
triads and their conjugate momenta

\begin{eqnarray}
(\, {}^3e_{(a)r}&,& {}^3{\tilde \pi}^r_{(a)}\, ) \mapsto
(\,\,\, {}^3{\tilde h}^r_{(a)}
={}^3e\, {}^3e^r_{(a)},\nonumber \\
&&2\, {}^3K_{(a)r}=
2[{}^3e^s_{(a)}\, {}^3K_{sr}+{1\over {4\, {}^3e}} {}^3{\tilde M}_{(a)(b)}\,
{}^3e_{(b)r}]=\nonumber \\
&=&{1\over 2}[{1\over k}{}^3G_{o(a)(b)(c)(d)}\, {}^3e_{(b)r}\,
{}^3e_{(c)u}\, {}^3{\tilde \pi}^u_{(d)}+{1\over {{}^3e}}{}^3{\tilde M}
_{(a)(b)}\, {}^3e_{(b)r}]\, ),
\label{III25}
\end{eqnarray}

\noindent which is used to make the transition to the complex Ashtekar
variables \cite{ash}

\begin{equation}
(\, {}^3{\tilde h}^r_{(a)},\quad\quad {}^3A_{(a)r}=2\,\, {}^3K_{(a)r}+i
\, {}^3\omega_{r(a)}\, ),
\label{III26}
\end{equation}

\noindent where ${}^3A_{(a)r}$ is a zero density whose real part (in this
notation) can be considered the
gauge potential of the Sen connection  and plays an
important role in the simplification of the functional form of the constraints
present in this approach; the
conjugate variable is a density 1 SU(2) soldering form.

\vfill\eject

\section{Comparison with ADM Canonical Metric Gravity.}

In this Section  we give a brief review of the Hamiltonian formulation
of ADM metric gravity (see Refs.\cite{witt,mtw,ish,ro,fm}) to express
its constraints in terms of those of Section III.

The ADM Lagrangian $S_{ADM}=\int d\tau \, L_{ADM}(\tau )=
\int d\tau d^3\sigma {\cal L}_{ADM}(\tau ,\vec \sigma )$
 given in Eq.(\ref{III1}) is expressed  in terms of the independent
variables N, $N_r={}^3g_{rs}N^s$, ${}^3g_{rs}$.

The Euler-Lagrange equations are

\begin{eqnarray}
L_N&=&{{\partial {\cal L}_{ADM}}\over {\partial N}}-\partial_{\tau}
{{\partial {\cal L}_{ADM}}\over {\partial \partial_{\tau}N}}-\partial_r
{{\partial {\cal L}_{ADM}}\over {\partial \partial_rN}}=\nonumber \\
&=&-\epsilon k
\sqrt{\gamma} [{}^3R-{}^3K_{rs}\, {}^3K^{rs}+({}^3K)^2]=-2\epsilon k\,
{}^4{\bar G}_{ll}\, {\buildrel \circ
\over =}\, 0,\nonumber \\
L^r_{\vec N}&=&{{\partial {\cal L}_{ADM}}\over {\partial N_r}}-\partial_{\tau}
{{\partial {\cal L}_{ADM}}\over {\partial \partial_{\tau}N_r}}-\partial_s
{{\partial {\cal L}_{ADM}}\over {\partial \partial_sN_r}}=\nonumber \\
&=&2\epsilon k
[\sqrt{\gamma}({}^3K^{rs}-{}^3g^{rs}\, {}^3K)]_{\, |s}=2k\, {}^4{\bar G}_l{}^r
\, {\buildrel \circ \over =}\, 0,\nonumber \\
L_g^{rs}&=& -\epsilon k \Big[
{{\partial}\over {\partial \tau}}[\sqrt{\gamma}({}^3K^{rs}-{}^3g
^{rs}\, {}^3K)]\, - N\sqrt{\gamma}({}^3R^{rs}-
{1\over 2} {}^3g^{rs}\, {}^3R)+\nonumber \\
&&+2N\, \sqrt{\gamma}({}^3K^{ru}\, {}^3K_u{}^s-{}^3K\, {}^3K^{rs})+{1\over 2}N
\sqrt{\gamma}[({}^3K)^2-{}^3K_{uv}\, {}^3K^{uv}){}^3g^{rs}+\nonumber \\
&&+\sqrt{\gamma} ({}^3g^{rs} N^{|u}{}_{|u}-N^{|r |s})\Big] =-\epsilon kN
\sqrt{\gamma}\, {}^4{\bar G}^{rs}\, {\buildrel \circ \over =}\,0,
\label{V1}
\end{eqnarray}

\noindent and correspond to the Einstein equations in the form ${}^4{\bar G}
_{ll}\, {\buildrel \circ \over =}\, 0$, ${}^4{\bar G}_{lr}\, {\buildrel \circ
\over =}\, 0$, ${}^4{\bar G}_{rs}\, {\buildrel \circ \over =}\, 0$,
respectively. As said after Eq.(\ref{a10}) the four contracted Bianchi
identities imply that only two of the equations $L^{rs}_g\, {\buildrel
\circ
\over =}\, 0$ are independent.

The canonical momenta (densities of weight -1) are

\begin{eqnarray}
&&{\tilde \Pi}^N(\tau ,\vec \sigma )={{\delta S_{ADM}}\over {\delta
\partial_{\tau}N(\tau ,\vec \sigma )}} =0,\nonumber \\
&&{\tilde \Pi}^r_{\vec N}(\tau ,\vec \sigma )={{\delta S_{ADM}}\over
{\delta \partial_{\tau} N_r(\tau ,\vec \sigma )}} =0,\nonumber \\
&&{}^3{\tilde \Pi}^{rs}(\tau ,\vec \sigma )={{\delta S_{ADM}}\over
{\delta \partial_{\tau} {}^3g_{rs}(\tau ,\vec \sigma )}}=\epsilon k\, [
\sqrt{\gamma}({}^3K^{rs}-{}^3g^{rs}\, {}^3K)](\tau ,\vec \sigma ),\nonumber \\
&&{}\nonumber \\
&&{}^3K_{rs}={{\epsilon}\over {k\sqrt{\gamma}}} [{}^3{\tilde \Pi}_{rs}-{1\over
2}{}^3g_{rs}\, {}^3{\tilde \Pi}],\quad\quad {}^3\tilde \Pi ={}^3g_{rs}\,
{}^3{\tilde \Pi}^{rs}=-2\epsilon k\sqrt{\gamma}\, {}^3K,
\label{IV2}
\end{eqnarray}

\noindent and satisfy the Poisson brackets

\begin{eqnarray}
&&\lbrace N(\tau ,\vec \sigma ),{\tilde \Pi}^N(\tau ,{\vec \sigma}^{'} )
\rbrace =\delta^3(\vec \sigma ,{\vec \sigma}^{'}),\nonumber \\
&&\lbrace N_r(\tau ,\vec \sigma ),{\tilde \Pi}^s_{\vec N}(\tau ,{\vec \sigma}
^{'} )\rbrace =\delta^s_r \delta^3(\vec \sigma ,{\vec \sigma}^{'}),\nonumber \\
&&\lbrace {}^3g_{rs}(\tau ,\vec \sigma ),{}^3{\tilde \Pi}^{uv}(\tau ,{\vec
\sigma}^{'}\rbrace = {1\over 2} (\delta^u_r\delta^v_s+\delta^v_r\delta^u_s)
\delta^3(\vec \sigma ,{\vec \sigma}^{'}).
\label{IV3}
\end{eqnarray}

Let us introduce  the Wheeler- DeWitt supermetric

\begin{equation}
{}^3G_{rstw}(\tau ,\vec \sigma )=[{}^3g_{rt}\, {}^3g_{sw}+{}^3g_{rw}\, {}^3g
_{st}-{}^3g_{rs}\, {}^3g_{tw}](\tau ,\vec \sigma ),
\label{IV4}
\end{equation}

\noindent whose inverse is defined by the equations

\begin{eqnarray}
{}&&{1\over 2} {}^3G_{rstw}\, {1\over 2} {}^3G^{twuv} ={1\over 2}(\delta^u_r
\delta^v_s+\delta^v_r\delta^u_s),\nonumber \\
{}^3G^{twuv}(\tau ,\vec \sigma )&=&[{}^3g^{tu}\, {}^3g^{wv}+{}^3g^{tv}\, {}^3g
^{wu}-2\, {}^3g^{tw}\, {}^3g^{uv}](\tau ,\vec \sigma ).
\label{IV5}
\end{eqnarray}

Then we get

\begin{eqnarray}
{}^3{\tilde \Pi}^{rs}(\tau ,\vec \sigma )&=&{1\over 2}\epsilon k \sqrt{\gamma}\,
{}^3G^{rsuv}(\tau ,\vec \sigma )\, {}^3K_{uv}(\tau ,\vec \sigma ),\nonumber \\
{}^3K_{rs}(\tau ,\vec \sigma )&=&{{\epsilon}\over {2k\sqrt{\gamma}}}\,
{}^3G_{rsuv}(\tau ,\vec \sigma )\, {}^3{\tilde \Pi}^{uv}(\tau ,\vec \sigma ),
\nonumber \\
&&[{}^3K^{rs}\, {}^3K_{rs}-({}^3K)^2](\tau ,\vec \sigma )=\nonumber \\
&&=k^{-2}[\gamma^{-1}({}^3{\tilde \Pi}^{rs}\, {}^3{\tilde \Pi}_{rs}-{1\over 2}
({}^3{\tilde \Pi})^2](\tau ,\vec \sigma )=(2k)^{-1}[\gamma^{-1}\, {}^3G_{rsuv}
\, {}^3{\tilde \Pi}^{rs}\, {}^3{\tilde \Pi}^{uv}](\tau ,\vec \sigma ),
\nonumber \\
\partial_{\tau}\, {}^3g_{rs}(\tau ,\vec \sigma )&=&[N_{r|s}+N_{s|r}-{{\epsilon
N}\over {k\sqrt{\gamma}}}\, {}^3G_{rsuv}\, {}^3{\tilde \Pi}^{uv}](\tau ,
\vec \sigma ).
\label{IV6}
\end{eqnarray}

Since ${}^3{\tilde \Pi}^{rs}\partial_{\tau}\, {}^3g_{rs}=$${}^3{\tilde \Pi}
^{rs} [N_{r | s}+N_{s | r}-{{\epsilon N}\over {k\sqrt{\gamma}}} {}^3G_{rsuv}
{}^3{\tilde \Pi}^{uv}]=$$-2 N_r {}^3{\tilde \Pi}^{rs}{}_{|
s}-{{\epsilon N}\over {k\sqrt{\gamma}}}\, {}^3G_{rsuv}\, {}^3{\tilde
\Pi}^{rs} {}^3{\tilde \Pi}^{uv}+ (2N_r\, {}^3{\tilde \Pi}^{rs})_{|
s}$, we obtain the canonical Hamiltonian \footnote{Since $N_r\,
{}^3{\tilde \Pi}^{rs}$ is a vector density of weight -1, we have
${}^3\nabla_s(N_r\, {}^3{\tilde \Pi}^{rs})=\partial_s(N_r\,
{}^3{\tilde \Pi}^{rs})$.}

\begin{eqnarray}
H_{(c)ADM}&=& \int_Sd^3\sigma \, [{\tilde \Pi}^N\partial_{\tau}N+{\tilde
\Pi}^r_{\vec N}\partial_{\tau}N_r+{}^3{\tilde \Pi}^{rs}\partial_{\tau}
{}^3g_{rs}](\tau ,\vec \sigma ) -L_{ADM}=\nonumber \\
&=&\int_Sd^3\sigma \, [\epsilon N(k\sqrt{\gamma}\, {}^3R-
{1\over {2k\sqrt{\gamma}}} {}^3G_{rsuv}{}^3{\tilde \Pi}^{rs} {}^3{\tilde
\Pi}^{uv})-2N_r\, {}^3{\tilde \Pi}^{rs}{}_{| s}](\tau ,\vec \sigma )+
\nonumber \\
&+&2\int_{\partial S}d^2\Sigma_s [N_r\, {}^3{\tilde \Pi}^{rs}\,\,
](\tau ,\vec \sigma ),
\label{IV7}
\end{eqnarray}

\noindent In the following discussion we shall omit the surface term.

The Dirac Hamiltonian is [the $\lambda (\tau ,\vec \sigma )$'s are arbitrary
Dirac multipliers]

\begin{equation}
H_{(D)ADM}=H_{(c)ADM}+\int d^3\sigma \, [\lambda_N\, {\tilde \Pi}^N + \lambda
^{\vec N}_r\, {\tilde \Pi}^r_{\vec N}](\tau ,\vec \sigma ).
\label{IV8}
\end{equation}

The $\tau$-constancy of the primary constraints
\footnote{$\partial_{\tau} {\tilde
\Pi}^N(\tau ,\vec \sigma )=\lbrace {\tilde \Pi}^N(\tau ,\vec \sigma ),H_{(D)
ADM}\rbrace \approx 0$, $\partial_{\tau} {\tilde \Pi}^r_{\vec N}(\tau ,\vec
\sigma )=\lbrace {\tilde \Pi}^r_{\vec N}(\tau ,\vec \sigma ),H_{(D)ADM}
\rbrace \approx 0$.} generates four secondary constraints (all 4 are densities
of weight -1) which correspond to the Einstein equations ${}^4{\bar
G}_{ll} (\tau ,\vec \sigma )\, {\buildrel \circ \over =}\, 0$,
${}^4{\bar G}_{lr} (\tau ,\vec \sigma )\, {\buildrel \circ \over =}\,
0$ [see after Eqs.(\ref{a10})]

\begin{eqnarray}
{\tilde {\cal H}}(\tau ,\vec \sigma )&=&\epsilon
[k\sqrt{\gamma}\, {}^3R-{1\over {2k
\sqrt{\gamma}}} {}^3G_{rsuv}\, {}^3{\tilde \Pi}^{rs}\, {}^3{\tilde \Pi}^{uv}]
(\tau ,\vec \sigma )=\nonumber \\
&=&\epsilon [\sqrt{\gamma}\, {}^3R-{1\over {k\sqrt{\gamma}}}({}^3{\tilde \Pi}
^{rs}\, {}^3{\tilde \Pi}_{rs}-{1\over 2}({}^3\tilde \Pi )^2)](\tau ,\vec
\sigma )=\nonumber \\
&=&\epsilon k \{ \sqrt{\gamma} [{}^3R-({}^3K_{rs}\,
{}^3K^{rs}-({}^3K)^2 )]\} (\tau ,\vec \sigma )\approx 0,\nonumber \\
 &&{}\nonumber \\
{}^3{\tilde {\cal H}}^r(\tau ,\vec \sigma )&=&-2\, {}^3{\tilde \Pi}^{rs}{}_{| s}
(\tau ,\vec \sigma )=-2[\partial_s\, {}^3{\tilde \Pi}^{rs}+{}^3\Gamma^r_{su}
{}^3{\tilde \Pi}^{su}](\tau ,\vec \sigma )=\nonumber \\
&=&-2\epsilon k \{ \partial_s[\sqrt{\gamma}({}^3K^{rs}-{}^3g^{rs}\, {}^3K)]+
{}^3\Gamma^r_{su}\sqrt{\gamma}({}^3K^{su}-{}^3g^{su}\, {}^3K) \}
(\tau ,\vec \sigma )\approx 0,
\label{IV9}
\end{eqnarray}

\noindent so that we have

\begin{equation}
H_{(c)ADM}= \int d^3\sigma [N\, {\tilde {\cal H}}+N_r\, {}^3{\tilde
{\cal H}}^r](\tau ,\vec \sigma )\approx 0,
\label{IV10}
\end{equation}

\noindent with ${\tilde {\cal H}}(\tau ,\vec \sigma )\approx 0$ called the
{\it superhamiltonian} constraint and ${}^3{\tilde {\cal H}}^r(\tau
,\vec \sigma )\approx 0$ called the {\it supermomentum} constraints.
See Ref.\cite{tei} for their interpretation as the generators of the
change of the canonical data ${}^3g_{rs}$, ${}^3{\tilde \Pi}^{rs}$,
under the normal and tangent deformations of the spacelike
hypersurface $\Sigma_{\tau}$ which generate $\Sigma_{\tau +d\tau}$
\footnote{One thinks to $\Sigma_{\tau}$ as determined by a cloud of
observers, one per space point; the idea of bifurcation and
reencounter of the observers is expressed by saying that the data on
$\Sigma_{\tau}$ (where the bifurcation took place) are propagated to
some final $\Sigma_{\tau + d\tau}$ (where the reencounter arises)
along different intermediate paths, each path being a monoparametric
family of surfaces that fills the sandwich in between the two
surfaces; embeddability of $\Sigma_{\tau}$ in $M^4$ becomes the
synonymous with path independence; see also Ref.\cite{gr13} for the
connection with the theorema egregium of Gauss.}.

In ${\tilde {\cal H}}(\tau ,\vec \sigma )\approx 0$ we can say that
the term $-\epsilon k \sqrt{\gamma}({}^3K_{rs}\, {}^3K^{rs}-{}^3K^2)$
is the kinetic energy and $\epsilon k\sqrt{\gamma}\, {}^3R$ the
potential energy: in any Ricci flat spacetime (i.e. one satisfying
Einstein's empty-space equations) the extrinsic and intrinsic scalar
curvatures of any spacelike hypersurface $\Sigma_{\tau}$ are both
equal to zero (also the converse is true\cite{jawh}).

All the constraints are first class, because the only non-identically zero
Poisson brackets correspond to the so called universal Dirac algebra
\cite{dirac}:

\begin{eqnarray}
\lbrace {}^3{\tilde {\cal H}}_r(\tau ,\vec \sigma ),{}^3{\tilde {\cal H}}_s
(\tau ,{\vec \sigma}^{'})\rbrace &=&{}\nonumber \\
&=&{}^3{\tilde {\cal H}}_r(\tau ,{\vec
\sigma}^{'} )\, {{\partial \delta^3(\vec \sigma ,{\vec \sigma}^{'})}\over
{\partial \sigma^s}} + {}^3{\tilde {\cal H}}_s(\tau ,\vec \sigma ) {{\partial
\delta^3(\vec \sigma ,{\vec \sigma}^{'})}\over {\partial \sigma^r}},
\nonumber \\
\lbrace {\tilde {\cal H}}(\tau ,\vec \sigma ),{}^3{\tilde {\cal H}}_r(\tau ,
{\vec \sigma}^{'})\rbrace &=& {\tilde {\cal H}}(\tau ,\vec \sigma )
{{\partial \delta^3(\vec \sigma ,{\vec \sigma}^{'})}\over {\partial \sigma^r}},
\nonumber \\
\lbrace {\tilde {\cal H}}(\tau ,\vec \sigma ),{\tilde {\cal H}}(\tau ,{\vec
\sigma}^{'})\rbrace &=&[{}^3g^{rs}(\tau ,\vec \sigma ) {}^3{\tilde {\cal H}}_s
(\tau ,\vec \sigma )+\nonumber \\
&+&{}^3g^{rs}(\tau ,{\vec \sigma}^{'}){}^3{\tilde
{\cal H}}_s(\tau ,{\vec \sigma}^{'})]{{\partial \delta^3(\vec \sigma ,{\vec
\sigma}^{'})}\over {\partial \sigma^r}},
\label{IV11}
\end{eqnarray}

\noindent with ${}^3{\tilde {\cal H}}_r={}^3g_{rs}\, {}^3{\tilde {\cal H}}^r$
as the combination of the supermomentum constraints satisfying the
algebra of 3-diffeomorphisms. In Ref.\cite{tei} it is shown that
Eqs.(\ref{IV11}) are sufficient conditions for the embeddability of
$\Sigma_{\tau}$ into $M^4$. In the second paper in Ref.\cite{kuchar}
it is shown that the last two lines of the Dirac algebra are the
equivalent in phase space of the contracted Bianchi identities
${}^4G^{\mu\nu}{}_{;\nu}\equiv 0$.

The Hamilton-Dirac equations are

\begin{eqnarray}
\partial_{\tau}N(\tau ,\vec \sigma )\, &{\buildrel \circ \over =}\,&
\lbrace N(\tau ,\vec \sigma ),H_{(D)ADM}
\rbrace =\lambda_N(\tau ,\vec \sigma ),\nonumber \\
\partial_{\tau}N_r(\tau ,\vec \sigma )\, &{\buildrel \circ \over =}\,&
\lbrace N_r(\tau ,\vec \sigma ),
H_{(D)ADM}\rbrace =\lambda^{\vec N}_r(\tau ,\vec \sigma ),\nonumber \\
\partial_{\tau}\, {}^3g_{rs}(\tau ,\vec \sigma )\, &{\buildrel \circ \over =}\,&
\lbrace {}^3g_{rs}(\tau ,
\vec \sigma ),H_{(D)ADM}\rbrace =[N_{r | s}+N_{s | r}-{{2\epsilon N}\over
{k\sqrt{\gamma}}}({}^3{\tilde \Pi}_{rs}-{1\over 2}{}^3g_{rs}\, {}^3{\tilde
\Pi})](\tau ,\vec \sigma )=\nonumber \\
&=&[N_{r|s}+N_{s|r}-2N\, {}^3K_{rs}](\tau ,\vec \sigma ),\nonumber \\
\partial_{\tau}\, {}^3{\tilde \Pi}^{rs}(\tau ,\vec \sigma )
\, &{\buildrel \circ \over =}\,& \lbrace {}^3
{\tilde \Pi}^{rs}(\tau ,\vec \sigma ),H_{(D)ADM}\rbrace =\epsilon [N\,
k\sqrt{\gamma} ({}^3R^{rs}-{1\over 2}{}^3g^{rs}\, {}^3R)](\tau ,\vec \sigma )-
\nonumber \\
&-&2\epsilon [{N\over {k\sqrt{\gamma}}}({1\over 2}{}^3{\tilde \Pi}\, {}^3{\tilde
\Pi}^{rs}-{}^3{\tilde \Pi}^r{}_u\, {}^3{\tilde \Pi}^{us})(\tau ,\vec \sigma )-
\nonumber \\
&-&{{\epsilon N}\over 2}
{{{}^3g^{rs}}\over {k\sqrt{\gamma}}}({1\over 2}{}^3{\tilde \Pi}^2-{}^3{\tilde
\Pi}_{uv}\, {}^3{\tilde \Pi}^{uv})](\tau ,\vec \sigma )+\nonumber \\
&+&{\cal L}_{\vec N}\, {}^3{\tilde \Pi}^{rs}(\tau ,\vec \sigma )+\epsilon
[k\sqrt{\gamma} (N^{| r | s}-{}^3g^{rs}\, N^{| u}{}_{| u})](\tau ,\vec \sigma ),
\nonumber \\
 &&\Downarrow \nonumber \\
\partial_{\tau}\, {}^3K_{rs}(\tau ,\vec \sigma )\, &{\buildrel \circ \over =}\,&
\Big( N [{}^3R_{rs}+{}^3K\, {}^3K_{rs}-2\, {}^3K_{ru}\, {}^3K^u{}_s]-
\nonumber \\
&-&N_{|s|r}+N^u{}_{|s}\, {}^3K_{ur}+N^u{}_{|r}\, {}^3K_{us}+N^u\, {}^3K_{rs|u}
\Big) (\tau ,\vec \sigma ),\nonumber \\
\partial_{\tau}\, \gamma (\tau ,\vec \sigma )\, &{\buildrel \circ \over =}\,&
\Big( 2 \gamma [-N\, {}^3K +N^u{}_{|u}]\Big) (\tau ,\vec \sigma ),
\nonumber \\
\partial_{\tau}\, {}^3K(\tau ,\vec \sigma )\, &{\buildrel \circ \over =}\,&
\Big( N [{}^3g^{rs}\, {}^3R_{rs} +({}^3K)^2] -N_{|u}{}^{|u}+N^u\, {}^3K_{|u}
\Big) (\tau ,\vec \sigma ),
\label{IV12}
\end{eqnarray}

\noindent with
${\cal L}_{\vec N}\, {}^3{\tilde \Pi}^{rs}=-\sqrt{\gamma}\,
{}^3\nabla_u({{N^u}\over {\sqrt{\gamma}}} {}^3{\tilde \Pi}^{rs})+
{}^3{\tilde \Pi}^{ur}\, {}^3\nabla_u N^s+{}^3{\tilde \Pi}^{us}\, {}^3\nabla_u
N^r$.

Use has been done of the following variation $\delta (\sqrt{\gamma}\,
{}^3R)(\tau ,\vec \sigma )=
\int d^3\sigma_1 \lbrace (\sqrt{\gamma}\, {}^3R)(\tau ,\vec \sigma ),
{}^3{\tilde \Pi}^{rs}(\tau ,{\vec \sigma}_1) \rbrace \delta \, {}^3g_{rs}
(\tau ,{\vec \sigma}_1)=\int d^3\sigma_1 \delta \, {}^3g_{rs}(\tau ,{\vec
\sigma}_1) \{ [-\sqrt{\gamma} ({}^3R^{rs}-{1\over 2} {}^3g^{rs}\, {}^3R)](\tau
,\vec \sigma ) \delta^3(\vec \sigma ,{\vec \sigma}_1)+[\sqrt{\gamma}\,
{}^3\Gamma^n_{lm}({}^3g^{rl}\, {}^3g^{sm}-{}^3g^{rs}\, {}^3g^{lm})](\tau ,{\vec
\sigma}_1){{\partial \delta^3(\vec \sigma ,{\vec \sigma}_1)}\over {\partial
\sigma^n}}+[\sqrt{\gamma} ({}^3g^{rl}\, {}^3g^{sm}-{}^3g^{rs}\, {}^3g^{lm})]
(\tau ,{\vec \sigma}_1) {{\partial^2 \delta^3(\vec \sigma ,{\vec \sigma}_1)}
\over {\partial \sigma^l\partial \sigma^m}} \}$.\hfill\break
\hfill\break

Let us remark that the canonical transformation [${}^4g_{AB}$ and
${}^4g^{AB}$ are given in Eqs.(\ref{a3})] ${\tilde \Pi}^N\, dN+{\tilde
\Pi}^r_{\vec N}\, dN_r+{}^3{\tilde \Pi}^{rs}\, d{}^3g_{rs} =
{}^4{\tilde \Pi}^{AB}\, d{}^4g
_{AB}$ defines the following momenta conjugated to ${}^4g_{AB}$

\begin{eqnarray}
{}^4{\tilde \Pi}^{\tau\tau}&=&{{\epsilon}\over {2N}} {\tilde \Pi}^N,\nonumber \\
{}^4{\tilde \Pi}^{\tau r}&=&{{\epsilon}\over 2} ({{N^r}\over N} {\tilde \Pi}
^N-{\tilde \Pi}^r_{\vec N}),\nonumber \\
{}^4{\tilde \Pi}^{rs}&=&\epsilon ({{N^rN^s}\over {2N}} {\tilde \Pi}^N-
{}^3{\tilde \Pi}^{rs}),
\nonumber \\
&&{}\nonumber \\
&&\lbrace {}^4g_{AB}(\tau ,\vec \sigma ),{}^4{\tilde \Pi}^{CD}(\tau ,{\vec
\sigma}^{'} )\rbrace ={1\over 2}(\delta^C_A\delta^D_B+\delta^D_A\delta^C_B)
\delta^3(\vec \sigma ,{\vec \sigma}^{'}),\nonumber \\
&&{}\nonumber \\
{\tilde \Pi}^N&=& {{2\epsilon}\over {\sqrt{\epsilon {}^4g^{\tau\tau}}}}
{}^4{\tilde \Pi}^{\tau\tau},\nonumber \\
{\tilde \Pi}^r_{\vec N}&=&2\epsilon {{{}^4g^{\tau r}}\over {{}^4g^{\tau\tau}}}
{}^4{\tilde \Pi}^{\tau\tau}-2\epsilon {}^4{\tilde \Pi}^{\tau r},\nonumber \\
{}^3{\tilde \Pi}^{rs}&=&\epsilon {{{}^4g^{\tau r} {}^4g^{\tau S}}\over
{({}^4g^{\tau\tau})^2}}\, {}^4{\tilde \Pi}^{\tau\tau} -\epsilon {}^4{\tilde
\Pi}^{rs},
\label{IV13}
\end{eqnarray}

\noindent which would emerge if the ADM action would be considered function
of ${}^4g_{AB}$ instead of N, $N_r$ and ${}^3g_{rs}$.

The standard ADM momenta ${}^3{\tilde \Pi}^{rs}$, defined in Eq.
(\ref{IV2}), may now be expressed in terms of the cotriads and their
conjugate momenta of the canonical formulation of tetrad gravity given
in Section III:

\begin{eqnarray}
{}^3{\tilde \Pi}^{rs}&=&\epsilon k\sqrt{\gamma} ({}^3K^{rs}-{}^3g^{rs}\, {}^3K)=
{{1}\over 4} [{}^3e^r_{(a)}\, {}^3{\tilde \pi}^s_{(a)}+{}^3e
^s_{(a)}\, {}^3{\tilde \pi}^r_{(a)}],\nonumber \\
&\Rightarrow& {}^3{\tilde \Pi}={}^3{\tilde \Pi}^{rs}\, {}^3g_{rs}=
-2\epsilon k\sqrt{\gamma} \, {}^3K=
{{1}\over 2} {}^3e_{(a)r}\, {}^3{\tilde \pi}^r_{(a)},\nonumber \\
&&{}\nonumber \\
&&\lbrace {}^3g_{rs}(\tau ,\vec \sigma )=
{}^3e_{(a)r}(\tau ,\vec \sigma )\, {}^3e_{(a)s}(\tau ,\vec \sigma ),
{}^3{\tilde \Pi}^{uv}(\tau ,{\vec \sigma}^{'})\rbrace ={1\over 2}(\delta^u_r
\delta^v_s+\delta^u_s \delta^v_r)\delta^3(\vec \sigma ,{\vec \sigma}^{'}),
\nonumber \\
&&\lbrace {}^3{\tilde \Pi}^{rs}(\tau ,\vec \sigma ),{}^3{\tilde \Pi}^{uv}(\tau ,
{\vec \sigma}^{'})\rbrace ={1\over {8}} \delta^3(\vec \sigma ,{\vec \sigma}
^{'}) \times \nonumber \\
&&{} [{}^3g^{ru}\, {}^3e^v_{(a)}\, {}^3e^s_{(b)}+{}^3g^{rv}\, {}^3e^u_{(a)}\,
{}^3e^s_{(b)}+{}^3g^{su}\, {}^3e^v_{(a)}\, {}^3e^r_{(b)}+{}^3g^{sv}\,
{}^3e^u_{(a)}\, {}^3e^r_{(b)}](\tau ,\vec \sigma )\cdot \nonumber \\
&& \cdot {}^3{\tilde M}_{(a)(b)}(\tau ,\vec \sigma )\approx 0.\nonumber \\
&&{}
\label{IV14}
\end{eqnarray}

\noindent The fact that in tetrad gravity the last Poisson brackets is only
weakly zero has been noted in Ref.\cite{hen2}.

Let us now consider the expression of the ADM supermomentum constraints in
tetrad gravity.
Since ${}^3e_{(b)u}\, {}^3{\tilde \Pi}^{us}={{1}\over 4}{}^3e_{(b)u}
[{}^3e^u_{(a)}\, {}^3{\tilde \pi}^s_{(a)}+{}^3e^s_{(a)}\, {}^3{\tilde \pi}^u
_{(a)}]={{1}\over 4}[{}^3{\tilde \pi}^s_{(b)}+{}^3e^s_{(a)}\, {}^3e
_{(b)u}\, {}^3{\tilde \pi}^u_{(a)}]={{1}\over 4}[{}^3{\tilde \pi}^s_{(b)}
+{}^3e^s_{(a)} ({}^3e_{(a)u}\, {}^3{\tilde \pi}^u_{(b)}+{}^3{\tilde M}_{(b)(a)}
)]={{1}\over 4}[2\, {}^3{\tilde \pi}^s_{(b)}-{}^3e^s_{(a)}\,
{}^3{\tilde M}_{(a)(b)}]$, we have

\begin{eqnarray}
&&{}^3{\tilde \Pi}^{rs}{}_{|s}=\partial_s\, {}^3{\tilde \Pi}^{rs}+{}^3\Gamma^r
_{su}\, {}^3{\tilde \Pi}^{us}=\nonumber \\
&&=\partial_s\, {}^3{\tilde \Pi}^{rs}+[\epsilon_{(a)(b)(c)}\, {}^3e^r_{(a)}\,
{}^3\omega_{s(c)}-\partial_s\, {}^3e^r_{(b)}] {}^3e_{(b)u}\, {}^3{\tilde \Pi}
^{us}=\nonumber \\
&&={{1}\over 4} ( \partial_s[{}^3e^r_{(a)}\, {}^3{\tilde \pi}^s_{(a)}
+{}^3e^s_{(a)}\, {}^3{\tilde \pi}^r_{(a)}]-\nonumber \\
&&-[\epsilon_{(a)(c)(b)}\, {}^3e^r_{(a)}\, {}^3\omega_{s(c)}+\partial_s\,
{}^3e^r_{(b)}]\cdot [2\, {}^3{\tilde \pi}^s_{(b)}-{}^3e^s_{(d)}\, {}^3{\tilde M}
_{(d)(b)}]\, )=\nonumber \\
&&={{1}\over 4} \{ \, {}^3e^r_{(a)} [\partial_s\, {}^3{\tilde \pi}^s_{(a)}-2
\epsilon_{(a)(b)(c)}\, {}^3\omega_{s(b)}\, {}^3{\tilde \pi}^s_{(c)}]-
{}^3{\tilde \pi}^s_{(a)} \partial_s\, {}^3e^r_{(a)}+\nonumber \\
&&+\partial_s({}^3e^s_{(a)}\, {}^3{\tilde \pi}^r_{(a)})+[\epsilon_{(a)(c)(b)}\,
{}^3e^r_{(a)}\, {}^3\omega_{s(c)}+\partial_s\, {}^3e^r_{(b)}]{}^3e^s_{(d)}\,
{}^3{\tilde M}_{(d)(b)}\, \} =\nonumber \\
&&={{1}\over 4} \{ 2\, {}^3e^r_{(a)} {\hat {\cal H}}_{(a)}+\partial_s
[{}^3e^s_{(a)}\, {}^3{\tilde \pi}^r_{(a)}-{}^3e^r_{(a)}\, {}^3{\tilde \pi}^s
_{(a)}]-\nonumber \\
&&-[\epsilon_{(a)(b)(c)}\, {}^3e^r_{(a)}\, {}^3\omega_{s(b)}+\partial_s\, {}^3e
^r_{(c)}]{}^3e^s_{(d)}\, {}^3{\tilde M}_{(c)(d)} \} .
\label{IV15}
\end{eqnarray}

\noindent Since ${}^3{\tilde \pi}^r_{(a)}={1\over 2} {}^3e^r_{(b)} [{}^3e
_{(b)u}\, {}^3{\tilde \pi}^u_{(a)}+{}^3e_{(a)u}\, {}^3{\tilde \pi}^u_{(b)}]-
{1\over 2} {}^3{\tilde M}_{(a)(b)}\, {}^3e^r_{(b)}$, we get $\partial_s [{}^3e
^s_{(a)}\, {}^3{\tilde \pi}^r_{(a)}- {}^3e^r_{(a)}\, {}^3{\tilde \pi}^s_{(a)}]
=\partial_s [{1\over 2} ({}^3e^s_{(a)}\, {}^3e^r_{(b)}-{}^3e^r_{(a)}\,
{}^3e^s_{(b)})({}^3e_{(b)u}\, {}^3{\tilde \pi}^u_{(a)}+{}^3e_{(a)u}\,
{}^3{\tilde \pi}^u_{(b)})-{}^3e^s_{(a)}\, {}^3e^r_{(b)}\, {}^3{\tilde M}
_{(a)(b)}]=-\partial_s [{}^3e^s_{(a)}\, {}^3e^r_{(b)}\, {}^3{\tilde M}_{(a)(b)}
]$,  the ADM metric supermomentum constraints (\ref{IV9}) are
satisfied in the following form

\begin{eqnarray}
{}^3{\tilde {\cal H}}^r&=&-2{}^3{\tilde \Pi}^{rs}{}_{|s}={{1}\over 2}
\{ \, -2\, {}^3e^r_{(a)} {\hat {\cal H}}_{(a)} +\partial_s [{}^3e^s_{(a)}\,
{}^3e^r_{(b)}\, {}^3{\tilde M}_{(a)(b)}]+\nonumber \\
&+&[\partial_s\, {}^3e^r_{(c)}-\epsilon_{(c)(b)(a)}\,  {}^3\omega_{s(b)}\,
{}^3e^r_{(a)}]{}^3e^s_{(d)}\, {}^3{\tilde M}_{(c)(d)} \, \} =\nonumber \\
&=&{{1}\over 2} \{ \, 2\, {}^3e^r_{(a)}\, {}^3e^s_{(a)}\, {}^3{\tilde \Theta}
_s+[{}^3e^r_{(a)}\, {}^3\omega_{s(b)}-{}^3e^r_{(b)}\, {}^3\omega_{s(a)}]{}^3e
^s_{(a)}\, {}^3{\tilde M}_{(b)}+\nonumber \\
&+&\epsilon_{(a)(b)(c)}\, {}^3e^r_{(b)} \partial_s[{}^3e^s_{(a)}\, {}^3{\tilde
M}_{(c)}] \, \} \approx 0.
\label{IV16}
\end{eqnarray}

\vfill\eject

\section{Conclusions.}

Motivated by the attempt to get a unified description and a canonical
reduction of the four interactions in the framework of Dirac-Bergmann
theory of constraint (the presymplectic approach), with this paper we
have begun an investigation of general relativity along these lines. A
complete analysis of the canonical reduction of this theory   using
constraint theory is still lacking, probably due to the fact that it
does not respect the requirement of manifest general covariance.
Instead, the presymplectic approach is the natural one to get an
explicit control on the degrees of freedom of theories described by
singular Lagrangians at the Hamiltonian level.

We have reviewed the kinematical framework for tetrad gravity, natural
for the coupling to fermion fields, on globally hyperbolic,
asymptotically flat at spatial infinity spacetimes whose 3+1
decomposition may be obtained with simultaneity spacelike
hypersurfaces $\Sigma_{\tau}$ diffeomorphic to $R^3$.

Then, we have given a new parametrization of arbitrary cotetrads in
terms of lapse and shift functions, of cotriads on $\Sigma_{\tau}$ and
of three boost parameters. Such parametrized cotetrads are put in the
ADM action for metric gravity to obtain the new Lagrangian for tetrad
gravity. In the Hamiltonian formulation, we obtain 14 first class
constraints, ten primary and four secondary ones, whose algebra is
studied.

A comparison with other formulations of tetrad gravity and with the Hamiltonian
ADM metric gravity has been done.

In future papers based on Ref.\cite{russo2}, we shall study the
Hamiltonian group of gauge transformations induced by the first class
constraints. Then, the multitemporal equations associated with the
constraints generating space rotations and space diffeomorphisms on
the cotriads will be studied and solved. The Dirac observables with
respect to thirteen of the fourteen constraints will be found in
3-orthogonal coordinates on $\Sigma_{\tau}$ and the associated
Shanmugadhasan canonical transformation will be done. The only left
constraint to be studied will be the superhamiltonian one. Some
interpretational problems (Dirac observables versus general
covariance) \cite{rove,be} will be faced, since they are deeply
different from their counterpart in ordinary gauge theories like
Yang-Mills one.

\vfill\eject

\appendix

\section{Notations.}

In this Appendix we shall introduce the notations needed to define the
ADM tetrads and triads used in this paper together with a review of
kinematical notations.

\subsection{Pseudo-Riemanniann Geometry.}

Let $M^4$ be a torsion-free, globally hyperbolic, asymptotically flat
pseudo-Riemannian (or Lorentzian) 4-manifold, whose non-degenerate
4-metric tensor ${}^4g_{\mu\nu}(x)$ has Lorentzian signature $\epsilon
(+,-,-,-)$ with $\epsilon =\pm 1$ according to particle physics and
general relativity conventions respectively; the inverse 4-metric is
${}^4g^{\mu\nu}(x)$ with
${}^4g^{\mu\rho}(x){}^4g_{\rho\nu}(x)=\delta^{\mu}_{\nu}$. We shall
denote with Greek letters $\mu ,\nu ,..$ ($\mu =0,1,2,3$), the world
indices and with Greek letters inside round brackets $(\alpha ),
(\beta ),..,$ flat Minkowski indices \footnote{With flat 4-metric
tensor ${}^4\eta_{(\alpha )(\beta )}=\epsilon (+,-,-,-)$ in Cartesian
coordinates.}; analogously, $a, b,..,$ and $(a), (b),..,$ [a=1,2,3],
will denote world and flat 3-space indices. We shall follow the
conventions of Refs.\cite{mtw,ciuf} for $\epsilon
=-1$ and those of Ref.\cite{wei} for $\epsilon =+1$ \footnote{I.e. the
conventions of standard textbooks; see also Ref.\cite{wald} for many
results (this book is consistent with Ref.\cite{mtw}, even if its
index conventions are different).}.

The coordinates of a chart of the atlas of $M^4$ will be denoted
$\lbrace x^{\mu}\rbrace$. $M^4$ is assumed to be orientable; its
volume element in any right-handed coordinate basis is $-\eta
\sqrt{{}^4g}\, d^4x$ \footnote{$\eta$ is a sign connected with the choice of the
orientation and ${}^4g=|det\, {}^4g_{\mu\nu}|$
; with $\eta =\epsilon$ we get
the choice of Ref.\cite{mtw} for $\epsilon =-1$ and of Ref.\cite{wei}
for $\epsilon =+1$.}. In the coordinate bases $e_{\mu}=\partial_{\mu}$
and $dx^{\mu}$ for vector fields ($TM^4$) and one-forms (or covectors;
$T^{*}M^4$) respectively, the unique metric-compatible Levi-Civita
affine connection has the symmetric Christoffel symbols
${}^4\Gamma^{\mu}_{\alpha\beta} =
{}^4\Gamma^{\mu}_{\beta\alpha}={1\over 2} {}^4g^{\mu\nu}
(\partial_{\alpha}\, {}^4g_{\beta\nu} + \partial_{\beta}\,
{}^4g_{\alpha\nu} - \partial_{\nu}\, {}^4g_{\alpha\beta})$ as
connection coefficients (${}^4\Gamma^{\mu}_{\mu\nu}=\partial_{\nu}
\sqrt{{}^4g}$) and the associated covariant derivative
is denoted ${}^4\nabla_{\mu}$ (or with a semicolon ``;"):
${}^4V^{\mu}{}_{;\nu}
={}^4\nabla_{\nu}\, {}^4V^{\mu}=\partial_{\nu}\, {}^4V^{\mu}+{}^4\Gamma^{\mu}
_{\nu\alpha}\, {}^4V^{\alpha}$, with the metric
compatibility condition being ${}^4\nabla_{\rho}\, {}^4g^{\mu\nu}=0$.

The Christoffel symbols are not tensors. If, instead of the chart of
$M^4$ with coordinates $\lbrace x^{\mu}\rbrace$, we choose another
chart of $M^4$, overlapping with the previous one, with coordinates
$\lbrace x^{{'}\mu}= x^{{'}\mu}(x)\rbrace$ ($x^{{'}\mu}(x)$ are smooth
functions), in the overlap of the two charts we have the following
transformation properties under general smooth coordinate
transformations or diffeomorphisms of $M^4$ ($Diff\, M^4$) of
${}^4g_{\alpha\beta}(x)$ and of ${}^4\Gamma^{\mu}
_{\alpha\beta}(x)$ respectively

\begin{eqnarray}
&&{}^4g^{'}_{\alpha\beta}(x^{'}(x)) = {{\partial x^{\mu}}\over {\partial
x^{{'}\alpha}}}\, {{\partial x^{\nu}}\over {\partial x^{{'}\beta}}}\,
{}^4g_{\mu\nu}(x),\nonumber \\
&&{}^4\Gamma^{{'}\mu}_{\alpha\beta}(x^{'}(x)) = {{\partial x^{{'}\mu}}\over
{\partial x^{\nu}}}\, {{\partial x^{\gamma}}\over {\partial x^{{'}\alpha}}}\,
{{\partial x^{\delta}}\over {\partial x^{{'}\beta}}}\, {}^4\Gamma^{\nu}
_{\gamma\delta}(x) + {{\partial^2x^{\nu}}\over {\partial x^{{'}\alpha} \partial
x^{{'}\beta} }}\, {{\partial x^{{'}\mu}}\over {\partial x^{\nu}}}.
\label{a1}
\end{eqnarray}

For a tensor density of
weight W, ${}^4{\cal T}^{\mu ...}{}_{\alpha ...}=({}^4g)^{-W/2}\, {}^4T
^{\mu ...}{}_{\alpha ...}$, we have ${}^4{\cal T}^{\mu ...}{}_{\alpha ..;\rho}=
({}^4g)^{-W/2}[({}^4g)^{W/2}\, {}^4{\cal T}^{\mu ...}{}_{\alpha ...}]_{;\rho}=
({}^4g)^{-W/2}\, {}^4T^{\mu ...}{}_{\alpha ..;\rho}=\partial_{\rho}\,
{}^4{\cal T}^{\mu ...}{}_{\alpha ...}+{}^4\Gamma^{\mu}_{\rho\nu}\,
{}^4{\cal T}^{\nu ...}{}_{\alpha ...}+\cdots -{}^4\Gamma^{\beta}_{\rho\alpha}
{}^4{\cal T}^{\mu ...}{}_{\beta ...} -\cdots +W\, {}^4\Gamma^{\sigma}
_{\sigma\rho}\, {}^4{\cal T}^{\mu ...}{}_{\alpha ...}$ \footnote{$\partial_{\rho}
({}^4g)^{-W/2}+W ({}^4g)^{-W/2}\, {}^4\Gamma^{\mu}_{\mu\rho}=0$.}.

The Riemann curvature tensor is \footnote{This is the definition of
Ref.\cite{mtw} for $\epsilon =-1$; for $\epsilon =+1$ it coincides
with minus the definition of Ref.\cite{wei}.}

\begin{eqnarray}
{}^4R^{\alpha}{}_{\mu\beta\nu}&=&{}^4\Gamma^{\alpha}_{\beta\rho}\, {}^4\Gamma
^{\rho}_{\nu\mu} - {}^4\Gamma^{\alpha}_{\nu\rho}\, {}^4\Gamma^{\rho}
_{\beta\mu} + \partial_{\beta}\, {}^4\Gamma^{\alpha}_{\mu\nu}\, -
\partial_{\nu}\, {}^4\Gamma^{\alpha}_{\beta\mu},\nonumber \\
 &&{}\nonumber \\
 {}^4R_{\alpha\mu\beta\nu}&=&{}^4g_{\alpha\gamma}\,
{}^4R^{\gamma} {}_{\mu\beta\nu}=
-{}^4R_{\alpha\mu\nu\beta} =- {}^4R_{\mu\alpha\beta\nu} = {}^4R_{\beta\nu
\alpha\mu},\nonumber \\
 &&{}\nonumber \\
 {}^4R_{\mu\nu} &=& {}^4R_{\nu\mu}
={}^4R^{\beta}{}_{\mu\beta\nu},\nonumber \\
 {}&&\nonumber \\
  {}^4R &=& {}^4g^{\mu\nu}\, {}^4R_{\mu\nu}={}^4R^{\mu\nu}{}_{\mu\nu},
\nonumber \\
 &&{}\nonumber \\
&&{}^4R^{\alpha}{}_{\mu\beta\nu}+{}^4R^{\alpha}{}_{\beta\nu\mu}+{}^4
R^{\alpha}{}_{\nu\mu\beta}\equiv 0,\nonumber \\
&&{}\nonumber \\
&&({}^4\nabla_{\gamma}\, {}^4R)^{\alpha}{}_{\mu\beta\nu}+ ({}^4\nabla_{\beta}\,
{}^4R)^{\alpha}{}_{\mu\nu\gamma}+({}^4\nabla_{\nu}\, {}^4R)^{\alpha}
{}_{\mu\gamma\beta}
\equiv 0,\nonumber \\
&&{}\nonumber \\
&&\Rightarrow \, ({}^4\nabla_{\gamma}\, {}^4R^{(ricci)})_{\mu\nu}+({}^4\nabla
_{\alpha}\, {}^4R)^{\alpha}{}_{\mu\nu\gamma}-({}^4\nabla_{\nu}\,
{}^4R^{(ricci)})_{\mu\gamma}
\equiv 0, \nonumber \\
&&\Rightarrow \, {}^4\nabla_{\mu}\, {}^4G^{\mu\nu}\equiv 0,\quad\quad
{}^4G_{\mu\nu}={}^4R_{\mu\nu} -{1\over 2} {}^4g_{\mu\nu}\, {}^4R,\quad
{}^4G=-{}^4R.
\label{a2}
\end{eqnarray}

\noindent We have also shown the Ricci tensor, the curvature scalar and
the first and second Bianchi identities for the curvature tensor with
their implications \footnote{${}^4G_{\mu\nu}$ is the Einstein tensor
and ${}^4\nabla_{\mu}\, {}^4G^{\mu\nu}
\equiv 0$ are  called contracted Bianchi identities.}.
There are 20 independent components of the Riemann tensor in four
dimensions due to its symmetry properties.

Let our globally hyperbolic spacetime $M^4$ be foliated with spacelike Cauchy
hypersurfaces $\Sigma_{\tau}$, obtained with the embeddings $i_{\tau}:\Sigma
\rightarrow \Sigma_{\tau} \subset M^4$, $\vec \sigma \mapsto
x^{\mu}=z^{\mu}(\tau ,\vec
\sigma )$, of a 3-manifold $\Sigma$ in $M^4$ \footnote{$\tau :M^4\rightarrow R$ is a
global, timelike, future-oriented function labelling the leaves of the
foliation; $x^{\mu}$ are local coordinates in a chart of $M^4$; $\vec
\sigma =\{ \sigma^r \}$, r=1,2,3, are local coordinates in a chart of
$\Sigma$, which is diffeomorphic to $R^3$; we shall use the notation
$\sigma^A=(\sigma^{\tau}=\tau ;\vec \sigma )$, $A=\{ \tau ,r\}$, and
$z^{\mu}(\sigma )=z^{\mu}(\tau ,\vec \sigma )$.}.

Let $n^{\mu}(\sigma )$ and $l^{\mu}(\sigma )= N(\sigma )
n^{\mu}(\sigma )$ be the controvariant timelike normal and unit normal
[${}^4g_{\mu\nu}(z(\sigma ))l^{\mu}(\sigma ) l^{\nu}(\sigma
)=\epsilon$] to $\Sigma_{\tau}$ at the point $z(\sigma )\in
\Sigma_{\tau}$. The positive function $N(\sigma ) > 0$ is the {\it lapse}
function: $N(\sigma )d\tau$ measures the proper time interval at
$z(\sigma )\in \Sigma_{\tau}$ between $\Sigma_{\tau}$ and
$\Sigma_{\tau +d\tau}$
. The {\it shift} functions $N^r(\sigma )$ are defined so that $N^r(\sigma )d\tau$
describes the horizontal shift on $\Sigma_{\tau}$ such that, if $z^{\mu}(\tau
+d\tau ,\vec \sigma +d\vec \sigma )\in \Sigma_{\tau +d\tau}$, then $z^{\mu}(\tau
+d\tau ,\vec \sigma +d\vec \sigma )\approx z^{\mu}(\tau ,\vec \sigma )+N(\tau ,
\vec \sigma )d\tau l^{\mu}(\tau ,\vec \sigma )+[d\sigma^r+N^r(\tau ,\vec
\sigma )d\tau ]{{\partial z^{\mu}(\tau ,\vec \sigma )}\over {\partial \sigma^r}
}$; therefore, we have ${{\partial z^{\mu}(\sigma )}\over {\partial \tau}}=
N(\sigma ) l^{\mu}(\sigma )+N^r(\sigma ) {{\partial z^{\mu}(\tau ,\vec
\sigma )}\over {\partial \sigma^r}}$ for the so called {\it evolution} vector. For the
covariant unit normal to $\Sigma_{\tau}$ we have $l_{\mu}(\sigma )={}^4g
_{\mu\nu}(z(\sigma )) l^{\nu}(\sigma )=N(\sigma ) \partial_{\mu}\tau{|}
_{x=z(\sigma )}$.

Instead of local coordinates $x^{\mu}$ for $M^4$,
we use local coordinates $\sigma^A$ on
$R\times \Sigma \approx M^4$ [$x^{\mu}=z^{\mu}(\sigma )$ with inverse $\sigma^A=
\sigma^A(x)$], i.e. a {\it $\Sigma_{\tau}$-adapted holonomic coordinate basis} for
vector fields $\partial_A={{\partial}\over {\partial \sigma^A}}\in
T(R\times \Sigma ) \mapsto b^{\mu}_A(\sigma ) \partial_{\mu} ={{\partial z
^{\mu}(\sigma )}\over {\partial \sigma^A}} \partial_{\mu} \in TM^4$, and for
differential one-forms $dx^{\mu}\in T^{*}M^4 \mapsto d\sigma^A=b^A
_{\mu}(\sigma )dx^{\mu}={{\partial \sigma^A(z)}\over {\partial z^{\mu}}} dx
^{\mu} \in T^{*}(R\times \Sigma )$. Let us note that in the flat Minkowski
spacetime the transformation coefficients $b^A_{\mu}(\sigma )$ and $b^{\mu}
_A(\sigma )$ become the flat orthonormal cotetrads $z^A_{\mu}(\sigma )=
{{\partial \sigma^A(x)}\over {\partial x^{\mu}}}{|}_{x=z(\sigma )}$ and
tetrads $z^{\mu}_A(\sigma )={{\partial z^{\mu}(\sigma )}\over {\partial \sigma
^A}}$ of Ref.\cite{lus1}.

The induced 4-metric and inverse 4-metric become in the new basis

\begin{eqnarray}
{}^4g(x) &=&  {}^4g_{\mu\nu}(x) dx^{\mu} \otimes dx^{\nu} =
{}^4g_{AB}(z(\sigma ))d\sigma^A \otimes d\sigma^B,\nonumber \\
&&{}\nonumber \\
{}^4g_{\mu\nu}&=&b^A_{\mu}\, {}^4g_{AB} b^B_{\nu} =\nonumber \\
&=&\epsilon \, (N^2-{}^3g_{rs}N^rN^s)\partial_{\mu}\tau \partial_{\nu}\tau -
\epsilon \, {}^3g_{rs}N^s(\partial_{\mu}\tau \partial_{\nu}\sigma^r+
\partial_{\nu}\tau \partial_{\mu}\sigma^r)-\epsilon \, {}^3g_{rs}
\partial_{\mu}\sigma^r \partial_{\nu}\sigma^s=\nonumber \\
&=& \epsilon \, l_{\mu} l_{\nu} -\epsilon \, {}^3g_{rs} (\partial_{\mu}
\sigma^r +N^r\, \partial_{\mu}\tau ) (\partial_{\nu}\sigma^s+N^s\,
\partial_{\nu}\tau ),\nonumber \\
&&{}\nonumber \\
&\Rightarrow&{}^4g_{AB}=\lbrace {}^4g_{\tau\tau}=
\epsilon (N^2-{}^3g_{rs}N^rN^s); {}^4g_{\tau r}=-
\epsilon \, {}^3g_{rs}N^s; {}^4g_{rs}=-\epsilon \, {}^3g_{rs}\rbrace =
\nonumber \\
&=&\epsilon [ l_Al_B-{}^3g_{rs}(\delta^r_A+N^r\delta^{\tau}_A)(\delta^s_B+
N^s\delta^{\tau}_B)], \nonumber \\
&&{}\nonumber \\
{}^4g^{\mu\nu}&=& b^{\mu}_A {}^4g^{AB} b^{\nu}_B=\nonumber \\
&=&{{\epsilon}\over {N^2}} \partial_{\tau}z^{\mu}\partial_{\tau}z^{\nu}-
{{\epsilon \, N^r}\over {N^2}} (\partial_{\tau}z^{\mu}\partial_rz^{\nu}+
\partial_{\tau}z^{\nu}\partial_rz^{\mu}) -\epsilon ({}^3g^{rs}-{{N^rN^s}\over
{N^2}})\partial_rz^{\mu}\partial_sz^{\nu}=\nonumber \\
&=& \epsilon [\, l^{\mu} l^{\nu} - \, {}^3g^{rs} \partial_rz^{\mu}
\partial_sz^{\nu}],\nonumber \\
&&{}\nonumber \\
&\Rightarrow&{}^4g^{AB}=\lbrace {}^4g^{\tau\tau}=
{{\epsilon}\over {N^2}}; {}^4g^{\tau r}=-{{\epsilon \, N^r}
\over {N^2}}; {}^4g^{rs}=-\epsilon ({}^3g^{rs} - {{N^rN^s}\over {N^2}})
\rbrace =\nonumber \\
&=&\epsilon [l^Al^B -{}^3g^{rs}\delta^A_r\delta^B_s],\nonumber \\
&&{}\nonumber \\
l^A&=&l^{\mu} b^A_{\mu} = N \, {}^4g^{A\tau}={{\epsilon}\over N} (1; -N^r),
\nonumber \\
l_A&=&l_{\mu} b_A^{\mu} = N \partial_A \tau =N \delta^{\tau}_A = (N; \vec 0).
\label{a3}
\end{eqnarray}

Here, we introduced the 3-metric  of $\Sigma_{\tau}$: $\,
{}^3g_{rs}=-\epsilon \, {}^4g_{rs}$ with signature (+++). If
${}^4\gamma^{rs}$ is the inverse of the spatial part of the 4-metric
(${}^4\gamma^{ru}\, {}^4g_{us}=\delta^r_s$), the inverse of the
3-metric is ${}^3g^{rs}=-\epsilon \, {}^4\gamma^{rs}$ (${}^3g^{ru}\,
{}^3g_{us}=\delta^r_s$). ${}^3g_{rs}(\tau ,\vec \sigma )$ are the
components of the {\it first fundamental form} of the Riemann
3-manifold $(\Sigma_{\tau},{}^3g)$ and we have

\bea
 ds^2&=&{}^4g_{\mu\nu} dx^{\mu} dx^{\nu}=
\epsilon (N^2-{}^3g_{rs}N^rN^s) (d\tau )^2-2\epsilon
\, {}^3g_{rs}N^s d\tau d\sigma^r -\epsilon \, {}^3g_{rs} d\sigma^rd\sigma^s=
 \nonumber \\
 &=&\epsilon \Big[ N^2(d\tau )^2 -{}^3g_{rs}(d\sigma^r+N^rd\tau )(d\sigma^s+
N^sd\tau )\Big],
\label{a4}
\eea

\noindent for the line element in $M^4$. We must have $\epsilon \, {}^4g_{oo} >0$,
$\epsilon \, {}^4g_{ij} < 0$, $\left| \begin{array}{cc} {}^4g_{ii}& {}^4g_{ij}
\\ {}^4g_{ji}& {}^4g_{jj} \end{array} \right| > 0$, $\epsilon \,
det\, {}^4g_{ij} > 0$.

If we define $g={}^4g=|\, det\, ({}^4g_{\mu\nu})\, |$ and $\gamma ={}^3g =|\,
det\, ({}^3g_{rs})\, |$, we also have

\begin{eqnarray}
&&N=\sqrt{ {{}^4g\over {{}^3g}} }={1\over { \sqrt{{}^4g^{\tau\tau}} }} =
\sqrt{{g\over {\gamma}}}=
\sqrt{{}^4g_{\tau\tau}-\epsilon \, {}^3g^{rs}\, {}^4g_{\tau r}
{}^4g_{\tau s} },\nonumber \\
&&N^r=-\epsilon \, {}^3g^{rs}\, {}^4g_{\tau s} =
-{{{}^4g^{\tau r}}\over {{}^4g^{\tau\tau}}}
,\quad N_r={}^3g_{rs}N^s=-\epsilon \,\, {}^4g_{rs}N^s=-\epsilon {}^4g_{\tau r}.
\label{a5}
\end{eqnarray}

Let us remark (see Ref.\cite{mol}) that in the study of space and time
measurements the equation $ds^2=0$ (use of light signals for the
synchronization of clocks) and the definition $d\bar \tau
=\sqrt{\epsilon \, {}^4g_{oo}} dx^o$ of proper time \footnote{$\sqrt{\epsilon
\, {}^4g_{oo}}$ determines the ratio between the rates of a standard
clock at rest and a coordinate clock at the same point.} imply the use
in $M^4$ of a 3-metric ${}^3{\tilde \gamma}_{rs}={}^4g_{rs}-
{{{}^4g_{or}\, {}^4g_{os}}\over {{}^4g_{oo}}}=-\epsilon
({}^3g_{rs}+{{N_rN_s}
\over {\epsilon \, {}^4g_{oo}}} )$ with the covariant shift functions $N_r=
{}^3g_{rs}N^s=-\epsilon \, {}^4g_{or}$, which are connected with the
conventionality of simultaneity \cite{simul} and with the direction
dependence of the velocity of light ($c(\vec n)=\sqrt{\epsilon \,
{}^4g_{oo}} / (1+N_rn^r)$ in direction $\vec n$).

In the standard (not Hamiltonian) description of the 3+1 decomposition
we utilize a {\it $\Sigma_{\tau}$-adapted non-holonomic non-coordinate
basis} [$\bar A=(l;r)$]

\begin{eqnarray}
{\hat b}^{\mu}_{\bar A}(\sigma ) &=&\lbrace {\hat b}^{\mu}_l(\sigma )=
\epsilon l^{\mu}
(\sigma ) =N^{-1}(\sigma ) [b^{\mu}_{\tau}(\sigma )- N^r(\sigma )b^{\mu}_r
(\sigma )];\nonumber \\
&&{\hat b}^{\mu}_r(\sigma ) = b^{\mu}_r(\sigma ) \rbrace ,\nonumber \\
{\hat b}^{\bar A}_{\mu}(\sigma ) &=& \lbrace {\hat b}^l_{\mu}(\sigma ) =
l_{\mu}(\sigma )= N(\sigma )b^{\tau}_{\mu}(\sigma )=N(\sigma )\partial_{\mu}
\tau (z(\sigma ));\nonumber \\
&&{\hat b}^r_{\mu}(\sigma ) = b^r_{\mu}(\sigma )+ N^r(\sigma )
b^{\tau}_{\mu}(\sigma ) \rbrace ,\nonumber \\
&&{}\nonumber \\
&&{\hat b}_{\mu}^{\bar A}(\sigma ) {\hat b}^{\nu}_{\bar A}(\sigma )=
\delta^{\nu}_{\mu},\quad {\hat b}^{\bar A}_{\mu}(\sigma ) {\hat b}^{\mu}
_{\bar B}(\sigma )=\delta^{\bar A}_{\bar B}, \nonumber \\
{}^4{\bar g}_{\bar A\bar B}(z(\sigma ))&=&{\hat b}^{\mu}_{\bar A}(\sigma )
{}^4g_{\mu\nu}(z(\sigma )) {\hat b}^{\nu}_{\bar B}(\sigma )=\nonumber \\
&&=\lbrace {}^4{\bar g}_{ll}(\sigma )=\epsilon ; {}^4{\bar g}_{lr}(\sigma )=0;
{}^4{\bar g}_{rs}(\sigma )=
{}^4g_{rs}(\sigma )=-\epsilon \, {}^3g_{rs}\rbrace ,\nonumber \\
{}^4{\bar g}^{\bar A\bar B}&=&\lbrace {}^4{\bar g}^{ll}=\epsilon ; {}^4{\bar g}
^{lr}=0; {}^4{\bar g}^{rs}={}^4\gamma^{rs}=-\epsilon {}^3g^{rs}\rbrace ,
\nonumber \\
&&{}\nonumber \\
&&X_{\bar A}={\hat b}^{\mu}_{\bar A}\partial_{\mu}=\{ X_l={1\over
N}(\partial_{\tau}- N^r\partial_r);\partial_r\},\nonumber \\
&&\theta^{\bar A}={\hat b}^{\bar A}_{\mu}dx^{\mu}=\{ \theta^l=Nd\tau ;
\theta^r=d\sigma^r+N^rd\tau \} ,\nonumber \\
&&{}\nonumber \\
&&\Rightarrow l_{\mu}(\sigma )b^{\mu}_r(\sigma )=0,\quad l^{\mu}(\sigma )b^r
_{\mu}(\sigma )=-N^r(\sigma )/N(\sigma ),\nonumber \\
&&{}\nonumber \\
l^{\bar A}&=& l^{\mu} {\hat b}_{\mu}^{\bar A} = (\epsilon ; l^r+N^rl^{\tau})=
(\epsilon ; \vec 0),\nonumber \\
l_{\bar A}&=& l_{\mu} {\hat b}^{\mu}_{\bar A} = (1; l_r) = (1; \vec 0).
\label{a6}
\end{eqnarray}

The non-holonomic basis in $\Sigma_{\tau}$-adapted coordinates is
${\hat b}_A^{\bar A}={\hat b}^{\bar A}_{\mu}b^{\mu}_A=
\{ {\hat b}^l_A=l_A;\, {\hat b}^r_A=\delta^r_A+N^r \delta^{\tau}_A
\}$, ${\hat b}^A_{\bar A}={\hat b}^{\mu}_{\bar A}b^A_{\mu}=
\{ {\hat b}^A_l=\epsilon l^A;\, {\hat b}^A_r=\delta^A_r \}$.

See Refs.\cite{in,mtw,ish} for the 3+1 decomposition of 4-tensors on
$M^4$. The {\it horizontal projector}
${}^3h^{\nu}_{\mu}=\delta^{\nu}_{\mu}-\epsilon \, l_{\mu} l^{\nu}$ on
$\Sigma_{\tau}$ defines the 3-tensor fields on $\Sigma_{\tau}$
starting from the 4-tensor fields on $M^4$. We have
${}^3h_{\mu\nu}={}^4g_{\mu\nu}-\epsilon l_{\mu}l_{\nu}=-
\epsilon \, {}^3g_{rs}(b^r_{\mu}+N^rb^{\tau}_{\mu})(b^s_{\mu}+N^sb^{\tau}
_{\mu})=-\epsilon \, {}^3g_{rs}{\hat b}^r_{\mu}{\hat b}^s_{\nu}$ and
for a 4-vector ${}^4V^{\mu}={}^4V^{\bar A}{\hat b}^{\mu}_{\bar A}=
{}^4V^l l^{\mu}+{}^4V^r{\hat b}^{\mu}_r$ we have ${}^3V^{\mu}={}^3V^r{\hat b}
^{\mu}_r={}^3h^{\mu}_{\nu}\, {}^4V^{\nu}$, ${}^3V^r={\hat b}^r_{\mu}\, {}^3V
^{\mu}$.

The 3-dimensional covariant derivative (denoted ${}^3\nabla$ or with
the subscript ``$|$") of a 3-dimensional tensor ${}^3T^{\mu_1..
\mu_p}{}_{\nu_1..\nu_q}$ of rank (p,q) is the  3-dimensional tensor
of rank (p,q+1)
${}^3\nabla_{\rho}\, {}^3T^{\mu_1..\mu_p}{}_{\nu_1..\nu_q}={}^3T^{\mu_1..\mu_p}
{}_{\nu_1..\nu_q | \rho}=
{}^3h^{\mu_1}_{\alpha_1}\cdots {}^3h^{\mu_p}_{\alpha_p}\, {}^3h^{\beta_1}
_{\nu_1}\cdots {}^3h^{\beta_q}_{\nu_q}\, {}^3h^{\sigma}_{\rho}\, {}^4\nabla
_{\sigma}\, {}^3T^{\alpha_1..\alpha_p}{}_{\beta_1..\beta_q}$. For (1,0) and
(0,1) tensors we have:
${}^3\nabla_{\rho} \, {}^3V^{\mu}={}^3V^{\mu}{}_{| \rho}= {}^3V^r
{}_{| s}\, {\hat b}^{\mu}_r {\hat b}^s_{\rho}$ ,
${}^3\nabla_s\, {}^3V^r={}^3V^r{}_{| s} =\partial_s\, {}^3V^r +
{}^3\Gamma^r_{su}\, {}^3V^u$ and
${}^3\nabla_{\rho}\, {}^3\omega_{\mu}={}^3\omega_{\mu | \rho}={}^3\omega
_{r | s}\, {\hat b}^r_{\mu} {\hat b}^s_{\rho}$,
${}^3\nabla_s\, {}^3\omega_r={}^3\omega_{r | s}=\partial_s\, {}^3\omega_r -
{}^3\Gamma^u_{rs} {}^3\omega_u$ respectively.

The 3-dimensional Christoffel symbols are
${}^3\Gamma^u_{rs}={\hat b}^u_{\mu}\, [{}^3
\nabla_{\rho}\, {\hat b}^{\mu}_r] {\hat b}^{\rho}_s=
{\hat b}^u_{\mu} {\hat b}^{\mu}_{r | \rho} {\hat b}^{\rho}
_s={1\over 2} {}^3g^{uv} (\partial_s\, {}^3g_{vr} +
\partial_r\, {}^3g_{vs} -\partial_v\, {}^3g_{rs})$ and
the metric compatibility \footnote{Levi-Civita connection on the
Riemann 3-manifold $(\Sigma_{\tau},{}^3g)$.} is ${}^3\nabla_{\rho}\,
{}^3g_{\mu\nu} ={}^3g_{\mu\nu | \rho}=0$
\footnote{$\,\,\, {}^3g_{\mu\nu}=-\epsilon
\, {}^3h_{\mu\nu}= {}^3g_{rs}{\hat b}^r_{\mu}{\hat b}^s_{\nu}$,
so that ${}^3{\bar g}_{\bar A\bar B}=\{ {}^3{\bar g}_{ll}
=0; {}^3{\bar g}_{lr}=0; {}^3{\bar g}_{rs}=-\epsilon \, {}^3g_{rs} \}$.}.
It is then possible to
define parallel transport on $\Sigma_{\tau}$.

The 3-dimensional curvature Riemann tensor is

\begin{eqnarray}
&&{}^3R^{\mu}{}_{\alpha\nu\beta}\, {}^3V^{\alpha}= {}^3V^{\alpha}{}_{| \beta
| \nu} - {}^3V^{\alpha}{}_{| \nu | \beta},\nonumber \\
&&\Rightarrow {}^3R^r{}_{suv}=\partial_u
\, {}^3\Gamma^r_{sv} -\partial_v\, {}^3\Gamma
^r_{su} + {}^3\Gamma^r_{uw}\,
{}^3\Gamma^w_{sv} - {}^3\Gamma^r_{vw}\, {}^3\Gamma^w_{su}.
\label{a7}
\end{eqnarray}

For 3-manifolds, the Riemann tensor has only 6 independent components
since the Weyl tensor vanishes: this gives the relation
${}^3R_{\alpha\mu\beta\nu}={1\over 2}({}^3R_{\mu\beta}\, {}^3g_{\alpha\nu}+
{}^3R_{\alpha\nu}\, {}^3g_{\mu\beta}-{}^3R_{\alpha\beta}\, {}^3g_{\mu\nu}-
{}^3R_{\mu\nu}\, {}^3g_{\alpha\beta})
-{1\over 6} ({}^3g_{\alpha\beta}\, {}^3g_{\mu\nu}-
{}^3g_{\alpha\nu}\, {}^3g_{\beta\mu})\, {}^3R$, which expresses the
Riemann tensor in terms of  the Ricci tensor.

The components of the {\it second fundamental form} of
$(\Sigma_{\tau},{}^3g)$ is the extrinsic curvature

\begin{equation}
{}^3K_{\mu\nu}={}^3K_{\nu\mu}=-{1\over 2}{\cal L}_l\, {}^3g_{\mu\nu}.
\label{a8}
\end{equation}

\noindent
We have ${}^4\nabla_{\rho} \, l^{\mu}=\epsilon \, {}^3a^{\mu}
l_{\rho}- {}^3K_{\rho}{}^{\mu}$, with the acceleration
${}^3a^{\mu}={}^3a^r {\hat b}
^{\mu}_r$ of the observers travelling along the congruence of timelike curves
with tangent vector $l^{\mu}$ given by ${}^3a_r=\partial_r\, ln\, N$. On
$\Sigma_{\tau}$ we have

\beq
{}^3K_{rs}={}^3K_{sr}={1\over {2N}}(N_{r|s}+N_{s|r}-{{\partial \,
{}^3g_{rs}}
\over {\partial \tau}}).
\label{a9}
\eeq

The information contained in the 20 independent components
${}^4R^{\mu} {}_{\nu\alpha\beta}$ of the curvature Riemann tensor of
$M^4$ is given in terms of 3-tensors on $\Sigma_{\tau}$ by the
following three projections
\footnote{See Ref.\cite{carter} for the geometry of embeddings; one
has ${}^4{\bar R}^r{}_{suv}={}^3{\bar R}^r{}_{suv}$.}

\begin{eqnarray}
{}^3h^{\mu}_{\rho}\, &{}^3&h^{\sigma}_{\nu}\, {}^3h^{\gamma}_{\alpha}\, {}^3h
^{\delta}_{\beta}\, {}^4R^{\rho}{}_{\sigma\gamma\delta}=
{}^4{\bar R}^r{}_{suv}{\hat b}
^{\mu}_r{\hat b}^s_{\nu}{\hat b}^u_{\alpha}{\hat b}^v_{\beta}=
{}^3R^{\mu}{}_{\nu
\alpha\beta}+{}^3K_{\alpha}{}^{\mu}\, {}^3K_{\beta\nu}-{}^3K_{\beta}{}^{\mu}
\, {}^3K_{\alpha\nu},\nonumber \\
&&GAUSS\, EQUATION,\nonumber \\
\epsilon &l_{\rho}&\, {}^3h^{\sigma}_{\nu}\, {}^3h^{\gamma}_{\alpha}\, {}^3h
^{\delta}_{\beta}\, {}^4R^{\rho}{}_{\sigma\gamma\delta}={}^4{\bar R}^l{}_{suv}
{\hat b}^s_{\nu}{\hat b}^u_{\alpha}{\hat b}^v_{\beta}=
{}^3K_{\alpha\nu |
\beta} - {}^3K_{\beta\nu | \alpha},\nonumber \\
&&CODAZZI-MAINARDI\, EQUATION,\nonumber \\
{}^4&R&_{\mu\sigma\gamma\delta}\, l^{\sigma}\, l^{\gamma}\, {}^3h^{\delta}
_{\nu}={}^4{\bar R}_{\mu llu}{\hat b}^u_{\nu}=
\epsilon ({\cal L}_l\, {}^3K_{\mu\nu}+{}^3K_{\mu}{}^{\rho}\, {}^3K
_{\rho\nu}+{}^3a_{\mu | \nu}+{}^3a_{\mu}\, {}^3a_{\nu}),\nonumber \\
&&RICCI\, EQUATION,\nonumber \\
&&{}\nonumber \\
&&with\quad\quad {\cal L}_l\, {}^3K_{\mu\nu}=l^{\alpha}\, {}^3K_{\mu\nu
;\alpha}-2\, {}^3K_{\mu}{}^{\alpha}\, {}^3K_{\alpha\nu}+2\epsilon \, {}^3a
^{\alpha}\, {}^3K_{\alpha (\nu}\, l_{\mu )} .
\label{a10}
\end{eqnarray}

After having expressed the 4-Riemann tensor components in  the
non-holonomic basis in terms of the 3-Riemann tensor on
$\Sigma_{\tau}$, the extrinsic curvature of $\Sigma_{\tau}$ and the
acceleration \footnote{For instance ${}^4R={}^3R+{}^3K_{rs}\,
{}^3K^{rs}-({}^3K)^2$.},  we can express ${}^4R_{\mu\nu}=\epsilon
{}^4{\bar R}_{ll} l_{\mu}l_{\nu}+\epsilon \, {}^4{\bar
R}_{lr}(l_{\mu}{\hat b}^r_{\nu}+l_{\nu} {\hat b}^r_{\mu})+{}^4{\bar
R}_{rs}{\hat b}^r_{\mu}{\hat b}^s_{\nu}$, ${}^4R$ and the Einstein
tensor ${}^4G_{\mu\nu}={}^4R_{\mu\nu}-{1\over 2}\, {}^4g_{\mu\nu}\,
{}^4R=\epsilon {}^4{\bar G}_{ll} l_{\mu}l_{\nu}+\epsilon \, {}^4{\bar
G}_{lr}(l_{\mu}{\hat b}^r_{\nu}+l_{\nu} {\hat b}^r_{\mu})+{}^4{\bar
G}_{rs}{\hat b}^r_{\mu}{\hat b}^s_{\nu}$. The vanishing of ${}^4{\bar
G}_{ll}$, ${}^4{\bar G}_{lr}$, corresponds to the four secondary
constraints (restrictions of Cauchy data) of the ADM Hamiltonian
formalism (see Section IV). The four contracted Bianchi identities,
${}^4G^{\mu\nu}{}_{;\nu}\equiv 0$, imply \cite{wald} that, if the
restrictions of Cauchy data are satisfied initially and the spatial
equations ${}^4G_{ij}\, {\buildrel \circ \over =}\, 0$ are satisfied
everywhere, then the secondary constraints are satisfied also at later
times \footnote{See Ref.\cite{cho,wald}
 for the initial value problem.}. The four contracted Bianchi
identities plus the four secondary constraints imply that only two
combinations of the Einstein equations ${}^4{\bar G}_{rs}\, {\buildrel
\circ \over =}\, 0$ are independent, namely contain the accelerations
(second time derivatives) of the two (non tensorial) independent
degrees of freedom of the gravitational field, and that  only these
two equations can be put in normal form \footnote{This was one of the
motivations behind the discovery of the Shanmugadhasan canonical
transformations \cite{sha}.}.

The {\it intrinsic geometry} of $\Sigma_{\tau}$ is defined by the {\it
Riemannian metric} ${}^3g_{rs}$ \footnote{It allows to evaluate the
length of space curves.}, the {\it Levi-Civita affine connection},
i.e. the Christoffel symbols ${}^3\Gamma^u_{rs}$, \footnote{For the
parallel transport of 3-dimensional tensors on $\Sigma_{\tau}$.} and
the {\it curvature Riemann tensor} ${}^3R^r{}_{stu}$ \footnote{For the
evaluation of the holonomy and for the geodesic deviation equation.}.
The {\it extrinsic geometry} of $\Sigma_{\tau}$ is defined by the {\it
lapse} N and {\it shift} $N^r$ fields, which describe the {\it
evolution} of $\Sigma_{\tau}$ in $M^4$, and by the {\it extrinsic
curvature} ${}^3K_{rs}$ \footnote{It is needed to evaluate how much a
3-dimensional vector goes outside $\Sigma_{\tau}$ under spacetime
parallel transport and to rebuild the spacetime curvature from the
3-dimensional one.}.

\subsection{Tetrads and Cotetrads on $M^4$.}

Besides the local dual coordinate bases ${}^4e_{\mu}=\partial_{\mu}$
and $dx^{\mu}$ for $TM^4$ and $T^{*}M^4$ respectively, we can
introduce special {\it non-coordinate} bases ${}^4{\hat E}_{(\alpha
)}={}^4{\hat E}^{\mu}
_{(\alpha )}(x)\partial_{\mu}$ and its dual ${}^4{\hat \theta}^{(\alpha )}=
{}^4{\hat E}^{(\alpha )}_{\mu}(x)dx^{\mu}$ \footnote{$i_{{}^4{\hat
E}_{(\alpha )}}\, {}^4{\hat \theta}^{(\beta )}={}^4E^{(\alpha
)}_{\mu}\, {}^4E^{\mu}_{(\beta )}=
\delta^{(\beta )}_{(\alpha )}$ $\Rightarrow {}^4\eta_{(\alpha )(\beta )}=
{}^4E^{\mu}_{(\alpha )}\, {}^4g_{\mu\nu}\, {}^4E^{\nu}_{(\beta )}$;
$(\alpha )=(0),(1),(2),(3)$  are numerical indices.} with the {\it
vierbeins or tetrads or (local) frames}  ${}^4{\hat E}^{\mu}
_{(\alpha )}(x)$, which are,
for each point $x^{\mu}\in M^4$, the matrix elements of matrices
$\lbrace {}^4{\hat E}^{\mu}_{(\alpha )}\rbrace \in GL(4,R)$; the set
of one-forms ${}^4{\hat \theta}^{(\alpha )}$ \footnote{With ${}^4{\hat
E}^{(\alpha )}
_{\mu}(x)$ being the dual {\it cotetrads}.} is also called {\it canonical} or
{\it soldering} one-form or {\it coframe}. Since a {\it frame}
${}^4{\hat E}$ at the point $x^{\mu}\in M^4$ is a linear
isomorphism\cite{blee} ${}^4{\hat E}: R^4\rightarrow T_xM^4$,
$\partial_{\alpha}\mapsto {}^4{\hat E}(\partial
_{\alpha})={}^4{\hat E}_{(\alpha )}$, a frame determines a basis ${}^4{\hat
E}_{(\alpha )}$ of $T_xM^4$ \footnote{The coframes ${}^4{\hat \theta}$
determine a basis ${}^4{\hat \theta}^{(\alpha )}$ of $T^{*}_xM^4$.}
and we can define a principal fiber bundle with structure group
GL(4,R), $\pi : L(M^4)\rightarrow M^4$ called the {\it frame bundle}
of $M^4$
\footnote{Its fibers are the sets of all the frames over the points $x^{\mu}\in
M^4$; it is an affine bundle, i.e. there is no (global when it exists)
cross section playing the role of the identity cross section of vector
bundles.}; if $\Lambda \in GL(4,R)$, then the free right action of
GL(4,R) on $L(M^4)$ is denoted $R_{\Lambda}({}^4{\hat E})={}^4{\hat
E}\circ \Lambda$, ${}^4{\hat E}_{(\alpha )}\mapsto {}^4{\hat E}
_{(\beta )}\, (\Lambda^{-1})^{(\beta )}{}_{(\alpha )}$. When $M^4$ is
{\it parallelizable} \footnote{I.e. $M^4$ admits four vector fields
which are independent in each point, so that the tangent bundle
$T(M^4)$ is trivial, $T(M^4)=M^4\times R^4$; this is not possible (no
hair theorem) for any compact manifold except a torus.}, as we shall
assume, then $L(M^4)=M^4\times GL(4,R)$ is a trivial principal bundle
\footnote{I.e. it admits a global cross section $\sigma : M^4\rightarrow
L(M^4)$, $x^{\mu}\mapsto {}^4_{\sigma}{\hat E}_{(\alpha )}(x)$.}. See
Ref.\cite{blee} for the differential structure on $L(M^4)$.

With the assumed pseudo-Riemannian manifold $(M^4,\, {}^4g)$, we can
use its metric ${}^4g_{\mu\nu}$ to define the {\it orthonormal frame
bundle} of $M^4$, $F(M^4)=M^4\times SO(3,1)$, with structure group
SO(3,1), of the orthonormal frames (or {\it non-coordinate basis} or
{\it orthonormal tetrads}) ${}^4E_{(\alpha )}={}^4E^{\mu}_{(\alpha
)}\partial_{\mu}$ of $TM^4$. The orthonormal tetrads and their duals,
the orthonormal cotetrads ${}^4E_{\mu}
^{(\alpha )}$ \footnote{${}^4\theta^{(\alpha )}={}^4E^{(\alpha )}_{\mu}dx^{\mu}$ are
the orthonormal coframes.}, satisfy the duality and orthonormality
conditions

\begin{eqnarray}
&&{}^4E^{(\alpha )}_{\mu}\, {}^4E^{\mu}_{(\beta )}=\delta^{(\alpha )}_{(\beta )}
,\quad\quad
{}^4E^{(\alpha )}_{\mu}\, {}^4E^{\nu}_{(\alpha )}=\delta^{\nu}_{\mu},
\nonumber \\
&&{}^4E^{\mu}_{(\alpha )}\, {}^4g_{\mu\nu}\, {}^4E^{\nu}_{(\beta )}=
{}^4\eta_{(\alpha )(\beta )},\quad\quad
{}^4E_{\mu}^{(\alpha )}\, {}^4g^{\mu\nu}\, {}^4E^{(\beta )}_{\nu} =
{}^4\eta^{(\alpha )(\beta )}.
\label{a11}
\end{eqnarray}

Under a rotation $\Lambda \in SO(3,1)$ ($\Lambda \, {}^4\eta \,
\Lambda^T= {}^4\eta$) we have ${}^4E^{\mu}_{(\alpha )}\mapsto
{}^4E^{\mu}_{(\beta )} (\Lambda^{-1})^{(\beta )}{}_{(\alpha )}$,
${}^4E^{(\alpha )}_{\mu}\mapsto
\Lambda^{(\alpha )}{}_{(\beta )}\, {}^4E^{(\beta )}_{\mu}$. Therefore, while
the indices $\alpha , \beta ...$ transform under general coordinate
transformations (the diffeomorphisms in $Diff\, M^4$), the indices
$(\alpha ), (\beta )...$ transform under Lorentz rotations. The
4-metric can be expressed in terms of orthonormal cotetrads or local
coframes in the non-coordinate basis

\begin{eqnarray}
&&{}^4g_{\mu\nu}={}^4E^{(\alpha )}_{\mu}\, {}^4\eta_{(\alpha )(\beta )}\,
{}^4E^{(\beta )}_{\nu},\quad\quad {}^4g^{\mu\nu}={}^4E^{\mu}_{(\alpha )}\,
{}^4\eta^{(\alpha )(\beta )}\, {}^4E^{\nu}_{(\beta )},\nonumber \\
&&{}^4g={}^4g_{\mu\nu}\, dx^{\mu} \otimes dx^{\nu} = {}^4\eta_{(\alpha )
(\beta )}\, \theta^{(\alpha )} \otimes \theta^{(\beta )}.
\label{a12}
\end{eqnarray}

For each vector ${}^4V^{\mu}$ and covector ${}^4\omega_{\mu}$ we have
the decompositions ${}^4V^{\mu}={}^4V^{(\alpha )}\,
{}^4E^{\mu}_{(\alpha )}$ (${}^4V^{(\alpha )}={}^4E^{(\alpha )}_{\mu}\,
{}^4V^{\mu}$), ${}^4\omega
_{\mu}={}^4E^{(\alpha )}_{\mu}\, {}^4\omega_{(\alpha )}$ (${}^4\omega
_{(\alpha )}={}^4E^{\mu}_{(\alpha )}\, {}^4\omega_{\mu}$).

In a non-coordinate basis we have

\begin{eqnarray}
&&[ {}^4E_{(\alpha )}, {}^4E_{(\beta )} ] = c_{(\alpha )(\beta )}{}^{(\gamma )}
\, {}^4E_{(\gamma )},\nonumber \\
&&c_{(\alpha )(\beta )}{}^{(\gamma )} = {}^4E^{(\gamma )}_{\nu} ({}^4E^{\mu}
_{(\alpha )}\, \partial_{\mu}\, {}^4E^{\nu}_{(\beta )} - {}^4E^{\mu}_{(\beta )}
\partial_{\mu}\, {}^4E^{\nu}_{(\alpha )}).
\label{a13}
\end{eqnarray}

Physically, in a coordinate system (chart) $x^{\mu}$ of $M^4$, a tetrad may be
considered as a collection of accelerated observers described by a congruence of
timelike curves with 4-velocity ${}^4E^{\mu}_{(o)}$; in each point $p\in
M^4$ consider a coordinate transformation to local inertial coordinates at p,
i.e. $x^{\mu} \mapsto X^{(\mu)}_p(x)$: then we have, in p, ${}^4E^{\mu}_{(\alpha
)}(p)={{\partial x^{\mu}(X_p(p))}\over {\partial X^{(\alpha )}_p}}$ and
${}^4E^{(\alpha )}_{\mu}(p)={{\partial X^{(\alpha )}_p(p))}\over {\partial
x^{\mu}}}$ and locally we have a freely falling observer.

All the connection one-forms $\omega$ are 1-forms on the orthonormal
frame bundle $F(M^4)=M^4\times SO(3,1)$. Since in general relativity
we consider only Levi-Civita connections associated with
pseudo-Riemannian 4-manifolds $(M^4,{}^4g)$, in $F(M^4)$ we consider
only $\omega_{\Gamma}$-horizontal subspaces $H_{\Gamma}$
\footnote{$TF(M^4)
= V_{\Gamma} +H_{\Gamma}$ as a direct sum, with $V_{\Gamma}$ the
vertical subspace isomorphic to the Lie algebra o(3,1) of SO(3,1).}.
Given a global cross section $\sigma : M^4\rightarrow F(M^4)=M^4\times
SO(3,1)$, the associated gauge potentials on $M^4$, ${}^4\omega
=\sigma^{*} \omega$, are the connection coefficients ${}^4\omega^{(T)}
=\sigma^{*}\omega$ in the
non-coordinate basis ${}^4E_{(\alpha )}$ \footnote{The second line
defines them through the covariant derivative in the non-coordinate
basis.}

\begin{eqnarray}
&&{}^4\omega^{(T)(\gamma )}_{(\alpha )(\beta )} =  {}^4E^{(\gamma )}_{\nu} {}^4E
^{\mu}_{(\alpha )} (\partial_{\mu}\, {}^4E^{\nu}_{(\beta )} +{}^4E^{\lambda}
_{(\beta )}\, {}^4\Gamma^{(T)\nu}_{\mu\lambda}) = {}^4E^{(\gamma )}_{\nu}
{}^4E^{\mu}_{(\alpha )}\, {}^4\nabla_{\mu}\, {}^4E^{\nu}_{(\beta )},
\nonumber \\
&&{}^4{\tilde \nabla}_{{}^4E_{(\alpha )}}\, {}^4E_{(\beta )} = {}^4\nabla
_{{}^4E_{(\alpha )}}\, {}^4E_{(\beta )} - {}^4\omega^{(T)(\gamma )}_{(\alpha )
(\beta )}\, {}^4E_{(\gamma )} =0.
\label{a14}
\end{eqnarray}

The components of the  Riemann tensors in the non-coordinate bases are
${}^4R^{(\alpha )} {}_{(\beta )(\gamma )(\delta )}= {}^4E_{(\gamma )}
({}^4\omega^{(T)(\alpha )}
_{(\delta )(\beta )}) - {}^4E_{(\delta )} ({}^4\omega^{(T)(\alpha )}
_{(\gamma )(\beta )}) + {}^4\omega^{(T)(\epsilon )}_{(\delta )(\beta )}\,
{}^4\omega^{(T)(\alpha )}_{(\gamma )(\epsilon )} - {}^4\omega^{(T)(\epsilon )}
_{(\gamma )(\beta )}\, {}^4\omega^{(T)(\alpha )}_{(\delta )(\epsilon )} -
c_{(\gamma )(\delta )}{}^{(\epsilon )}\, {}^4\omega^{(T)(\alpha )}
_{(\epsilon )(\beta )}$. The connection
(gauge potential) one-form is ${}^4\omega^{(T)(\alpha )}{}_{(\beta
)}={}^4\omega
^{(T)(\alpha )}_{(\gamma )(\beta )}\, {}^4\theta^{(\gamma )}$ \footnote{It is called
improperly {\it spin connection}, while its components are called {\it
Ricci rotation coefficients}.} and the curvature (field strength)
2-form is ${}^4\Omega^{(T)(\alpha )}{}_{(\beta )} = {1\over 2}
{}^4\Omega
^{(T)(\alpha )}
{}_{(\beta )(\gamma )(\delta )}\, {}^4\theta^{(\gamma )} \wedge {}^4\theta
^{(\delta )}$.

With the Levi-Civita connection \footnote{It has zero torsion 2-form
${}^4T^{(\alpha )}={1\over 2} T^{(\alpha )}{}_{(\beta )(\gamma )}\,
{}^4\theta
^{(\beta )} \wedge {}^4\theta^{(\gamma )}=0$, namely
${}^4T^{(\alpha )}{}_{(\beta )
(\gamma )}= {}^4\omega^{(T)(\alpha )}
_{(\beta )(\gamma )} - {}^4\omega^{(T)(\alpha )}_{(\gamma )(\beta )} -
c_{(\beta )(\gamma )}{}^{(\alpha )}=0$.}, in a non-coordinate basis
the spin connection takes the form

\begin{eqnarray}
{}^4\omega^{(\alpha )}{}_{(\beta )}&=&{}^4\omega^{(\alpha )}
_{(\gamma )(\beta )}\, {}^4\theta^{(\gamma )}={}^4\omega^{(\alpha )}_{\mu
(\beta )} dx^{\mu},\nonumber \\
{}^4\omega_{(\alpha )(\gamma )(\beta )}&=&{}^4\eta_{(\alpha )(\delta )}\,
{}^4E^{(\delta )}_{\nu}\, {}^4E^{\mu}_{(\gamma )}\, {}^4\nabla_{\mu}\,
{}^4E^{\nu}_{(\beta )}={}^4\eta_{(\alpha )(\delta )}
{}^4\omega^{(\delta )}_{(\gamma )(\beta )},\nonumber \\
{}^4\omega^{(\alpha )}_{\mu (\beta )}&=&{}^4\omega^{(\alpha )}_{(\gamma )
(\beta )}\, {}^4E^{(\gamma )}_{\mu}={}^4E^{(\alpha )}_{\nu}\, {}^4\nabla_{\mu}
\, {}^4E^{\nu}_{(\beta )}={}^4E^{(\alpha )}_{\nu}[\partial_{\mu}\, {}^4E^{\nu}
_{(\beta )}+{}^4\Gamma^{\nu}_{\mu\rho}\, {}^4E^{\rho}_{(\beta )}],
\nonumber \\
\Rightarrow {}^4&\Gamma^{\mu}_{\rho\sigma}&={1\over 2}[{}^4E^{(\beta )}
_{\sigma}({}^4E^{\mu}_{(\alpha )}\, {}^4E^{(\gamma )}_{\rho}\, {}^4\omega
^{(\alpha )}_{(\gamma )(\beta )}-\partial_{\rho}\, {}^4E^{\mu}_{(\beta )})+
\nonumber \\
&+&{}^4E^{(\beta )}_{\rho}({}^4E^{\mu}_{(\alpha )}\,
{}^4E^{(\gamma )}_{\sigma}\,
{}^4\omega^{(\alpha )}_{(\gamma )(\beta )}-\partial_{\sigma}\, {}^4E^{\mu}
_{(\beta )})],
\label{a15}
\end{eqnarray}

\noindent and
the metric compatibility ${}^4\nabla_{\rho}\, {}^4g_{\mu\nu}=0$ becomes the
following condition

\begin{equation}
{}^4\omega_{(\alpha )(\beta )}= {}^4\eta_{(\alpha )(\delta )}\, {}^4\omega
^{(\delta )}{}_{(\beta )} = {}^4\eta_{(\alpha )(\delta )} {}^4\omega^{(\delta )}
_{(\gamma )(\beta )}\, {}^4\theta^{(\gamma )} = {}^4\omega_{(\alpha )(\gamma )
(\beta )}\, {}^4\theta^{(\gamma )} = - {}^4\omega_{(\beta )(\alpha )}
\label{a16}
\end{equation}

\noindent or ${}^4\omega_{(\alpha )(\gamma )(\beta )}=-{}^4\omega_{(\beta )
(\gamma )(\alpha )}$ \footnote{${}^4\omega_{(\alpha )(\gamma )(\beta
)}$ are the {\it Ricci rotation coefficients}, only 24 of which are
independent.}

Given a vector ${}^4V^{\mu}={}^4V^{(\alpha )}\, {}^4E^{\mu}_{(\alpha )}$ and a
covector ${}^4\omega_{\mu}={}^4\omega_{(\alpha )}\, {}^4E^{(\alpha )}_{\mu}$,
we define the covariant derivative of the components ${}^4V^{(\alpha )}$ and
${}^4\omega_{(\alpha )}$ as ${}^4\nabla_{\nu}\, {}^4V^{\mu}={}^4V^{\mu}{}_{;
\nu}\equiv [{}^4\nabla_{\nu}\, {}^4V^{(\alpha )}]\, {}^4E^{\mu}_{(\alpha )}=
{}^4V^{(\alpha )}{}_{; \nu}\, {}^4E^{\mu}_{(\alpha )}$ and ${}^4\nabla_{\nu}\,
{}^4\omega_{\mu}={}^4\omega_{\mu ; \nu}\equiv [{}^4\nabla_{\nu}\, {}^4\omega
_{(\alpha )}]\, {}^4E^{(\alpha )}_{\mu}={}^4\omega_{(\alpha ) ; \nu}\, {}^4E
^{(\alpha )}_{\mu}$, so that

\begin{eqnarray}
{}^4V^{\mu}_{; \nu}&=&\partial_{\nu}\, {}^4V^{(\alpha )}\, {}^4E^{\mu}
_{(\alpha )}+{}^4V^{(\alpha )}\, {}^4E^{\mu}_{(\alpha ) ; \nu},\nonumber \\
&&\Rightarrow {}^4V^{(\alpha )}{}_{; \nu}=\partial_{\nu}\, {}^4V^{(\alpha )}+
{}^4\omega^{(\alpha )}_{\nu (\beta )}\, {}^4V^{(\beta )},\nonumber \\
{}^4\omega_{\mu ; \nu}&=&\partial_{\nu}\, {}^4\omega_{(\alpha )}\, {}^4E
^{(\alpha )}_{\mu}+{}^4\omega_{(\alpha )}\, {}^4E^{(\alpha )}_{\mu ; \nu},
\nonumber \\
&&\Rightarrow {}^4\omega_{(\alpha ) ; \nu}=\partial_{\nu}\, {}^4\omega_{(\alpha
)}-{}^4\omega_{(\beta )}\, {}^4\omega^{(\beta )}_{\nu (\alpha )}.
\label{a17}
\end{eqnarray}

Therefore,  for the {\it internal tensors} ${}^4T^{(\alpha
)...}{}_{(\beta )...}$, the spin connection ${}^4\omega^{(\alpha
)}_{\mu (\beta )}$ is a gauge potential associated with a gauge group
SO(3,1). For internal vectors ${}^4V^{(\alpha )}$ at $p\in M^4$ the
cotetrads ${}^4E
^{(\alpha )}_{\mu}$ realize a {\it soldering} of this internal vector space at p
with the tangent space $T_pM^4$: ${}^4V^{(\alpha )}={}^4E^{(\alpha )}_{\mu}\,
{}^4V^{\mu}$. For tensors with mixed world and internal indices, like
tetrads and cotetrads, we could define a generalized covariant derivative
acting on both types of indices ${}^4{\tilde \nabla}_{\nu}\, {}^4E^{\mu}
_{(\alpha )}=\partial_{\nu}\, {}^4E^{\mu}_{(\alpha )}+{}^4\Gamma^{\mu}_{\nu
\rho}\, {}^4E^{\rho}_{(\alpha )}-{}^4E^{\mu}_{(\beta )}\, {}^4\omega^{(\beta )}
_{\nu (\alpha )}$: then ${}^4\nabla_{\nu}\, {}^4V^{\mu}={}^4\nabla_{\nu}\,
{}^4V^{(\alpha )}\, {}^4E^{\mu}_{(\alpha )}+{}^4V^{(\alpha )}\, {}^4{\tilde
\nabla}_{\nu}\, {}^4E^{\mu}_{(\alpha )}\equiv {}^4\nabla_{\nu}\, {}^4V
^{(\alpha )}\, {}^4E^{\mu}_{(\alpha )}$ implies ${}^4{\tilde \nabla}_{\nu}\,
{}^4E^{\mu}_{(\alpha )}=0$ (or ${}^4\nabla_{\nu}\,
{}^3E^{\mu}_{(\alpha )}={}^4E
^{\mu}_{(\beta )}\, {}^4\omega^{(\beta )}_{\nu (\alpha )}$)
which is nothing else that the definition (\ref{a15}) of the spin
connection ${}^4\omega^{(\alpha )}_{\mu (\beta )}$.

We have

\begin{eqnarray}
[ {}^4E_{(\alpha )}, {}^4E_{(\beta )}] &=& c_{(\alpha )(\beta )}{}^{(\gamma )}
{}^4E_{(\gamma )} = {}^4\nabla_{{}^4E_{(\alpha )}}\, {}^4E_{(\beta )} -
{}^4\nabla_{{}^4E_{(\beta )}}\, {}^4E_{(\alpha )} =\nonumber \\
&=& ({}^4\omega^{(\gamma )}_{(\alpha )(\beta )} - {}^4\omega^{(\gamma )}
_{(\beta )(\alpha )})\, {}^4E_{(\gamma )},
\label{a18}
\end{eqnarray}

\begin{eqnarray}
 {}^4\Omega_{\mu\nu}{}^{(\alpha )}{}_{(\beta )}&=&
{}^4E^{(\gamma )}_{\mu}\, {}^4E^{(\delta )}_{\nu}\, {}^4\Omega
^{(\alpha )}{}_{(\beta )(\gamma )(\delta )}={}^4R^{\rho}{}_{\sigma\mu\nu}\,
{}^4E^{(\alpha )}_{\rho}\, {}^4E^{\sigma}_{(\beta )}=\nonumber \\
&=&\partial_{\mu}
{}^4\omega^{(\alpha )}_{\nu (\beta )}-\partial_{\nu}\, {}^4\omega^{(\alpha )}
_{\mu (\beta )} +{}^4\omega^{(\alpha )}_{\mu (\gamma )}\, {}^4\omega^{(\gamma )}
_{\nu (\beta )} - {}^4\omega^{(\alpha )}_{\nu (\gamma )}\, {}^4\omega
^{(\gamma )}_{\mu (\beta )},\nonumber \\
 &&{}\nonumber \\
{}^4R^{\alpha}{}_{\beta\mu\nu}&=&{}^4E^{\alpha}_{(\gamma )}\,
{}^4E^{(\delta )}
_{\beta}\, {}^4\Omega_{\mu\nu}{}^{(\gamma )}{}_{(\delta )}.
\label{a19}
\end{eqnarray}

Let us remark that Eqs.(\ref{a14}) and (\ref{a15}) imply
${}^4\Gamma^{\rho}
_{\mu\nu}={}^4\triangle^{\rho}_{\mu\nu}+{}^4\omega^{\rho}_{\mu\nu}$ with
${}^4\omega^{\rho}_{\mu\nu}={}^4E^{\rho}_{(\alpha )}\, {}^4E^{(\beta )}_{\nu}
\, {}^4\omega^{(\alpha )}_{\mu (\beta )}$ and ${}^4\triangle^{\rho}
_{\mu\nu}={}^4E^{\rho}_{(\alpha )}\, \partial_{\mu}\, {}^4E^{(\alpha )}_{\nu}$;
the Levi-Civita connection (i.e. the Christoffel symbols) turn out to be
decomposed in a flat connection ${}^4\triangle^{\rho}_{\mu\nu}$ (it produces
zero Riemann tensor as was already known to Einstein\cite{dav}) and in a
tensor, like in the Yang-Mills case\cite{lusa}.

\subsection{Triads and Cotriads on $\Sigma_{\tau}$.}

On $\Sigma_{\tau}$ with local coordinate system $\{ \sigma^r \}$ and
Riemannian metric ${}^3g_{rs}$ of signature (+++) we can introduce
orthonormal frames ({\it triads})
${}^3e_{(a)}={}^3e_{(a)}^r{{\partial}\over {\partial \sigma^r}}$,
a=1,2,3, and coframes ({\it cotriads}) ${}^3\theta^{(a)}=
{}^3e^{(a)}_r d\sigma^r$ satisfying

\begin{eqnarray}
&&{}^3e^r_{(a)}\, {}^3g_{rs}\, {}^3e^s_{(b)}=\delta_{(a)(b)},\quad\quad
{}^3e^{(a)}_r\, {}^3g^{rs}\, {}^3e^{(b)}_s=\delta^{(a)(b)},\nonumber \\
&&{}^3e^r_{(a)}\, \delta^{(a)(b)}\, {}^3e^s_{(b)}={}^3g^{rs},\quad\quad
{}^3e_r^{(a)}\, \delta_{(a)(b)}\, {}^3e_s^{(b)}={}^3g_{rs}.
\label{a20}
\end{eqnarray}

\noindent and consider the orthonormal frame bundle $F(\Sigma_{\tau})$
over $\Sigma_{\tau}$ with structure group SO(3). See Ref.\cite{gr13} for
geometrical properties of triads.

The 3-dimensional spin connection 1-form ${}^3\omega^{(a)}_{r(b)}d\sigma^r$ is

\begin{eqnarray}
{}^3\omega^{(a)}_{r(b)}&=&{}^3\omega^{(a)}_{(c)(b)}\, {}^3e^{(c)}_r={}^3e_s
^{(a)}\, {}^3\nabla_r\, {}^3e^s_{(b)}=\nonumber \\
&=&{}^3e^{(a)}_s\, {}^3e^s_{(b) | r}={}^3e^{(a)}_s [\partial_r\, {}^3e^s_{(b)}+
{}^3\Gamma^s_{ru}\, {}^3e^u_{(b)}],\nonumber \\
&&{}\nonumber \\
{}^3\omega_{(a)(b)}&=&\delta_{(a)(c)}\, {}^3\omega^{(c)}_{r(b)} d\sigma^r=-
{}^3\omega_{(b)(a)},\quad\quad
{}^3\omega_{r(a)}={1\over 2}\epsilon_{(a)(b)(c)}\, {}^3\omega_{r(b)(c)},
\nonumber \\
{}^3\omega_{r(a)(b)}&=&\epsilon_{(a)(b)(c)}\, {}^3\omega_{r(c)}=[{\hat R}^{(c)}
{}^3\omega_{r(c)}]_{(a)(b)}=[{}^3\omega_r]_{(a)(b)},\nonumber \\
&&{}\nonumber \\
&&[{}^3e_{(a)},
{}^3e_{(b)}]=({}^3\omega^{(c)}_{(a)(b)}-{}^3\omega^{(c)}_{(b)(a)}){}^3e_{(c)},
\label{a21}
\end{eqnarray}

\noindent where $\epsilon_{(a)(b)(c)}$ is the standard Euclidean antisymmetric
tensor and $({\hat R}^{(c)})_{(a)(b)}=\epsilon_{(a)(b)(c)}$ is the adjoint
representation of SO(3) generators.

Given vectors and covectors ${}^3V^r={}^3V^{(a)}\, {}^3e^r_{(a)}$,
${}^3V_r= {}^3V_{(a)}\, {}^3e^{(a)}_r$, we have \footnote{Remember
that ${}^3\nabla_s\, {}^3e^r
_{(a)}={}^3e^r_{(b)}\, {}^3\omega^{(b)}_{s (a)}$.}

\begin{eqnarray}
{}^3\nabla_s\, &{}^3V^r&={}^3V^r{}_{| s}\equiv {}^3V^{(a)}_{| s}\,
{}^3e^r_{(a)},\nonumber \\
&&\Rightarrow {}^3V^{(a)}{}_{| s}=\partial_s\, {}^3V^{(a)}+
{}^3\omega^{(a)}_{s(b)}
\, {}^3V^{(b)}=\partial_s\, {}^3V^{(a)}+\delta^{(a)(c)}\epsilon_{(c)(b)(d)}{}^3
\omega_{s(d)}{}^3V^{(b)},\nonumber \\
{}^3\nabla_s\, &{}^3V_r&={}^3V_{r | s}={}^3V_{(a) | s} {}^3e^{(a)}_r,
\nonumber \\
&&\Rightarrow {}^3V_{(a) | s}=\partial_s\, {}^3V_{(a)}-{}^3V_{(b)}\, {}^3\omega
^{(b)}_{s(a)}=\partial_s\, {}^3V_{(a)}-{}^3V_{(b)}\delta^{(b)(c)}\,
\epsilon_{(c)(a)(d)}{}^3\omega_{s(d)}.
\label{a22}
\end{eqnarray}

For the field strength and the curvature tensors we have

\begin{eqnarray}
{}^3\Omega^{(a)}{}_{(b)(c)(d)}&=&{}^3e_{(c)}({}^3\omega^{(a)}_{(d)(b)})-{}^3e
_{(d)}({}^3\omega^{(a)}_{(c)(b)})+\nonumber \\
&+&{}^3\omega^{(n)}_{(d)(b)}\, {}^3\omega^{(a)}_{(c)(n)}-{}^3\omega^{(n)}
_{(c)(b)}\, {}^3\omega^{(a)}_{(d)(n)}-({}^3\omega^{(n)}_{(c)(d)}-{}^3\omega
^{(n)}_{(d)(c)}){}^3\omega^{(a)}_{(a)(b)}=\nonumber \\
&=&{}^3e^{(a)}_r\, {}^3R^r{}_{stw}\, {}^3e^s_{(b)}\, {}^3e^t_{(c)}\, {}^3e^w
_{(d)},\nonumber \\
&&{}\nonumber \\
{}^3\Omega_{rs}{}^{(a)}{}_{(b)}&=&{}^3e^{(c)}_r\, {}^3e^{(d)}_s\, {}^3\Omega
^{(a)}{}_{(b)(c)(d)}={}^3R^t{}_{wrs}\, {}^3e_t^{(a)}\, {}^3e^w_{(b)}=
\nonumber \\
&=&\partial_r\, {}^3\omega^{(a)}_{s(b)} -\partial_s\, {}^3\omega^{(a)}_{r(b)}
+{}^3\omega^{(a)}_{r(c)}\, {}^3\omega^{(c)}_{s(b)} -{}^3\omega^{(a)}_{s(c)}\,
{}^3\omega^{(c)}_{r(b)}=\nonumber \\
&=&\delta^{(a)(c)}\, {}^3\Omega_{rs(c)(b)}=\delta^{(a)(c)}\, \epsilon_{(c)(b)
(d)}\, {}^3\Omega_{rs(d)},\nonumber \\
&&{}\nonumber \\
{}^3\Omega_{rs(a)}&=&{1\over 2}\epsilon_{(a)(b)(c)}\, {}^3\Omega_{rs(b)(c)}=
\partial_r\, {}^3\omega_{s(a)}-\partial_s\, {}^3\omega_{r(a)} -\epsilon
_{(a)(b)(c)}\, {}^3\omega_{r(b)}\, {}^3\omega_{s(c)},\nonumber \\
&&{}\nonumber \\
{}^3R^r{}_{stw}&=&
\epsilon_{(a)(b)(c)}\, {}^3e^r_{(a)}\, \delta_{(b)(n)}\,
{}^3e^{(n)}_s\, {}^3\Omega_{tw(c)},\nonumber \\
{}^3R_{rs}&=&
\epsilon_{(a)(b)(c)}\, {}^3e^u_{(a)}\, \delta_{(b)(n)}\, {}^3e
^{(n)}_r\, {}^3\Omega_{us(c)},\nonumber \\
{}^3R&=&
\epsilon_{(a)(b)(c)}\, {}^3e^r_{(a)}\, {}^3e^s_{(b)}\, {}^3\Omega
_{rs(c)}.
\label{a23}
\end{eqnarray}

The first Bianchi identity (\ref{a2})
${}^3R^t{}_{rsu}+{}^3R^t{}_{sur}+{}^3R^t{}
_{urs}\equiv 0$ implies the cyclic identity ${}^3\Omega_{rs(a)}\, {}^3e^s_{(a)}
\equiv 0$.

Under local SO(3) rotations R [$R^{-1}=R^t$] we have

\begin{eqnarray}
{}^3\omega^{(a)}_{r(b)} &\mapsto& [R\, {}^3\omega_r\, R^T-R \partial_r\, R^T]
^{(a)}{}_{(b)},\nonumber \\
{}^3\Omega_{rs}{}^{(a)}{}_{(b)} &\mapsto& [R\, {}^3\Omega_{rs}\, R^T]^{(a)}
{}_{(b)}.
\label{a24}
\end{eqnarray}

Since the flat metric $\delta_{(a)(b)}$ has signature (+++), we have
${}^3V^{(a)}=\delta^{(a)(b)}\, {}^3V_{(b)}={}^3V_{(a)}$ and one can
simplify the notations by using only lower (a) indices:
${}^3e^{(a)}_r={}^3e_{(a)r}$. For instance, we have

\begin{eqnarray}
{}^3\Gamma^u_{rs}&=&{}^3\Gamma^u_{sr}=
{1\over 2}\, {}^3e^u_{(a)} \Big[ \partial_r\, {}^3e_{(a)s}+
\partial_s\, {}^3e_{(a)r}+\nonumber \\
&+&{}^3e^v_{(a)} \Big( {}^3e_{(b)r}(\partial_s\, {}^3e_{(b)v}-\partial_v\,
{}^3e_{(b)s})+{}^3e_{(b)s}(\partial_r\, {}^3e_{(b)v}-\partial_v\, {}^3e_{(b)r})
\Big) \Big] =\nonumber \\
&=&{1\over 2}\epsilon_{(a)(b)(c)}\, {}^3e^u_{(a)}({}^3e_{(b)r}\, {}^3\omega
_{s(c)}+{}^3e_{(b)s}\, {}^3\omega_{r(c)})-{1\over 2}({}^3e_{(a)r}\partial_s\,
{}^3e^u_{(a)}+{}^3e_{(a)s}\partial_r\, {}^3e^u_{(a)}),\nonumber \\
{}^3\omega_{r(a)(b)}&=&-{}^3\omega_{r(b)(a)}=
{1\over 2}\Big[ {}^3e^s_{(a)}(\partial_r\, {}^3e_{(b)s}-
\partial_s\, {}^3e_{(b)r})+\nonumber \\
&+&{}^3e^s_{(b)}(\partial_s\, {}^3e_{(a)r}-\partial_r\, {}^3e_{(a)s})+{}^3e^u
_{(a)}\, {}^3e^v_{(b)}\, {}^3e_{(c)r}(\partial_v\, {}^3e_{(c)u}-\partial_u\,
{}^3e_{(c)v}) \Big] =\nonumber \\
&=&{1\over 2} \Big[ {}^3e_{(a)u} \partial_r\, {}^3e^u_{(b)}-{}^3e_{(b)u}
\partial_r\, {}^3e^u_{(a)}+{}^3\Gamma^u_{rs} ({}^3e_{(a)u}\, {}^3e^s_{(b)}-
{}^3e_{(b)u}\, {}^3e^s_{(a)})\Big] ,\nonumber \\
{}^3\omega_{r(a)}&=&{1\over 2} \epsilon_{(a)(b)(c)} \Big[ {}^3e^u_{(b)}
(\partial_r\, {}^3e_{(c)u}-\partial_u\, {}^3e_{(c)r})+\nonumber \\
&+&{1\over 2}\, {}^3e^u_{(b)}\, {}^3e^v_{(c)}\, {}^3e_{(d)r}(\partial_v\,
{}^3e_{(d)u}-\partial_u\, {}^3e_{(d)v})\Big] ,\nonumber \\
{}^3\Omega_{rs(a)}&=&{1\over 2} \epsilon_{(a)(b)(c)} \Big[ \partial_r\,
{}^3e^u_{(b)}\partial_s\, {}^3e_{(c)u}-\partial_s\, {}^3e^u_{(b)}\partial_r\,
{}^3e_{(c)u}+\nonumber \\
&+&{}^3e^u_{(b)}(\partial_u\partial_s\, {}^3e_{(c)r}-\partial_u\partial_r\,
{}^3e_{(c)s})+\nonumber \\
&+&{1\over 2}\Big( {}^3e^u_{(b)}\, {}^3e^v_{(c)}(\partial_r\, {}^3e_{(d)s}-
\partial_s\, {}^3e_{(d)r})(\partial_v\, {}^3e_{(d)u}-\partial_u\, {}^3e_{(d)v})
+\nonumber \\
&+&({}^3e_{(d)s}\partial_r-{}^3e_{(d)r}\partial_s)[{}^3e^u_{(b)}\, {}^3e^v
_{(c)}(\partial_v\, {}^3e_{(d)u}-\partial_u\, {}^3e_{(d)v})]\Big) \Big] -
\nonumber \\
&-&{1\over 8}[\delta_{(a)(b_1)}\epsilon_{(c_1)(c_2)(b_2)}+\delta_{(a)(b_2)}
\epsilon_{(c_1)(c_2)(b_1)}+\delta_{(a)(c_1)}\epsilon_{(b_1)(b_2)(c_2)}+\delta
_{(a)(c_2)}\epsilon_{(b_1)(b_2)(c_1)}]\times \nonumber \\
&&{}^3e^{u_1}_{(b_1)}\, {}^3e^{u_2}_{(b_2)}\Big[ (\partial_r\, {}^3e_{(c_1)u_1}
-\partial_{u_1}\, {}^3e_{(c_1)r})(\partial_s\, {}^3e_{(c_2)u_2}-\partial_{u_2}\,
{}^3e_{(c_2)s})+\nonumber \\
&+&{1\over 2}\Big( {}^3e^{v_2}_{(c_2)}\, {}^3e_{(d)s}(\partial_r\, {}^3e
_{(c_1)u_1}-\partial_{u_1}\, {}^3e_{(c_1)r})(\partial_{v_2}\, {}^3e_{(d)u_2}-
\partial_{u_2}\, {}^3e_{(d)v_2})+\nonumber \\
&+&{}^3e^{v_1}_{(c_1)}\, {}^3e_{(d)r}(\partial_s\, {}^3e_{(c_2)u_2}-\partial
_{u_2}\, {}^3e_{(c_2)s})(\partial_{v_1}\, {}^3e_{(d)u_1}-\partial_{u_1}\,
{}^3e_{(d)v_1})\Big) +\nonumber \\
&+&{1\over 4}\, {}^3e^{v_1}_{(c_1)}\, {}^3e^{v_2}_{(c_2)}\, {}^3e_{(d_1)r}\,
{}^3e_{(d_2)s}(\partial_{v_1}\, {}^3e_{(d_1)u_1}-\partial_{u_1}\, {}^3e
_{(d_1)v_1})(\partial_{v_2}\, {}^3e_{(d_2)u_2}-\partial_{u_2}\, {}^3e
_{(d_2)v_2}) \Big] ,\nonumber \\
{}^3\Omega_{rs(a)(b)}&=&\epsilon_{(a)(b)(c)}\, {}^3\Omega_{rs(c)},\nonumber \\
{}^3R_{rsuv}&=&\epsilon_{(a)(b)(c)}\, {}^3e_{(a)r}\, {}^3e_{(b)s}\, {}^3\Omega
_{uv(c)},\nonumber \\
{}^3R_{rs}&=&{1\over 2} \epsilon_{(a)(b)(c)}\, {}^3e^u_{(a)} \Big[ {}^3e_{(b)r}
\, {}^3\Omega_{us(c)}+{}^3e_{(b)s}\, {}^3\Omega_{ur(c)}\Big] ,\nonumber \\
{}^3R&=&\epsilon_{(a)(b)(c)}\, {}^3e^r_{(a)}\, {}^3e^s_{(b)}\, {}^3\Omega
_{rs(c)}.
\label{a25}
\end{eqnarray}

\subsection{Action Principles.}

Let us finish this Appendix with a review of some action principles
used for general relativity. In {\it metric gravity}, one uses the
generally covariant {\it Hilbert action}  depending on the 4-metric
and its first and second derivatives \footnote{G is Newton
gravitational constant; $U\subset M^4$ is a subset of spacetime; we
use units with $x^o=ct$.}

\begin{equation}
S_H={{c^3}\over {16\pi G}}\, \int_U\, d^4x\, \sqrt{{}^4g}\, {}^4R= \int_U
d^4x\, {\cal L}_H.
\label{a26}
\end{equation}

\noindent The variation of $S_H$ is ($d^3\Sigma_{\gamma}=d^3\Sigma l_{\gamma}$)

\begin{eqnarray}
&&\delta S_H = \delta S_E +\Sigma_H = -{{c^3}\over {16\pi G}} \int_U
d^4x\, \sqrt{{}^4g}\, {}^4G^{\mu\nu} \delta {}^4g_{\mu\nu} +\Sigma_H,
\nonumber \\
&&{}\nonumber \\
&&\Sigma_H ={{c^3}\over {16\pi G}} \int_{\partial U} {d^3\Sigma}_{\gamma}
\sqrt{{}^4g}\, ({}^4g^{\mu\nu} \delta^{\gamma}_{\delta} - {}^4g^{\mu\gamma}
\delta^{\nu}_{\delta}) \delta \, {}^4\Gamma^{\delta}_{\mu\nu}=\nonumber \\
&&={{c^3}\over {8\pi G}} \int_{\partial U} d^3\Sigma \, \sqrt{{}^3\gamma}\,\,
\delta \, {}^3K,\nonumber \\
&&{}\nonumber \\
&&\delta {}^4\Gamma^{\delta}_{\mu\nu} ={1\over 2} {}^4g^{\delta\beta}
[{}^4\nabla_{\mu} \delta \, {}^4g_{\beta\nu} +{}^4\nabla_{\nu} \delta \,
{}^4g_{\beta\mu} -{}^4\nabla_{\beta} \delta \, {}^4g_{\mu\nu}].
\label{a27}
\end{eqnarray}

\noindent where ${}^3\gamma_{\mu\nu}$ is the metric induced on $\partial U$ and
$l_{\mu}$ is the outer unit covariant normal to $\partial U$. The trace of the
extrinsic curvature ${}^3K_{\mu\nu}$ of $\partial U$ is ${}^3K=-l^{\mu}
{}_{;\mu}$.
The surface term $\Sigma_H$ takes care of the second derivatives of the 4-metric
and to get Einstein equations ${}^4G_{\mu\nu}={}^4R_{\mu\nu}-
{1\over 2}{}^4g_{\mu\nu}\, {}^4R \, {\buildrel \circ \over =}\, 0$
one must take constant certain normal
derivatives of the 4-metric on the boundary of $U$ [${\cal L}_l\, ({}^4g
_{\mu\nu}-l_{\mu}l_{\nu})=0$] to have $\delta S_H=0$ \cite{yo}.

The term $\delta S_E$ in Eq.(\ref{a25}) means the variation of the
action $S_E$, which is the (not generally covariant) {\it Einstein
action} depending only on the 4-metric and its first derivatives
\footnote{$\delta S_E=0$ gives ${}^4G_{\mu\nu}\, {\buildrel \circ \over
=}\, 0$ if ${}^4g
_{\mu\nu}$ is held fixed on $\partial U$.}

\begin{eqnarray}
S_E&=& \int_U d^4x\, {\cal L}_E=
{{c^3}\over {16\pi G}}\, \int_U\, d^4x\, \sqrt{{}^4g}\,
{}^4g^{\mu\nu}({}^4\Gamma^{\rho}_{\nu\lambda}\, {}^4\Gamma^{\lambda}_{\rho\mu}
- {}^4\Gamma^{\lambda}_{\lambda\rho}\, {}^4\Gamma^{\rho}_{\mu\nu})=\nonumber \\
&=&S_H-{{c^3}\over {16\pi G}} \int_Ud^4x\, \partial_{\lambda} [\sqrt{{}^4g}
({}^4g^{\mu\nu}\, {}^4\Gamma^{\lambda}_{\mu\nu}-{}^4g^{\lambda\mu}\, {}^4\Gamma
^{\rho}_{\rho\mu})],\nonumber \\
{}&&\nonumber \\
\delta S_E&=&{{c^3}\over {16\pi G}} \int_U d^4x\, ({{\partial {\cal L}_E}\over
{\partial \, {}^4g^{\mu\nu}}}-\partial_{\rho}\, {{\partial {\cal L}_E}\over
{\partial \partial_{\rho}\, {}^4g^{\mu\nu}}})\, \delta \, {}^4g^{\mu\nu}=
-{{c^3}\over {16\pi G}} \int_U d^4x\, \sqrt{{}^4g}\, {}^4G_{\mu\nu} \delta
\, {}^4g^{\mu\nu}.
\label{a28}
\end{eqnarray}

We shall not consider the first-order {\it Palatini action}; see for
instance Ref.\cite{ro}, where there is also  a review of the
variational principles of the connection-dependent formulations of
general relativity.

In Ref.\cite{yo} (see also Ref.\cite{in}), it is shown that the {\it
DeWitt-ADM action}\cite{witt,adm} for a 3+1 decomposition of $M^4$ can
be obtained from $S_H$ in the following way
\footnote{$\sqrt{{}^4g} {}^4R
=-\epsilon \sqrt{{}^4g}({}^3R+{}^3K_{\mu\nu}{}^3K^{\mu\nu}-
({}^3K)^2)-2\epsilon \,
\partial_{\lambda}(\sqrt{{}^4g} ({}^3K l^{\lambda}+a^{\lambda}))$,
with $a^{\lambda}$ the 4-acceleration ($l^{\mu}a_{\mu}=0$); the
4-volume U is $[\tau_f,\tau_i]\times S$.}

\begin{eqnarray}
&&S_H = S_{ADM}+\Sigma_{ADM}, \nonumber \\
&&S_{ADM} = -\epsilon
{{c^3}\over {16\pi G}} \int_{U} d^4x\, \sqrt{{}^4g}
[{}^3R+{}^3K_{\mu\nu}\, {}^3K^{\mu\nu} -({}^3K)^2],\nonumber \\
&&\Sigma_{ADM}=-\epsilon {{c^3}\over {8\pi G}} \int d^4x \partial_{\alpha}
[\sqrt{{}^4g} ({}^3K l^{\alpha} +l^{\beta} l^{\alpha}{}_{;\beta})]=\nonumber \\
&&=-\epsilon {{c^3}\over {8\pi G}} \Big[
\int_S d^3\sigma \, [\sqrt{\gamma}\,\, {}^3K](\tau ,\vec \sigma ){|}^{\tau_f}
_{\tau_i}+\nonumber \\
&&+\int^{\tau_f}_{\tau_i}d\tau \int_{\partial S}d^2\Sigma^r [{}^3\nabla_r
(\sqrt{\gamma}N)- {}^3K N_r](\tau ,\vec \sigma )\Big] ,\nonumber \\
&&{}\nonumber \\
&&\delta S_{ADM}=-\epsilon {{c^3}\over {16\pi G}} \int d\tau d^3\sigma
\sqrt{\gamma}\Big[ 2\, {}^4{\bar G}_{ll} \delta N+{}^4{\bar G}_l{}^r\delta N_r-
{}^4{\bar G}^{rs} \delta \, {}^3g_{rs}\Big] (\tau ,\vec \sigma )+\nonumber \\
&&+\delta S_{ADM} {|}_{{}^4G_{\mu\nu}=0}-\epsilon \int^{\tau_f}_{\tau_i}d\tau
\int_{\partial U}d^3\Sigma_r [N_{|s}\delta \, {}^3g^{rs}-N\delta \, {}^3g
^{rs}{}_{|s}](\tau ,\vec \sigma ),\nonumber \\
&&\delta S_{ADM} {|}_{{}^4G_{\mu\nu}=0} =-\epsilon
 {{c^3}\over {16\pi G}} \int_{\partial
U} d^3\sigma \, {}^3{\tilde \Pi}^{\mu\nu} \delta {}^3\gamma_{\mu\nu},\nonumber\\
&& {}^3{\tilde \Pi}^{\mu\nu}=\sqrt{\gamma}
({}^3K^{\mu\nu}-{}^3g^{\mu\nu}\, {}^3K)=
{{16\pi G}\over {c^3}} \epsilon {\hat b}^{\mu}_r{\hat b}^{\nu}_s\,
{}^3{\tilde \Pi}^{rs},
\label{a29}
\end{eqnarray}

\noindent so that $\delta S_{ADM}=0$ gives ${}^4G_{\mu\nu}\, {\buildrel \circ
\over =}\, 0$ if one holds
fixed the intrinsic 3-metric ${}^3\gamma_{\mu\nu}$ on the boundary
\footnote{${}^3{\tilde  \Pi}^{\mu\nu}$ is the ADM momentum with world indices,
whose form in a 3+1 splitting is given in Section IV.}. This action is
not generally covariant, but it is quasi-invariant under the 8 types
of gauge transformations generated by the ADM first class constraints
(see Appendix A of Ref.\cite{russo3}). As shown in
Refs.\cite{dew,reg,yo,hh} in this way one obtains a well defined
gravitational energy. However, in so doing one still neglects some
boundary terms. Following Ref.\cite{hh}, let us assume that, given a
subset $U\subset M^4$ of spacetime, $\partial U$ consists of two
slices, $\Sigma_{\tau_i}$ (the initial one) and $\Sigma_{\tau_f}$ (the
final one) with outer normals $-l^{\mu}(\tau_i,\vec \sigma )$ and
$l^{\mu}(\tau_f,
\vec \sigma )$ respectively, and of a surface $S_{\infty}$ near space infinity
with outer unit (spacelike) normal $n^{\mu}(\tau ,\vec \sigma )$
tangent to the slices \footnote{So that the normal $l^{\mu}(\tau ,\vec
\sigma )$ to every slice is asymptotically tangent to $S_{\infty}$.}.
The 3-surface $S_{\infty}$ is foliated by a family of 2-surfaces
$S^2_{\tau ,
\infty}$ coming from its intersection with the slices $\Sigma_{\tau}$
\footnote{Therefore, asymptotically $l^{\mu}(\tau ,\vec \sigma )$ is normal to
the corresponding $S^2_{\tau ,\infty}$.}. The vector
$b^{\mu}_{\tau}=z^{\mu}
_{\tau}=   N l^{\mu}+N^rb^{\mu}_r$ is not in general tangent to $S
_{\infty}$. It is assumed that there are no inner boundaries (see Ref.
\cite{hh} for their treatment), so that the slices $\Sigma_{\tau}$ do
not intersect and are complete. This does not rule out the existence
of horizons, but it implies that, if horizons form, one continues to
evolve the spacetime inside the horizon as well as outside. Then, in
Ref.\cite{hh} it is shown that one gets \footnote{${}^2K$ the trace of
the 2-dimensional extrinsic curvature of the 2-surface $S^2_{\tau
,\infty}= S_{\infty}\cap \Sigma_{\tau}$; to get this result one
assumes that the lapse function $N(\tau ,\vec \sigma )$ on
$\Sigma_{\tau}$ tends asymptotically to a function $N_{(as)}(\tau
)\,\,$ and that the term on $\partial S$ vanishes due to the boundary
conditions.}

\begin{eqnarray}
\Sigma_{ADM}&=&-\epsilon {{c^3}\over {8\pi G}} [\int_{\Sigma_{\tau_f}} d^3\Sigma
-\int_{\Sigma_{\tau_i}} d^3\Sigma ]\, N\, \sqrt{\gamma} \, {}^3K=\nonumber \\
&=&-\epsilon
{{c^3}\over {8\pi G}} \int_{\tau_i}^{\tau_f} d\tau \, N_{(as)}(\tau ) \int
_{S^2_{\tau ,\infty}} d^2\Sigma \, \sqrt{\gamma}\, {}^2K.
\label{a30}
\end{eqnarray}

In Einstein metric gravity the gravitational field, described by the
4-metric ${}^4g_{\mu\nu}$ depends on 2, and not 10, physical degrees
of freedom in each point; this is not explicitly evident if one starts
with the Hilbert action, which is invariant under $Diff\, M^4$, a
group with only four generators. Instead in ADM canonical gravity (see
Section IV) there are in each point 20 canonical variables and 8 first
class constraints, implying the determination of 8 canonical variables
and the arbitrariness of the 8 conjugate ones. At the Lagrangian
level, only 6 of the ten Einstein equations are independent, due to
the contracted Bianchi identities, so that four components of the
metric tensor ${}^4g_{\mu\nu}$ (the lapse and shift functions) are
arbitrary not being determined by the equation of motion. Moreover,
the four combinations ${}^4{\bar G}_{ll}\, {\buildrel \circ \over =}\,
0$, ${}^4{\bar G}_{lr}\, {\buildrel \circ \over
=}\, 0$, of the Einstein equations do not depend on the second time derivatives
or accelerations (they
are restrictions on the Cauchy data and become the secondary first class
constraints of the ADM canonical theory): the general theory \cite{sha} implies
that four generalized velocities (and therefore other four components of the
metric) inherit the arbitrariness of the lapse and shift functions. Only two
combinations of the Einstein equations depend on the accelerations
(second time derivatives) of the
two (non tensorial) independent degrees of freedom of the gravitational field
and are
genuine equations of motion. Therefore, the ten components of every 4-metric
${}^4g_{\mu\nu}$, compatible with the Cauchy data, depend on 8 arbitrary
functions not determined by the Einstein equations.

Instead, in {\it tetrad
gravity}\cite{weyl,dirr,schw,kib,tetr,char,maluf,hen1,hen2,hen3,hen4},
in which ${}^4g_{\mu\nu}$ is no more the independent variable, the new
independent 16 variables are a set of cotetrads ${}^4E^{(\alpha )}
_{\mu}$ so that ${}^4g_{\mu\nu}={}^4E^{(\alpha )}_{\mu}\, {}^4\eta_{(\alpha )
(\beta )}\, {}^4E^{(\beta )}_{\nu}$. Tetrad gravity
has not only the invariance under
$Diff\, M^4$ but also under local Lorentz transformations on $TM^4$ [acting
on the flat indices $(\alpha )$]. An action principle with these local
invariances is obtained by replacing the 4-metric in the Hilbert action
$S_H$ with its expression in terms of the cotetrads. The action acquires the
form

\begin{equation}
S_{HT}={{c^3}\over {16\pi G}} \int_{U} d^4x\, {}^4\tilde E\, {}^4E^{\mu}
_{(\alpha )}\, {}^4E^{\nu}_{(\beta )}\, {}^4\Omega_{\mu\nu}{}^{(\alpha )
(\beta )},
\label{a31}
\end{equation}

\noindent where ${}^4\tilde E=det\, ({}^4E_{\mu}^{(\alpha )})=\sqrt{{}^4g}$
and ${}^4\Omega_{\mu\nu}{}^{(\alpha )(\beta )}$ is the spin 4-field strength
. One has

\begin{eqnarray}
\delta S_{HT}&=& {{c^3}\over {16\pi G}} \int_U d^4x\, {}^4\tilde E\,
{}^4G_{\mu\nu}\, {}^4E^{\mu}_{(\alpha )}\, {}^4\eta^{(\alpha )(\beta )}\,
\delta \, {}^4E^{\nu}_{(\beta )}+\nonumber \\
&+&{{c^3}\over {8\pi G}} \int_U d^4x\, \partial_{\mu} [{}^4\tilde E\,
({}^4E^{(\rho )}_{\nu} \delta ({}^4g^{\mu\lambda}\, {}^4\nabla_{\lambda}
{}^4E^{\nu}_{(\rho )})-{}^4\eta^{(\rho )(\sigma )}\, {}^4E^{\nu}_{(\rho )}
\delta ({}^4\nabla_{\nu}\, {}^4E^{\mu}_{(\sigma )}))].
\label{a32}
\end{eqnarray}

Again $\delta S_{HT}=0$ produces Einstein equations if complicated derivatives
of the tetrads vanish at the boundary.

Tetrad gravity with action $S_{HT}$, in which the elementary natural
Lagrangian object is the soldering or canonical one-form (or
orthogonal coframe) $\theta^{(\alpha )}={}^4E^{(\alpha )}_{\mu}
dx^{\mu}$, is gauge invariant simultaneously under diffeomorphisms
($Diff\, M^4$) and Lorentz transformations [SO(3,1)]. Instead in phase
space (see Section III)  only two of the 16 components of the cotetrad
${}^4E^{(\alpha )}_{\mu}(x)$ are physical degrees of freedom in each
point, since  the 32 canonical variables present in each point are
restricted by 14 first class constraints, so that the 16 components of
a cotetrad compatible with the Cauchy data depend on 14 arbitrary
functions not determined by the equation of motion.

In Ref.\cite{char}, by using ${}^4\tilde E {}^4E^{\mu}_{(\alpha )}\,
{}^4E^{\nu}_{(\beta )}\, {}^4\Omega_{\mu\nu} {}^{(\alpha )(\beta
)}=2\, {}^4\tilde E\, {}^4E^{\mu}_{(\alpha )}\, {}^4E^{\nu}_{(\beta )}
[{}^4\omega_{\mu}\, {}^4\omega_{\nu} - {}^4\omega_{\nu}
{}^4\omega_{\mu}]^{(\alpha )(\beta )}+2\, \partial_{\mu}({}^4\tilde
E\, {}^4E^{\mu}_{(\alpha )}\, {}^4E^{\nu}_{(\beta )}\,
{}^4\omega_{\nu}{}^{(\alpha )(\beta )})$, the analogue of $S_E$, i.e.
the (not locally Lorentz invariant, therefore not expressible only in
terms of the 4-metric) {\it Charap action}, is defined as

\begin{equation}
S_C=-{{c^3}\over {8\pi G}}\int_U d^4x\, {}^4\tilde E\, {}^4E^{\mu}
_{(\alpha )}\, {}^4E^{\nu}_{(\beta )} ({}^4\omega_{\mu}\, {}^4\omega_{\nu}-
{}^4\omega_{\nu}\, {}^4\omega_{\mu})^{(\alpha )(\beta )}.
\label{a33}
\end{equation}

\noindent Its variation $\delta S_C$ vanishes if $\delta \, {}^4E^{\mu}
_{(\alpha )}$ vanish at the boundary and the Einstein equations hold. However
its Hamiltonian formulation gives too complicated first class constraints
to be solved.

Instead in Refs.\cite{hen1,hen2,hen3,hen4} it was {\it implicitly}
used the metric ADM action $S_{ADM}[{}^4g_{\mu\nu}]$ with the metric
expressed in terms of cotetrads in the Schwinger time gauge\cite{schw}
as independent Lagrangian variables $S_{ADMT} [{}^4E^{(\alpha
)}_{\mu}]$. This is the action we shall study in this paper after
having expressed arbitrary cotetrads in terms of
$\Sigma_{\tau}$-adapted ones in the next Section.

\vfill\eject

\end{document}